\documentclass[iop,usenatbib]{emulateapj}

\usepackage[english]{babel}
\usepackage{graphicx}
\usepackage{fleqn}
\usepackage{amssymb}
\usepackage{amsmath}
\usepackage{booktabs}
\usepackage{threeparttable}
\usepackage[breaklinks,colorlinks,citecolor=blue]{hyperref}

\usepackage{txfonts} 

\renewcommand{\S}{Section}
\newcommand{\F}{Fig.}

\newcommand{\ve}[1]{\boldsymbol{#1}}
\newcommand{\unit}[1]{\hat{\boldsymbol{#1}}}

\newcommand{\Ec}{\mathcal{E}} 
\newcommand{\Rc}{\mathcal{R}}

\newcommand{\msun}{\mathrm{M}_\odot}

\newcommand{\complexi}{\mathrm{i}}

\usepackage{etoolbox}
\newtoggle{ApJFigs}
\togglefalse{ApJFigs}

\begin{document}

\title{Relaxation near supermassive black holes driven by nuclear spiral arms: anisotropic hypervelocity stars, S-stars and tidal disruption events}
\author{Adrian S. Hamers}
\affil{Institute for Advanced Study, School of Natural Sciences, Einstein Drive, Princeton, NJ 08540, USA}
\email{hamers@ias.edu}

\author{Hagai B. Perets}
\affil{Technion - Israel Institute of Technology, Haifa 32000, Israel}

\begin{abstract} 
Nuclear spiral arms are small-scale transient spiral structures found in the centers of galaxies. Similarly to their galactic-scale counterparts, nuclear spiral arms can perturb the orbits of stars. In the case of the Galactic Center (GC), these perturbations can affect the orbits of stars and binaries in a region extending to several hundred parsecs around the supermassive black hole (MBH), causing diffusion in orbital energy and angular momentum. This diffusion process can drive stars and binaries to close approaches with the MBH, disrupting single stars in tidal disruption events (TDEs), or disrupting binaries, leaving a star tightly bound to the MBH, and an unbound star escaping the galaxy, i.e., a hypervelocity star (HVS). Here, we consider diffusion by nuclear spiral arms in galactic nuclei, specifying to the Milky Way GC. We determine nuclear spiral arm-driven diffusion rates using test-particle integrations, and compute disruption rates. Our TDE rates are up to $20\%$ higher compared to relaxation by single stars. For binaries, the enhancement is up to a factor of $\sim 100$, and our rates are comparable to the observed numbers of HVSs and S-stars. Our scenario is complementary to relaxation driven by massive perturbers. In addition, our rates depend on the inclination of the binary with respect to the Galactic plane. Therefore, our scenario provides a novel potential source for the observed anisotropic distribution of HVSs. Nuclear spiral arms may also be important for accelerating the coalescence of binary MBHs, and for supplying nuclear star clusters with stars and gas.
\end{abstract}


\section{Introduction}
\label{sect:introduction}
The Galactic Center (GC) is well known for harboring Sgr A*, a supermassive black hole (MBH) of $M_\bullet = 4 \times 10^6 \, \msun$ \citep{2005ApJ...628..246E,2008ApJ...689.1044G,2009ApJ...692.1075G,2016ApJ...830...17B}. The GC region lies in the central region of the Galactic Bulge ($\sim 0.3-3$ kpc from the GC), which consists mainly of a population of old ($\sim 10\,\mathrm{Gyr}$) stars. The inner 300 pc are known as the Nuclear Bulge (e.g., \citealt{1996A&ARv...7..289M}), which, in addition to old stars, also contains young stars in dense environments \citep{1999A&A...348..768P,1999A&A...348..457M,2004ApJ...601..319F}. In particular, there are two massive clusters, the young ($\sim 5$ Myr old) Quintuplet cluster ($\sim 10^4 \, \msun$) at $\sim 30$ pc from the center (e.g., \citealt{1999ApJ...525..750F,1999ApJ...514..202F}) and the even younger Arches cluster ($\sim 10^4\,\msun$) at $\sim 30$ pc from the center (e.g., \citealt{1999ApJ...525..750F}). Furthermore, the central parsec contains, in addition to old stars, a population of several hundred young O and B stars (e.g., \citealt{2014A&A...566A..47S}; see \citealt{2005PhR...419...65A,2010RvMP...82.3121G} for reviews).

These young stellar populations are indicative of recent star formation several million years ago, and this is consistent with the presence of a large supply of molecular gas in the inner $\sim 200 \,\mathrm{pc}$ (the central molecular zone; \citealt{1996Natur.382..602S}). On scales of several hundred pc, gas can be perturbed by torques from nuclear bars (e.g., \citealt{1989Natur.338...45S,2004ApJ...617L.115E}), driving inflows to the central regions \citep{2000ApJ...528..677E}. Simulations show that these inflows can result in transient features in the gas density, closely resembling spiral arms \citep{2000ApJ...528..677E,2002MNRAS.329..502M,2004MNRAS.354..892M,2005ApJ...620..197A,2009ApJ...691.1525N,2009ApJ...693..586T,2012ApJ...747...60K,2015ApJ...806..150L,2017arXiv170403665R}. These nuclear spiral arms are indeed observed in 50 to 80 per cent of both active and quiescent galaxies \citep{1999MNRAS.302L..33L,1999AJ....117.2676R,1999AJ....118.2646M,2002ApJ...569..624P,2003ApJS..146..353M,2003ApJ...589..774M}. In particular, for NGC 1097 \citep{2006ApJ...641L..25F,2009ApJ...702..114D,2010ApJ...723..767V} and NGC 6951 \citep{2007ApJ...670..959S}, streaming motions along the nuclear spiral arms have been mapped. 

Nuclear spiral arms can be considered as small-scale variants of galactic-scale spiral arms. Whereas galactic-scale spiral arms consist of both stars and gas, nuclear spiral arms are observed to consist of gas only (e.g., \citealt{2006ApJ...641L..25F,2009ApJ...702..114D,2010ApJ...723..767V,2007ApJ...670..959S}). Galactic-scale spiral arms can perturb stars orbiting in the Galaxy (\citealt{1979ApJ...233..857G}; \citealt[S6.2.6]{2008gady.book.....B}). If the spiral-structure features are transient, then these perturbations can drive heating of stars in the Galactic disk, i.e., increasing the velocity dispersion in the radial and transverse directions, while not affecting velocities in the vertical direction \citep{1967ApJ...150..461B}. This has been invoked \citep{1984ApJ...282...61S,2004MNRAS.350..627D} to explain the observed age-velocity-dispersion relation in the Galactic disk and the ratio of the vertical-to-radial velocity dispersion, although a combination with heating by giant molecular clouds (GMCs; \citealt{1951ApJ...114..385S,1953ApJ...118..106S}) is likely required to best match the observations \citep{1987ApJ...322...59C,1990MNRAS.245..305J,1992MNRAS.257..620J}. 

Similarly, nuclear-scale spiral arms can perturb the orbits of stars in certain regions in the central few hundred pc around the MBH. In particular, nuclear spiral structures can drive stars onto orbits bringing them very close to the MBH, triggering strong interactions. 

These interactions include the disruption of a single star by the tidal force of the MBH resulting in a luminous outburst known as a tidal disruption event (TDE; \citealt{1975Natur.254..295H}), or the tidal disruption of a stellar binary \citep{1988Natur.331..687H}. In the latter case, one of the stars in the binary remains bound to the MBH in a close and eccentric orbit, whereas the other star is ejected with high speeds of $\sim 1000 \, \mathrm{km \,s^{-1}}$. This scenario can explain both the S-stars in the GC, orbiting the MBH in close and eccentric orbits \citep{2003ApJ...594..812G,2005ApJ...628..246E,2008ApJ...689.1044G,2009ApJ...692.1075G}, and hypervelocity stars (HVSs), stars observed to be moving with unusually high velocities and no longer bound to the Galaxy (e.g., \citealt{2005ApJ...622L..33B}; see \citealt{2015ARA&A..53...15B} for a review). HVSs can also be produced by the disruption of clusters of stars such as globular clusters \citep{2016MNRAS.456.2457A,2015MNRAS.454.2677C} and young clusters like the Arches cluster \citep{2017MNRAS.467..451F}.

A pertinent issue in the disruptions of single stars and binaries by the MBH is how these objects are (continuously) injected to loss cone orbits of the MBH, i.e., to orbits with small pericenter distances (see, e.g., \citealt{2013degn.book.....M}). Such perturbations can come from other stars; the resulting disruption rates for single stars in earlier studies were found to be $\sim10^{-5} \, \mathrm{yr^{-1}}$ \citep{1999MNRAS.306...35S,1999MNRAS.309..447M}, whereas rates of $\sim 10^{-4} \, \mathrm{yr^{-1}}$ are found in more recent calibrated studies \citep{2004ApJ...600..149W,2016MNRAS.455..859S}. For stellar binaries, the rates arising from relaxation by stars are too low to explain the observed number of S-stars and HVSs \citep{2003ApJ...599.1129Y}. However, relaxation can also be driven by massive structures in the GC such as GMCs and massive stellar clusters, which can enhance the rate to be consistent with the observed numbers \citep{2007ApJ...656..709P,2009ApJ...690..795P}. 

Given the likely presence of nuclear spiral arms in the GC and in galactic centers in general, this motivates an investigation of the possibility that nuclear spiral structures within the regions of the central few hundred pc around the central MBHs of galaxies can drive stars and binaries onto loss-cone orbits. This provides a channel, in addition to GMCs and massive stellar clusters, to supply the MBH with stars and binaries on close orbits, and hence produce TDEs, S-star-like stars and HVSs. 

In addition to enhancing event rates, binary disruptions driven by nuclear spiral arms may result in different signatures of HVSs that can help pinpoint their origin. Plausibly, nuclear spiral arm structures are highly flattened and lie within the plane of the Galaxy. This suggests that relaxation driven by nuclear spiral arms is a strong function of the inclination of the orbit of the (barycenter of the) stellar binary with respect to the Galactic disk, and the latter should be reflected in the inclination distribution of both the captured and ejected stars. For the captured stars, relaxation by the surrounding stellar cluster rapidly randomizes the orbital orientation and modifies the eccentricity distribution to be less eccentric; this is consistent with the S-stars \citep{2009ApJ...702..884P,2013ApJ...763L..10A,2014ApJ...794..106A,2014MNRAS.443..355H}. The ejected stars, on the other hand, do not experience orientation randomization, and the inclination dependence of relaxation of binaries by nuclear spiral arms should be reflected in the HVS inclination distribution. The latter is indeed not completely consistent with a random orientation with respect to the GC \citep{2014ApJ...787...89B}.

In this paper, we investigate relaxation driven by nuclear spiral arm structures, and consider the implications for TDEs, S-star-like stars and HVSs. In \S s \,\ref{sect:orbit} and \,\ref{sect:dif}, we carry out test-particle integrations in an assumed potential of the inner region of the GC and transient nuclear spiral arms, and investigate diffusion driven by the latter. These results are used in \S\,\ref{sect:rates} to compute the resulting disruption rates for both single stars and binaries. We discuss our results in \S\,\ref{sect:discussion}, and conclude in \S\,\ref{sect:conclusions}.

\section{Orbits around the GC in the presence of nuclear spiral arms}
\label{sect:orbit}
Our methodology consists of two main parts. First, we integrate orbits in the GC environment considering also the contribution to the potential due to nuclear spiral arms (\S\,\ref{sect:orbit}), and we extract relaxation time-scales and diffusion coefficients from these integrations (\S\,\ref{sect:dif}). Subsequently, we use our results to compute the disruption rates by the MBH for both single and binary stars (\S\,\ref{sect:rates}). 

\subsection{Potential}
\label{sect:orbit:pot}
We model the potential of the inner several hundred pc of the GC with the following components. Throughout \S\,\ref{sect:orbit}, we adopt a cylindrical coordinate system $(R,\phi,z)$ with the GC as the origin; the distance to the GC is given by $r \equiv \sqrt{R^2 + z^2}$.

\subsubsection{MBH}
\label{sect:orbit:pot:mbh}
For the MBH, we adopt the potential
\begin{align}
\label{eq:phi_MBH}
\Phi_\bullet(R,\phi,z) = - \frac{GM_\bullet}{\sqrt{R^2+z^2 + \epsilon_\mathrm{soft}^2}},
\end{align}
where $G$ is the gravitational constant, and with an MBH mass $M_\bullet = 4 \times 10^6 \, \msun$ adopted from the MBH in the Milky Way GC \citep{2005ApJ...628..246E,2008ApJ...689.1044G,2009ApJ...692.1075G,2016ApJ...830...17B}. For numerical reasons, the MBH potential is softened with a (conservative) softening length of $\epsilon_\mathrm{soft} = 10^{-3} \, \mathrm{pc}$. Our aim is not to accurately model the dynamics in the innermost region around the MBH (i.e., collisional relaxation by stars near the sphere of influence), but our interest is in larger regions extending to several hundred pc around the MBH. Therefore, such softening is justified. Note that the potential equation~(\ref{eq:phi_MBH}), equivalent to a Plummer potential \citep{1911MNRAS..71..460P}, gives rise to steady precession of the orbit around the MBH, and does not lead to diffusion of the orbital angular momentum nor energy.

\subsubsection{Bulge}
\label{sect:orbit:pot:bulge}
The bulge is modeled with a potential similar to that of \citet{2001ApJ...554.1059K}, i.e.,
\begin{align}
\Phi_\mathrm{bulge}(R,\phi,z) = \Phi_0 \left (1 + \frac{a}{a_0} \right )^{2+\beta}
\end{align}
with
\begin{align}
a = \left ( \frac{R^2 \cos^2(\phi)}{\gamma_x^2} + \frac{R^2 \sin^2(\phi)}{\gamma_y^2} + \frac{z^2}{\gamma_z^2} \right )^{1/2}.
\end{align}
Here, we adopt $\beta = -1.8$, $a_0 = 1\,\mathrm{pc}$ and $\Phi_0 = 3\times 10^4 \, \mathrm{pc^2 \, Myr^{-2}}$ from \citet{2001ApJ...554.1059K}. 

In contrast to \citealt{2001ApJ...554.1059K}, who assumed a triaxial and rotating bulge potential with $\gamma_y=0.9$ and $\gamma_z=0.8$ in most of their models, we assume a spherically symmetric potential, i.e., we set $\gamma_x=\gamma_y=\gamma_z=1$. In the latter case, orbits experience precession only, and the energy and angular momentum are conserved. Any deviations from energy and angular-momentum conservation can then be ascribed to perturbations from nuclear spiral arms. Triaxial potentials, although interesting on their own, are beyond the scope of this paper. 

\begin{figure}
\center
\iftoggle{ApJFigs}{
\includegraphics[scale = 0.48, trim = 5mm -2mm 0mm 0mm]{v_c_test03.eps}
}{
\includegraphics[scale = 0.48, trim = 5mm -2mm 0mm 0mm]{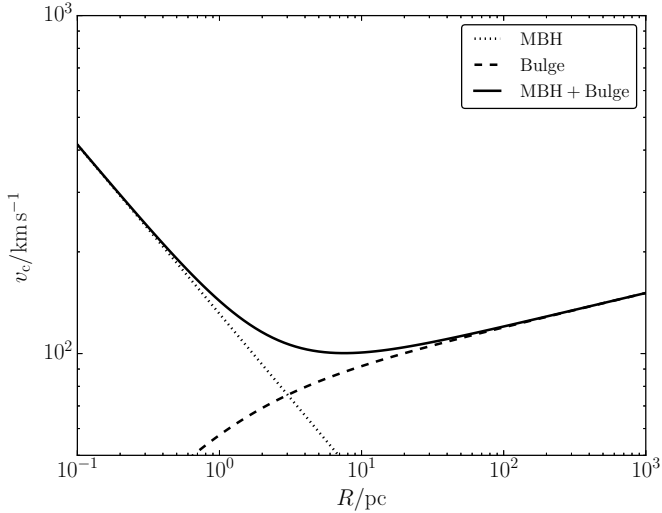}
}
\caption{\small The circular-speed curve $v_\mathrm{c}(R)$ from the potential $\Phi_\bullet + \Phi_\mathrm{bulge}$ (solid line) as a function of $R$ (with $\phi=z=0$, and setting $\gamma_x=\gamma_y=\gamma_z=1$). The dotted (dashed) line applies to including the MBH (bulge) potential only. }
\label{fig:v_c_test03}
\end{figure}

In \F\,\ref{fig:v_c_test03}, we show the circular-speed curve $v_\mathrm{c}(R)$ from the potential $\Phi_\bullet + \Phi_\mathrm{bulge}$ as a function of $R$ (with $\phi=0$ and $z=0$, and setting $\gamma_x=\gamma_y=\gamma_z=1$). The inner few pc are dominated by the MBH (dotted lines), in which case 
\begin{align}
v_\mathrm{c}^2 = G M_\bullet/R.
\end{align}
The outer regions are dominated by the bulge, for which, in our model,
\begin{align}
\nonumber v_\mathrm{c}^2 &= \Phi_0 (2+\beta) (R/a_0) \left ( 1 + R/a_0 \right)^{1+\beta} \\
&\approx \Phi_0(2+\alpha) (R/a_0)^{2+\beta},
\end{align}
where the last line applies if $R\gg a_0$, i.e., in that case, $v_c \propto R^{1+\beta/2} = R^{0.1}$. The circular speed in our model is $\sim 100 \, \mathrm{km\,s^{-1}}$ at $\sim$ 10 pc from the center, and rises to $\sim 150\,\mathrm{km\,s^{-1}}$ at $\sim$ 1 kpc. This is consistent with, e.g., the potential Model I in S2.7 of \citet{2008gady.book.....B}. At larger radii, our model clearly does not give a realistic representation of the Galaxy potential: the circular-speed curve of the latter flattens around $220\,\mathrm{km\,s^{-1}}$ (see, e.g., fig. 2.20 of \citealt{2008gady.book.....B}), whereas our model gives an unbounded increase of $v_\mathrm{c}$. Here, we are only interested in the central few hundred pc around the MBH; therefore, this inconsistency should not affect our conclusions.

\subsubsection{Nuclear spiral arms}
\label{sect:orbit:pot:spiral}
\paragraph{General potential}
\label{sect:orbit:pot:spiral:gen}
Let $\Sigma_\mathrm{s}(R,\phi,t)$ be the surface density associated with spiral structure at the coordinate $(R,\phi)$ at time $t$. We assume that the spiral structure is located within an infinitely thin disk $z=0$. Following \S\,6.2.2b of \citet{2008gady.book.....B}, we write $\Sigma_\mathrm{s}$ in a form separating the variations in density along a spiral arm, described by $\tilde{\Sigma}_\mathrm{s}(R,t)$, and the variations experienced while passing between arms, i.e.,
\begin{align}
\label{eq:sigma_s_def}
\Sigma_\mathrm{s}(R,\phi,t) = \tilde{\Sigma}_\mathrm{s}(R,t) \exp \left [ \complexi m \phi + \complexi f(R,t) \right ]
\end{align}
(see also, e.g., \citealt{2004MNRAS.354..883M}). Here, the surface density has been written in complex form ($\complexi \equiv \sqrt{-1}$). The number of spiral arms is given by $m>0$, and the shape function $f(R,t)$ describes the spiral shape. Along a spiral arm,
\begin{align}
\label{eq:f_def}
m \phi + f(R,t) = \mathrm{constant}
\end{align}
(e.g., \S\,6.1.3c of \citealt{2008gady.book.....B}), and, therefore, the radial surface-density dependence along the spiral arms is described by $\tilde{\Sigma}_\mathrm{s}(R,t)$. The spiral wave number $k$ is defined from the shape function as
\begin{align}
\label{eq:k_def}
k(R,t) \equiv \frac{\partial f (R,t)}{\partial R},
\end{align}
and is related to the winding angle $\alpha$ according to
\begin{align}
\label{eq:alpha_def}
\cot(\alpha) \equiv \left | R \frac{\partial \phi}{\partial R}\right | = \left | \frac{k R}{m} \right |,
\end{align}
where we used equation~(\ref{eq:f_def}) after the last equality.

To find the potential at any point $(R,\phi,z)$ and time $t$ due to the surface density $\Sigma_\mathrm{s}(R,\phi,t)$, we follow \S\,6.2.2b of \citet{2008gady.book.....B} and expand $\Sigma_\mathrm{s}(R,\phi,t)$ around a local point $(R_0,\phi_0)$, neglecting any variations with respect to $\phi$ and variations in $\tilde{\Sigma}_\mathrm{s}(R,t)$, i.e., $f(R,t) \approx f(R_0,t) + k(R_0,t)(R-R_0)$ and $\tilde{\Sigma}_\mathrm{s}(R,t) \approx \tilde{\Sigma}_\mathrm{s}(R_0,t)$. This expansion is appropriate for tightly-wound spiral arms ($|k|R\gg 1$) because the short wavelengths, $\lambda = 2\pi/|k|\ll R$, imply that only the surface density close to a point $(R_0,\phi_0)$ contributes to the potential, whereas the contribution from the more distant surface density cancels out. Therefore, 
\begin{align}
\label{eq:sigma_s_exp}
\Sigma_\mathrm{s}(R,\phi,t) \approx \Sigma_\mathrm{a}(t) \exp \left [ \complexi k(R_0,t)(R-R_0) \right ],
\end{align}
with
\begin{align}
\Sigma_\mathrm{a}(t) = \tilde{\Sigma}_\mathrm{s}(R_0,t) \exp\left [ \complexi m \phi_0 + \complexi f(R_0,t) \right ].
\end{align}
Equation~(\ref{eq:sigma_s_exp}) represents a plane wave in a razor-thin homogenous disk with the wave vector $\bf{k}$ oriented radially. The potential due to this distribution is \citep[S5.6.1]{2008gady.book.....B}
\begin{align}
\nonumber \Phi_\mathrm{plane\,wave} &\approx - \frac{2\pi G \Sigma_\mathrm{a}(t)}{|k(R_0,t)|} \exp \left [ \complexi k(R_0,t) (R-R_0) \right ] \\
&\qquad \times \exp \left [ - | k(R_0,t) z | \right ].
\end{align}
Here, $G$ is the gravitational constant. Setting $R_0=R$ and $\phi_0=\phi$, this gives the approximate potential due to surface density equation~(\ref{eq:sigma_s_def}),
\begin{align}
\label{eq:phi_spiral_gen}
\nonumber \Phi_\mathrm{s}(R,\phi,z,t) &\approx - \frac{2\pi G \tilde{\Sigma}_\mathrm{s}(R,t)}{|k(R,t)|} \exp\left [ \complexi m \phi + \complexi f(R,t) \right ] \\
&\qquad \times \exp \left [- | k(R,t) z | \right ].
\end{align}
The error made in equation~(\ref{eq:phi_spiral_gen}) is $\mathcal{O}( |kR|^{-1})$ \citep[S6.2.2b]{2008gady.book.....B}; for tightly-wound spirals, $|kR|^{-1} \ll 1$. Also, in equation~(\ref{eq:phi_spiral_gen}) we assumed that $\tilde{\Sigma}_\mathrm{s}(R,t)$ is a slowly varying function of $R$. 

\paragraph{Logarithmic spirals}
\label{sect:orbit:pot:spiral:log}
Nuclear spiral arms are observed to be approximately logarithmic (e.g., \citealt{2010ApJ...723..767V}). For a logarithmic spiral, $R \propto \exp(\phi)$. From equation~(\ref{eq:f_def}), this implies that the shape function $f \propto \ln(R)$. Furthermore, from equation~(\ref{eq:k_def}), $k \propto 1/R$ implying that $\alpha$ is constant (cf. equation~\ref{eq:alpha_def}). Therefore, the wave number can be written as
\begin{align}
|k(R,t)| = \frac{m \cot(\alpha)}{R}.
\end{align}
Integrating equation~(\ref{eq:k_def}), the shape function is given in terms of the constant $\alpha$ and $m$ by
\begin{align}
\label{eq:f_log_1}
f(R,t) = m \cot(\alpha) \ln (R) + f_0(t),
\end{align}
where $f_0(t)$ is a function of time only. We define the radius $R_0$ such that the logarithmic term in equation~(\ref{eq:f_log_1}) vanishes (i.e., the value of $R_0$ only determines the overall phase), and introduce a pattern speed $\Omega_\mathrm{s}$ at which the spiral pattern rotates. Equation~(\ref{eq:f_log_1}) can then be written as
\begin{align}
\label{eq:f_log_2}
f(R,t) = m \cot(\alpha) \ln (R/R_0) - m \Omega_\mathrm{s} t.
\end{align}
Substituting the expressions for $k$ and $f$ into equation~(\ref{eq:phi_spiral_gen}) we find, for logarithmic spirals,
\begin{align}
\label{eq:phi_spiral_log}
&\nonumber \Phi_\mathrm{s}(R,\phi,z,t) \approx - 2\pi G \, \tilde{\Sigma}_\mathrm{s}(R,t) \frac{R \tan(\alpha)}{m} \\
\nonumber &\quad \times \exp\left [ \complexi m \left \{ \phi + \cot(\alpha) \ln(R/R_0) - \Omega_\mathrm{s} t \right \} \right ] \\
&\quad \times \exp \left [ - |z| \frac{m \cot(\alpha)}{R} \right ].
\end{align}
The $z$-scale height of equation~(\ref{eq:phi_spiral_log}) is $H_z = R \tan(\alpha)/m$, i.e., equation~(\ref{eq:phi_spiral_log}) implies a constant $H_z/R = \tan(\alpha)/m$. For logarithmic spirals, the fractional error in the potential is $|k R|^{-1} = \tan(\alpha)/m$. For example, for $m=2$ spirals with $\alpha=10^\circ$, $\tan(\alpha)/m \approx 0.088$. 

\begin{figure}
\center
\iftoggle{ApJFigs}{
\includegraphics[scale = 0.48, trim = 5mm -2mm 0mm 0mm]{spiral_plot.eps}
}{
\includegraphics[scale = 0.48, trim = 5mm -2mm 0mm 0mm]{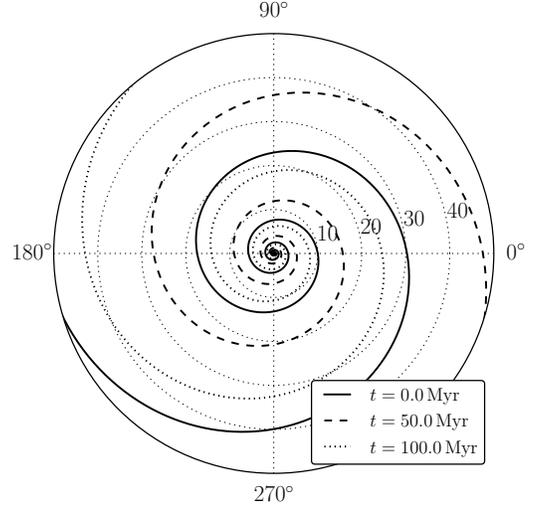}
}
\caption{\small Isosurface density curves for logarithmic spirals, assuming $\tilde{\Sigma}_\mathrm{s}$ is a constant, $m=2$, $\alpha=10^\circ$, $R_0=10\,\mathrm{pc}$, and $\Omega_\mathrm{s} = 50 \, \mathrm{km\,s^{-1}\,kpc^{-1}}$. The three different curves correspond to different times. The numbers near the dotted circles indicate the radius in pc. }
\label{fig:spiral_plot}
\end{figure}

For illustration, we show in \F\,\ref{fig:spiral_plot} isosurface density curves for logarithmic spirals by combining equations~(\ref{eq:sigma_s_def}) and (\ref{eq:f_log_2}). We assume that $\tilde{\Sigma}_\mathrm{s}$ is a constant, and plot curves at different times $t$ for which $\Sigma_\mathrm{s} = \tilde{\Sigma}_\mathrm{s}$, setting $m=2$, $\alpha=10^\circ$, $R_0=10\,\mathrm{pc}$ and $\Omega_\mathrm{s} = 50 \, \mathrm{km\,s^{-1}\,kpc^{-1}}$. Note that $R_0$ only sets the phase of the spiral arms; the arms have no intrinsic scale. 

\paragraph{Transient spirals}
\label{sect:orbit:pot:spiral:trans}
The spiral surface density as described above is assumed to last for a certain amount of time, i.e., the nuclear spiral arms are transient and occur in episodes, referred to as nuclear spiral arm events. The total surface density due to nuclear spirals at a given time $t$ is given by summing over all events, i.e., 
\begin{align}
\tilde{\Sigma}_\mathrm{s}(R,t) = \sum_{i=0}^{N_\mathrm{s}} \tilde{\Sigma}_{\mathrm{s,0},i}(R) \exp \left [ - \frac{(t-t_{\mathrm{s},i})^2}{2 \sigma_{\mathrm{s},i}^2 } \right ].
\end{align}
Here, $i$ sums over the spiral arm events, and $\tilde{\Sigma}_{\mathrm{s,0},i}(R)$, $t_{\mathrm{s},i}$, and $\sigma_{\mathrm{s},i}$ are the radial surface density profile, central time and approximate duration, respectively, of the $i^\mathrm{th}$ spiral arm event.

\begin{figure}
\center
\iftoggle{ApJFigs}{
\includegraphics[scale = 0.48, trim = 5mm -2mm 0mm 0mm]{masses.eps}
}{
\includegraphics[scale = 0.48, trim = 5mm -2mm 0mm 0mm]{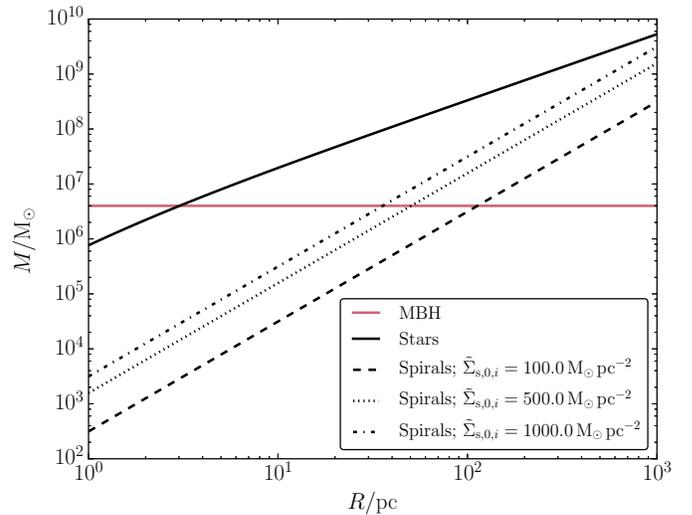}
}
\caption{\small The enclosed masses associated with the stars (solid line) and with nuclear spiral arms (dashed, dotted and dot-dashed lines) in our models as a function of $R$. The red horizontal line indicates our adopted value of the MBH mass.}
\label{fig:masses}
\end{figure}

In \F\,\ref{fig:masses}, we show how gravitationally relevant nuclear spiral arms are compared to the MBH and other stars, by plotting the associated mass, $\pi R^2 \tilde{\Sigma}_{\mathrm{s},0,i}$, as a function of $R$. Here, we compute the stellar mass from the stellar density implied by the bulge potential (\S\,\ref{sect:orbit:pot:bulge}), assuming a (single) stellar mass of $M_\star = 1\,\msun$. We include three values of $\tilde{\Sigma}_{\mathrm{s},0,i}$ which we adopt in the numerical integrations of \S\,\ref{sect:dif} below (see \S\,\ref{sect:dif:IC} for a motivation of these values). The mass fraction in nuclear spirals arms compared to stars is $\sim 10^{-3}$ to $\sim 10^{-2}$ at $R=10\,\mathrm{pc}$ depending on the surface density; at $R=10^2\,\mathrm{pc}$, the mass fractions are $\sim 10^{-2}$ to $10^{-1}$. This already shows that nuclear spiral arms are important at relatively large distances from the MBH ($R\gtrsim 10\,\mathrm{pc}$), which we also find in the integrations in \S\,\ref{sect:dif}.

\subsection{Orbit integrations}
\label{sect:orbit:int}
We integrate the equations of motion in cylindrical coordinates for the position vector $\ve{r} = R \unit{R} + z \unit{z}$, $\ve{\ddot{r}} = -\ve{\nabla} \Phi$, in the total potential $\Phi = \Phi_\bullet + \Phi_\mathrm{bulge} + \Phi_\mathrm{s}$, i.e.,
\begin{subequations}
\label{eq:EOM}
\begin{align}
\ddot{R} &= R \dot{\phi}^2 - \frac{\partial{\Phi}}{\partial R}; \\
\ddot{\phi} &= -2 \frac{\dot{R}}{R}  \dot{\phi} - \frac{1}{R} \frac{\partial \Phi}{\partial \phi}; \\
\ddot{z} &= - \frac{\partial \Phi}{\partial z},
\end{align}
\end{subequations}
where dots denote derivatives with respect to time. Here, when modeling the motion of a binary in the GC, we neglect the quadrupole moment of the stellar binary, i.e., we treat the binary as a point mass, and $\ve{r}$ should be interpreted as the binary barycenter position.

\begin{figure}
\center
\iftoggle{ApJFigs}{
\includegraphics[scale = 0.45, trim = 0mm -4mm 0mm 0mm]{ver_test01.eps}
}{
\includegraphics[scale = 0.45, trim = 0mm -4mm 0mm 0mm]{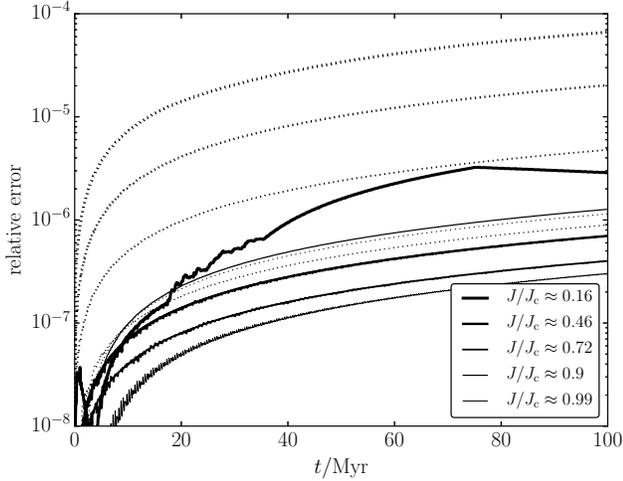}
}
\caption{\small Relative energy errors (solid lines) and angular-momentum errors (dashed lines) as a function of time for an orbit initially at $(R,\phi,z) = (10\,\mathrm{pc},0,0)$ with the spherical bulge and MBH potential. The different line widths correspond to different initial values of $J/J_\mathrm{c}$ indicated in the legend, where $J$ is the orbital angular momentum with respect to the origin, and $J_\mathrm{c}=J_\mathrm{c}(\Ec)$ is the angular momentum of a circular orbit. }
\label{fig:ver_test01}
\end{figure}

Equations~(\ref{eq:EOM}) can be cast in a system of ordinary differential equations (ODEs), which we solve numerically using \textsc{odeint} from the \textsc{Python} \textsc{Scipy} library. The latter is an interface to the \textsc{LSODA} \textsc{Fortran} routine, which uses variable time-steps to integrate the system of ODEs. The routine automatically and dynamically detects between stiff and nonstiff methods; for stiff cases it uses the backward differentiation formula method (with a dense or banded Jacobian), and for nonstiff cases it uses the Adams method. Error control within the solver is determined by the input parameters $\textsc{rtol}$ and $\textsc{atol}$, such that the error in each ODE variable $y_i$ is less than or equal to $\textsc{rtol} \times \mathrm{abs}(y_i) + \textsc{atol}$. We set the relative and absolute tolerance parameters to $\textsc{rtol}=10^{-9}$ and $\textsc{atol}=10^{-9}$, respectively. 

In the case without nuclear spiral arms and a spherical (non-rotating) bulge, orbits experience precession only, and the energy and angular momentum are conserved. In \F\,\ref{fig:ver_test01}, we show relative energy errors (solid lines) and angular momentum errors (dashed lines) as a function of time for an orbit initially at $(R,\phi,z) = (10\,\mathrm{pc},0,0)$. The integration time is 100 Myr, which corresponds to $\approx 455$ orbits (the orbital period is $\approx 0.22 \, \mathrm{Myr}$). Different line widths correspond to different initial values of $J/J_\mathrm{c}$, where $J$ is the orbital angular momentum with respect to the origin, and $J_\mathrm{c}$ is the angular momentum of a circular orbit (see also \S\,\ref{sect:dif}). As expected, the lower $J/J_\mathrm{c}$ (the more eccentric the orbit), the larger the relative energy and angular-momentum errors. Note that the highly eccentric orbits experience rapid precession (with the argument of periapsis changing of the of order of $\sim90^\circ$ per orbit). 

After 100 Myr, which is also the time-span used in the integrations in \S\,\ref{sect:dif}, the energy errors remain $\lesssim 3\times 10^{-6}$, and the angular-momentum errors remain $\lesssim 10^{-4}$. We are interested in the statistical properties for an ensemble of orbits. A typical change in $J/J_\mathrm{c}$ in the simulations of \S\,\ref{sect:dif} is $\sim 0.1$, which is $\gg 10^{-4}$. Therefore, we believe these errors are sufficiently small for our purposes. 

\begin{figure}
\center
\iftoggle{ApJFigs}{
\includegraphics[scale = 0.36, trim = 15mm 0mm 0mm 0mm]{examples_E_test01.eps}
}{
\includegraphics[scale = 0.36, trim = 15mm 0mm 0mm 0mm]{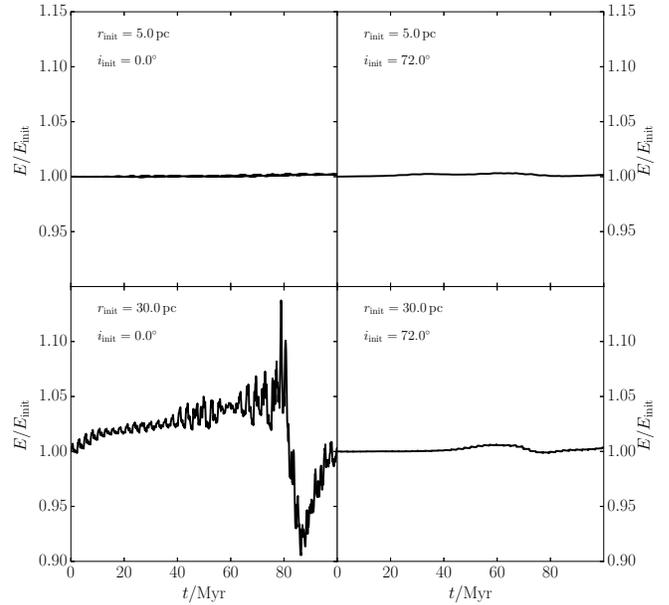}
}
\caption{\small Examples of orbit integrations, showing the energy (normalized to the initial energy) as a function of time for four different initial orbital parameters $r_\mathrm{init}$ and $i_\mathrm{init}$, indicated in the panels. Refer to \S\,\ref{sect:orbit:example} for details on the other assumed parameters. }
\label{fig:examples_E_test01}
\end{figure}

\begin{figure}
\center
\iftoggle{ApJFigs}{
\includegraphics[scale = 0.36, trim = 15mm 0mm 0mm 0mm]{examples_AM_test01.eps}
}{
\includegraphics[scale = 0.36, trim = 15mm 0mm 0mm 0mm]{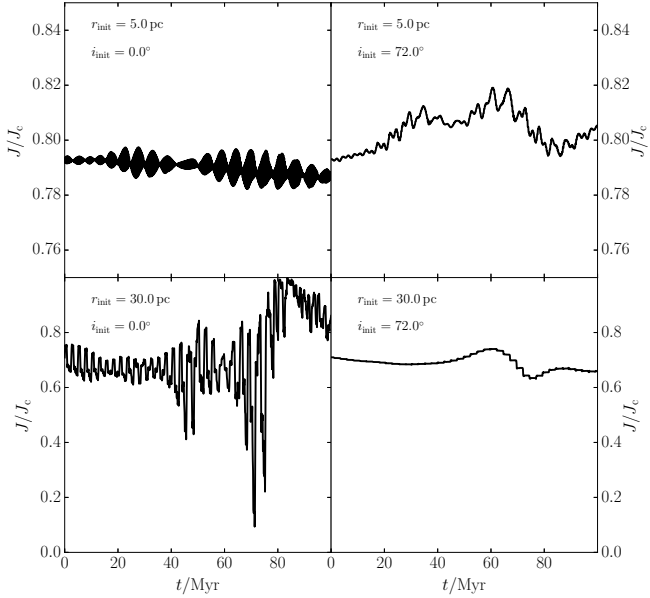}
}
\caption{\small Similar to \F\,\ref{fig:examples_E_test01}, here showing the angular momentum, normalized to $J_\mathrm{c}(\Ec)$, as a function of time. }
\label{fig:examples_AM_test01}
\end{figure}

\begin{figure}
\center
\iftoggle{ApJFigs}{
\includegraphics[scale = 0.36, trim = 15mm 0mm 0mm 0mm]{examples_i_test01.eps}
}{
\includegraphics[scale = 0.36, trim = 15mm 0mm 0mm 0mm]{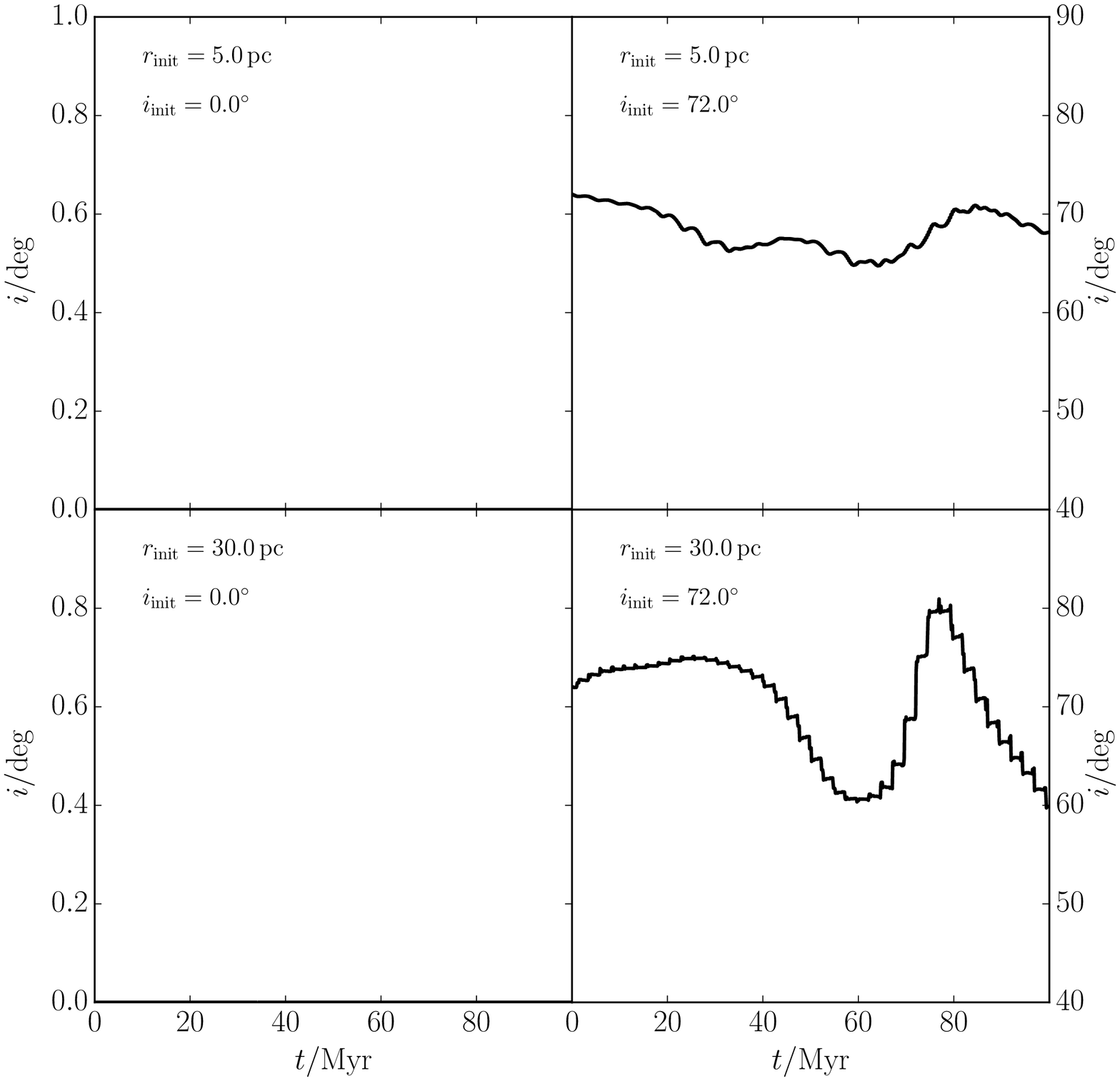}
}
\caption{\small Similar to \F\,\ref{fig:examples_E_test01}, here showing the orbital inclination as a function of time. }
\label{fig:examples_i_test01}
\end{figure}

\subsection{Example evolution}
\label{sect:orbit:example}
We give a number of examples of orbit integrations to illustrate the ability of transient nuclear spiral structures to diffuse the orbits of objects (single stars and binaries) in regions around the central few hundred pc of the GC. In Figs \,\ref{fig:examples_E_test01}, \ref{fig:examples_AM_test01} and \ref{fig:examples_i_test01}, we plot the energy (normalized to the initial energy), angular momentum (normalized to $J_\mathrm{c}$, cf. \S\,\ref{sect:dif}) and orbital inclination, respectively, for four orbits at different initial radii and initial orbital inclinations. The orbital inclination is measured with respect to the $z$ axis; the nuclear spiral arms are assumed to be confined to the plane $z=0$ (cf. \S\,\ref{sect:orbit:pot:spiral}). In the four panels, we take two values of the initial distance to the center, $r_\mathrm{init}=5\,\mathrm{pc}$ and $r_\mathrm{init}=30\,\mathrm{pc}$, and two values of the initial inclination, $i_\mathrm{init} = 0^\circ$ and $i_\mathrm{init}=72^\circ$. 

For the nuclear spiral arm parameters, we assume a constant spiral gas surface density, i.e., $\tilde{\Sigma}_{\mathrm{s,0},i}$ is independent of radius, and take $\tilde{\Sigma}_{\mathrm{s,0},i} = 500 \, \msun\,\mathrm{pc^{-2}}$. The number of spiral arm events is $N_\mathrm{s} = 10$, with $t_{\mathrm{s},i}$ at fixed equal intervals between 0 and 100 Myr, and $\sigma_{\mathrm{s},i} = 10 \, \mathrm{Myr}$ (i.e., 10 spiral arm events, each lasting about 10 Myr). For each spiral arm event, the spiral pattern speed is set to $50 \, \mathrm{km \, s^{-1} \, pc^{-1}}$, $m=2$ arms are assumed, and the winding angle is $\alpha=10^\circ$. We refer to \S\,\ref{sect:dif:IC} for motivation of these parameter choices. 

Close to the MBH (cf. the top panels in Figs \,\ref{fig:examples_E_test01}, \ref{fig:examples_AM_test01}), the potential is dominated by the MBH, and spiral structure leads to only small variations in the energy and angular momentum. Further away from the MBH (cf. the bottom panels in Figs \,\ref{fig:examples_E_test01} and \ref{fig:examples_AM_test01}), nuclear spiral arms can have a large effect on both energy and angular momentum, with the largest variations occurring in $J/J_\mathrm{c}$. There is also a strong dependence on the initial inclination: for $r_\mathrm{init}=30\,\mathrm{pc}$, $J/J_\mathrm{c}$ changes drastically over time if the orbit lies in the same plane as the spiral arms ($i_\mathrm{init}=0^\circ$), whereas the changes are much smaller if $i_\mathrm{init}=72^\circ$. In the latter case, $E$ and $J$ change substantially only when the orbit crosses the $z=0$ plane and spiral arms happen to be active near this moment. 

\F\,\ref{fig:examples_i_test01} shows how the inclination changes over time. If the orbit is initially coplanar, it remains coplanar. In the highly inclined case, the inclination is affected by the nuclear spiral arms, with $i$ changing `impulsively' at each passage of the Galactic plane, increasing and decreasing in a periodic fashion. In our example, the inclination changes remain modest, with overall changes of less than $\approx 10^\circ$.

\section{Diffusion by nuclear spiral arms}
\label{sect:dif}
Adopting the potential model described in \S\,\ref{sect:orbit}, we carry out a large number (1,200,000) of orbital integrations representing single stars and binaries in the GC (\S\,\ref{sect:dif:IC}). From these integrations, we determine diffusion coefficients (\S\,\ref{sect:dif:results}), which are then used to compute disruption rates in \S\,\ref{sect:rates}. 

Throughout, we use the orbital angular momentum $J \equiv |\ve{r}\times \dot{\ve{r}}|$ and the (negative) energy $\Ec \equiv -\frac{1}{2}\dot{\ve{r}}^2 + \psi(r)$ to describe the relaxation process. Here, $\psi(r) \equiv - \Phi(r)$ is the negative of the potential; note that $\Ec$ is defined with a minus sign, contrary to convention. The orbital angular momentum is normalized to the orbital angular momentum of a circular orbit, $J_\mathrm{c}$. The latter is determined from the potential via 
\begin{align}
\label{eq:J_c_def}
J_\mathrm{c} = r_\mathrm{c} v_\mathrm{c},
\end{align}
where $v_\mathrm{c}$ is the speed of a circular orbit at radius $r_\mathrm{c}$, i.e.,
\begin{align}
\label{eq:v_c_def}
v_\mathrm{c}^2/r_\mathrm{c} =  | \ve{\nabla} \Phi |_{r=r_\mathrm{c}}, 
\end{align}
and $r_\mathrm{c}$ is determined implicitly by setting the radial speed $v_\mathrm{r}$ to zero (cf. \S\,\ref{sect:rates:meth}), i.e.,
\begin{align}
\label{eq:r_c_def}
2[\psi(r_\mathrm{c}) - \Ec] - J_\mathrm{c}^2/r_\mathrm{c}^2  = 2[\psi(r_\mathrm{c}) - \Ec] - r_\mathrm{c}  | \nabla \Phi  |_{r=r_\mathrm{c}} = 0.
\end{align}

\begin{table}
\begin{tabular}{lp{2.2cm}cc}
\toprule
Symbol & Description & Value(s) &  \\
\midrule
\multicolumn{3}{c}{Grid parameters} \\
\midrule
$i$              & Initial orbital inclination w.r.t. the Galactic plane              & $0^\circ$-$180^\circ$ (20 values; linear) \\
$N_\mathrm{s}$ & Number of spiral arm events & [1, 10] \\
$\tilde{\Sigma}_{\mathrm{s},0,i}$ & Spiral arm gas surface density & [100,500,1000] $\msun\,\mathrm{pc^{-2}}$ \\
\midrule
\multicolumn{3}{c}{Fixed parameters} \\
\midrule
$r_\mathrm{init}$ & Initial distance to the center & 10-1000 pc (20 values; log.) \\
$\dot{\phi}_\mathrm{init}$ & Initial $\phi$ time derivative & $(0.1-1) \,\frac{v_\mathrm{c,init}}{r_\mathrm{init}}$ (5 values; linear) \\
$t_\mathrm{int}$ & Integration time & $100\,\mathrm{Myr}$ \\
$m$ & Number of spiral arms (per event) & 2 \\
$\Omega_\mathrm{s}$ & Spiral arm pattern speed & $50\,\mathrm{km\,s^{-1}\, kpc^{-1}}$ \\
$\sigma_{\mathrm{s},i}$ & Spiral arm event temporal Gaussian width & $50\,\mathrm{Myr}$ ($N_\mathrm{s}=1$) \\
& & $10\,\mathrm{Myr}$ ($N_\mathrm{s}=10$) \\
$t_{\mathrm{s},i}$ & Spiral arm event times & $50 \, \mathrm{Myr}$ ($N_\mathrm{s}=1$) \\
& & $10-90\,\mathrm{Myr}$ ($N_\mathrm{s}=10$) \\
$R_0$ & Spiral arm phase parameter & $x$ pc \\
\bottomrule
\end{tabular}
\caption{ Overview of the initial conditions for the Monte-Carlo integrations in \S\,\ref{sect:dif}. The values for $i$, $r_\mathrm{init}$ and $\dot{\phi}_\mathrm{init}$ have linear, logarithmic, and linear spacings, respectively. In the bottom row, $x$ is a random number between 0 and 1. }
\label{table:IC}
\end{table}

\subsection{Initial conditions}
\label{sect:dif:IC}
An overview of the initial conditions is given in Table\,\ref{table:IC}. We determine energy and angular-momentum relaxation time-scales and diffusion coefficients on a grid of three parameters: the inclination $i$, the number of spiral arm events $N_\mathrm{s}$, and the nuclear spiral arm gas surface density $\tilde{\Sigma}_{\mathrm{s},0,i}$. For the latter, we assume, for simplicity, a surface density that is independent of radius $R$. Also, for simplicity, the initial orbits are realized with $\dot{R}_\mathrm{init} = \dot{z}_\mathrm{init} = 0$. In this case, the orbital inclination $i$ is given by
\begin{align}
\cos i = \frac{\ve{J} \cdot \unit{z}}{|\ve{J} \cdot \unit{z}|} = \mathrm{sign}\left(\dot{\phi} \right) \frac{R}{r}.
\end{align}
Also, note that the nodal orientation of the orbit with respect to the plane $z=0$ is described by $\phi$, which is sampled randomly (see below).

For each grid parameter combination $(i,N_\mathrm{s},\tilde{\Sigma}_{\mathrm{s},0,i})$, we take 20 values of the initial distance to the center $r_\mathrm{init}$ ranging between 10 and 1000 pc with logarithmic spacing, and 5 values of $\dot{\phi}_\mathrm{init}$, the initial $\phi$ time derivative. The latter values are taken to be $(0.1-1) \,v_\mathrm{c,init}/r_\mathrm{init}$, with five values sampled linearly; these values correspond to different values of the initial angular momentum. For each resulting parameter combination, i.e., $(i,N_\mathrm{s},\tilde{\Sigma}_{\mathrm{s},0,i},r_\mathrm{init},\dot{\phi}_\mathrm{init})$, we carry out $N_\mathrm{MC}=100$ integrations for a duration of $t_\mathrm{int}=100 \,\mathrm{Myr}$, each with a different initial phase angle $\phi$ sampled randomly in the range $[0,2\pi)$. This approach results in a total of 1,200,000 integrations. 

The nuclear spiral arm parameters are currently not well constrained. For the active galaxy Arp 102B, \citet{2011ApJ...736...77F} found a two-armed spiral structure (i.e., $m=2$) in the inner kpc. Measurements of the velocity field in NGC 1097 \citep{2010ApJ...723..767V} indicate a logarithmic spiral with two arms and a pitch angle of $\alpha = 52^\circ \pm 4^\circ$. The spiral overdensity in electrons was found to be $\Delta n_e \sim 50 \, \mathrm{cm^{-3}}$ at radii of several hundred pc (cf. Fig. 4 of \citealt{2010ApJ...723..767V}), and the radius-to-height ratio at similar radii was inferred to be $h/R\sim 0.25$. The electron overdensity corresponds to a gas overdensity of $\Delta \rho \sim 1.36 \, m_\mathrm{p} \Delta n_e$, where $m_\mathrm{p}$ is the proton mass, and the factor 1.36 takes into account the presence of helium \citep{2010ApJ...723..767V}. This gives a gas surface overdensity of $\Sigma \sim \Delta \rho \, h$, where $h$ is the height associated with the nuclear spiral arm gas, i.e., 
\begin{align}
\Sigma \sim 1.36 \, \Delta n_e \, m_\mathrm{p} 0.25 R \approx 42 \, \msun\,\mathrm{pc^{-2}} \, \left ( \frac{R}{100\,\mathrm{pc}} \right ).
\end{align}

An estimated theoretical upper limit on $\Sigma$ can be obtained by requiring stability with respect to gravitational collapse, i.e., by setting the Toomre $Q$ parameter to unity \citep{1964ApJ...139.1217T}. Approximately, this implies $Q=c_\mathrm{s} \kappa/(\pi G \Sigma) = 1$, where $c_\mathrm{s}$ is the sound speed, here taken to be a constant $c_\mathrm{s}=10\,\mathrm{km \, s^{-1}}$, and $\kappa$ is the epicyclic frequency, here computed as a function of radius $R$ using Eq. (3.79a) from \citealt{2008gady.book.....B} with the potential $\Phi(R) = \Phi_\bullet(R) + \Phi_\mathrm{bulge}(R)$ and setting $z=0$, $\epsilon_\mathrm{soft}=0$, and $\gamma_x = \gamma_y = \gamma_z = 1$ (cf. \S s\,\ref{sect:orbit:pot:mbh} and \ref{sect:orbit:pot:bulge}). The surface density associated with $Q=1$ is then given by
\small
\begin{align}
\label{eq:sigma_estimate_Q}
\nonumber \Sigma &= \frac{c_\mathrm{s}}{\pi G} \left [ \frac{GM_\bullet}{R^3} + \frac{\Phi_0}{a_0^2} (2+\beta)\left(1+\frac{R}{a_0} \right )^\beta \left(4 + \beta + \frac{3a_0}{R} \right ) \right ]^{1/2} \\
\quad &\sim  1.3 \times 10^3 \, \msun\,\mathrm{pc^{-2}} \,\left ( \frac{R}{100 \, \mathrm{pc}} \right )^{-0.9},
\end{align}
\normalsize
where the last line applies if $R\gg 1\,\mathrm{pc}$, and we substituted the values from \S\,\ref{sect:orbit:pot}. 

In the simulations of \citet{2000ApJ...528..677E}, the pitch angle was found to be a function of radius, with $\alpha \sim 10^\circ-20^\circ$ between $\sim 200$ and 400 pc from the center. In other simulations by \citet{2012ApJ...747...60K}, typical spiral arm surface densities were $\sim 100 \, \msun\,\mathrm{pc^{-2}}$, and the spiral arm structures lasted on the order of 100 Myr. In the recent simulations of \citet{2017arXiv170403665R}, two-armed spirals were found in their GC models, with surface densities of $\sim 100 \, \msun\,\mathrm{pc^{-2}}$.

Here, we take a constant $m=2$ and a pitch angle of $\alpha=10^\circ$. Note that the approximation of tightly-wound spirals, made in the potential (cf. \S\,\ref{sect:orbit:pot:spiral}), breaks down for large $\alpha$. For the surface density $\tilde{\Sigma}_{\mathrm{s},0,i}$, we take three values, 100, 500 and 1000 $\msun\,\mathrm{pc^{-2}}$. We consider the lower value as typical, whereas the upper value is close to the maximum value for gravitational stability (cf. equation~\ref{eq:sigma_estimate_Q}), and should be interpreted as an extreme case. The number of spiral arm events is taken to be either $N_\mathrm{s}=1$ or 10. For $N_\mathrm{s}=1$, we assume there is one spiral arm event at $t=50\,\mathrm{Myr}$, with a characteristic width of $\sigma_{\mathrm{s},i} = 50 \, \mathrm{Myr}$. For $N_\mathrm{s}=10$, the spiral event times are set between 10 and 90 Myr with linear spacing, and $\sigma_{\mathrm{s},i} = 10 \, \mathrm{Myr}$. The spiral arm pattern speed $\Omega_\mathrm{s}$ is set to $\Omega_\mathrm{s}=50 \, \mathrm{km \, s^{-1} \, kpc^{-1}}$. 

\subsection{Data reduction}
\label{sect:dif:data}
From the integrations, we extract time series of the energy $\Ec$ and orbital angular momentum $J$. The latter is normalized to the angular momentum of a circular orbit, $J_\mathrm{c}$ (cf. equation~\ref{eq:J_c_def}). For each realization $i$, we record the energy and angular-momentum changes, $\Delta \Ec_i$ and $\Delta[J/J_\mathrm{c}(\Ec)]_i$, respectively, at the end of the integration time with respect to the initial time, i.e., for a time interval of $\Delta t = t_\mathrm{int} = 100 \, \mathrm{Myr}$. Subsequently, the energy and angular-momentum diffusion coefficients are computed from the rms changes according to
\begin{subequations}
\label{eq:D_J_E_def}
\begin{align}
\label{eq:D_E_def}
\mathcal{D}_\Ec &= \Delta t^{-1} \frac{1}{N_\mathrm{MC}} \sum_{i=1}^{N_\mathrm{MC}} \left ( \Delta \Ec_i/\Ec_{0,i} \right )^2; \\
\label{eq:D_J_def}
\mathcal{D}_J &= \Delta t^{-1}   \frac{1}{N_\mathrm{MC}} \sum_{i=1}^{N_\mathrm{MC}} \left [ \Delta (J/J_\mathrm{c}) \right ]_i^2.
\end{align}
\end{subequations}
The underlying assumption in equations~(\ref{eq:D_J_E_def}) is that diffusion in both energy and angular momentum occurs in a random-walk process, i.e., $\Ec$ and $J/J_\mathrm{c}$ grow proportionally to $\Delta t^{1/2}$. 

In determining the diffusion coefficients, we exclude integrations in which the test particle becomes essentially unbound from the GC. We consider this to be the case if the final distance from the center $r_\mathrm{final}$ (i.e., after 100 Myr) satisfies $r_\mathrm{final}>1000\,\mathrm{pc}$. This can occur if the spiral arm perturbations are strong, particularly for the highest surface density considered, $\tilde{\Sigma}_{\mathrm{s},0,i} = 1000 \, \msun\, \mathrm{pc^{-2}}$. In the case of an absence of sufficient data points among the 100 Monte-Carlo integrations ($N\leq 40$), the diffusion coefficients are set to zero, i.e., no relaxation. 

\begin{figure*}
\center
\iftoggle{ApJFigs}{
\includegraphics[scale = 0.46, trim = 15mm 0mm 0mm 0mm]{t_r_J_radii_various_inclinations_index_par4_0_index_par5_0_test06.eps}
\includegraphics[scale = 0.46, trim = 15mm 0mm 0mm 0mm]{t_r_J_radii_various_inclinations_index_par4_0_index_par5_2_test06.eps}
\includegraphics[scale = 0.46, trim = 15mm 0mm 0mm 0mm]{t_r_J_radii_various_inclinations_index_par4_1_index_par5_0_test06.eps}
\includegraphics[scale = 0.46, trim = 15mm 0mm 0mm 0mm]{t_r_J_radii_various_inclinations_index_par4_1_index_par5_2_test06.eps}
}{
\includegraphics[scale = 0.46, trim = 15mm 0mm 0mm 0mm]{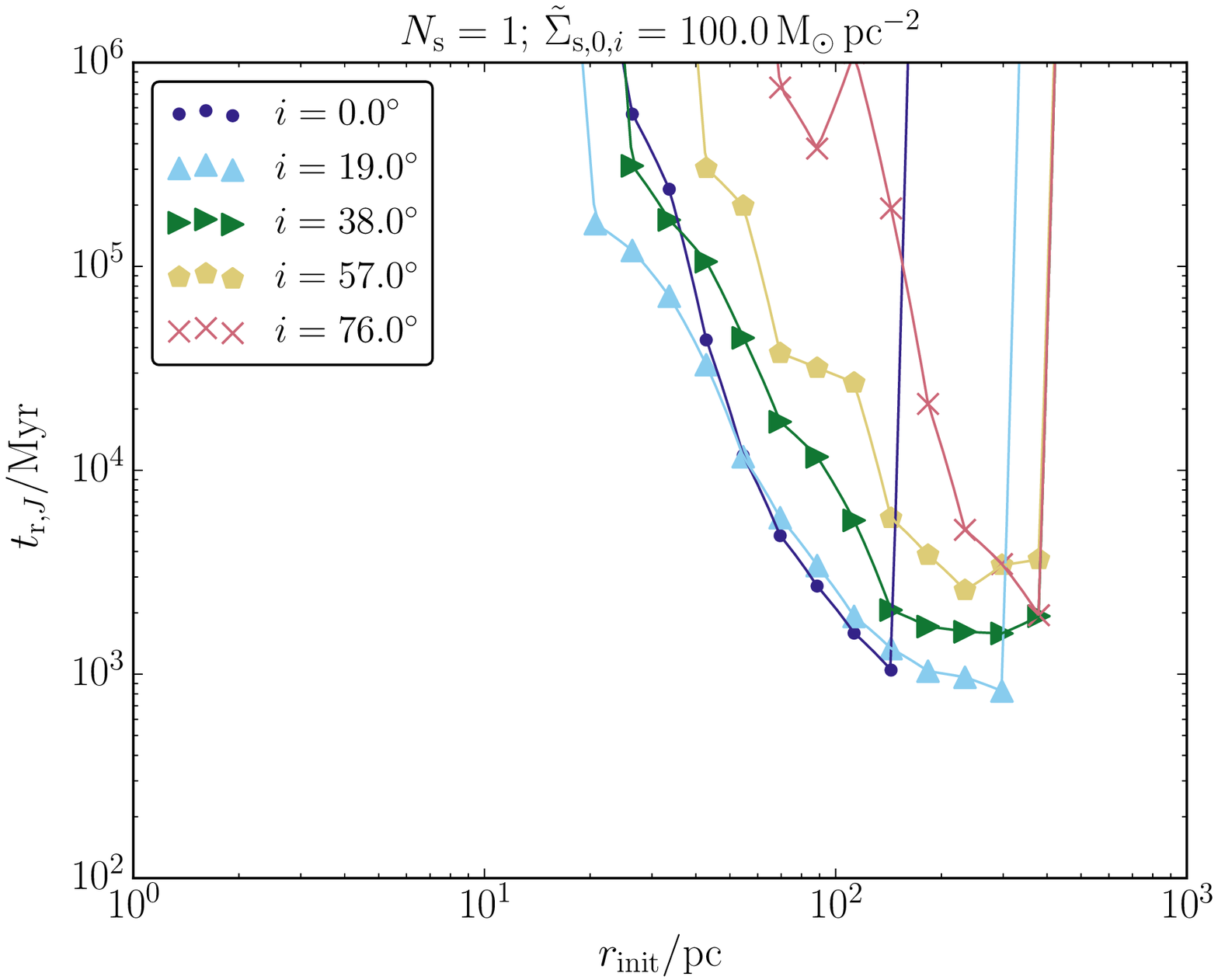}
\includegraphics[scale = 0.46, trim = 15mm 0mm 0mm 0mm]{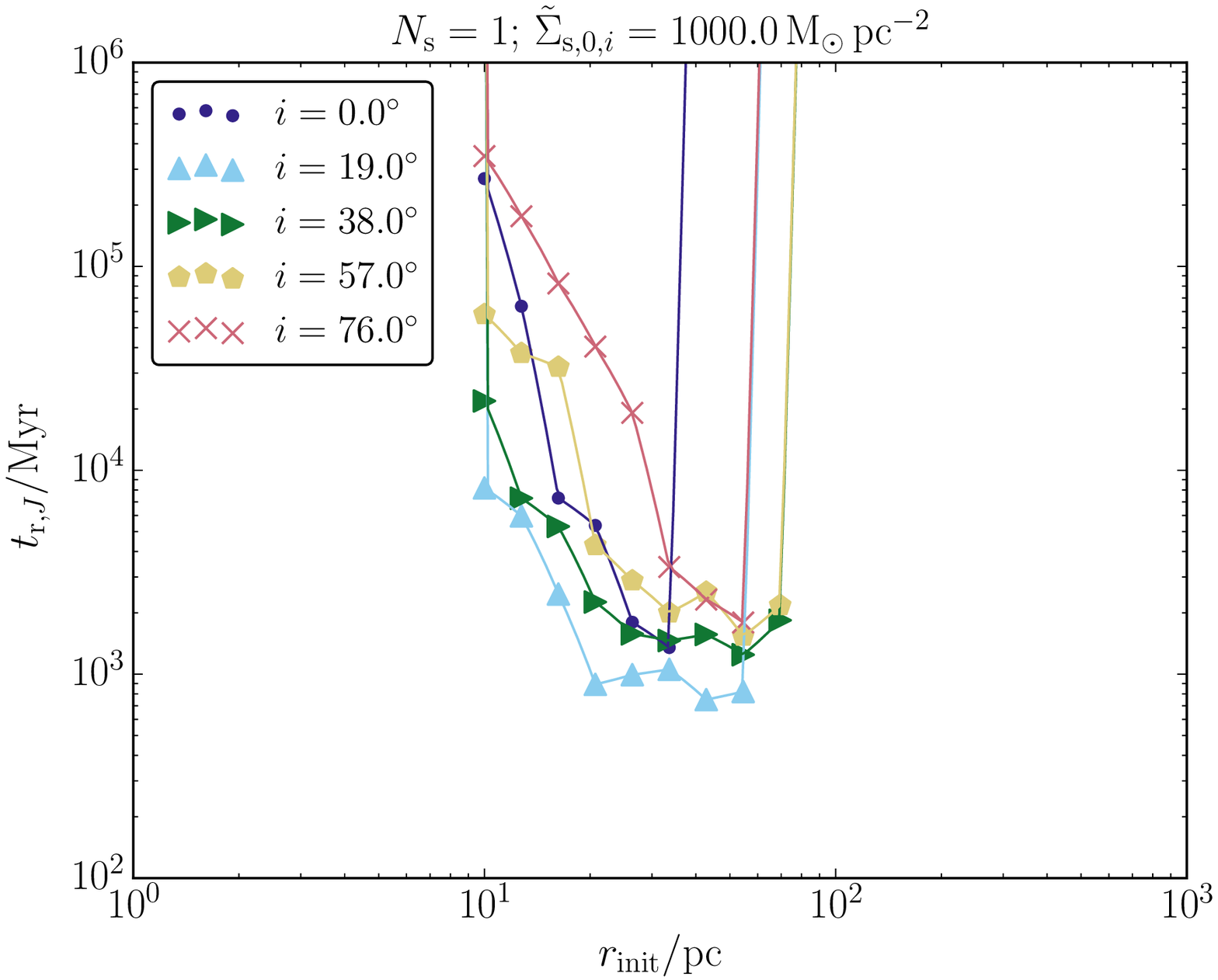}
\includegraphics[scale = 0.46, trim = 15mm 0mm 0mm 0mm]{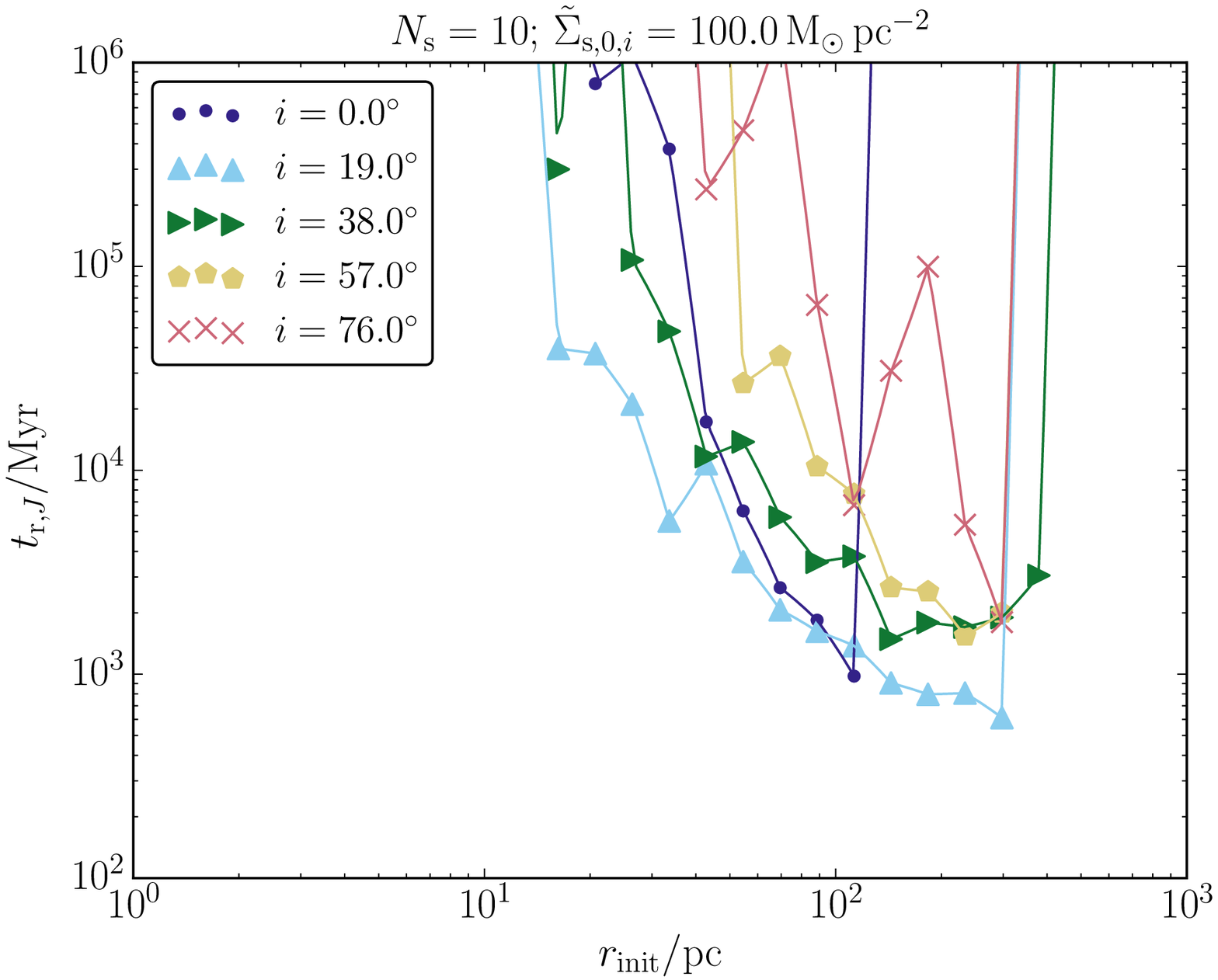}
\includegraphics[scale = 0.46, trim = 15mm 0mm 0mm 0mm]{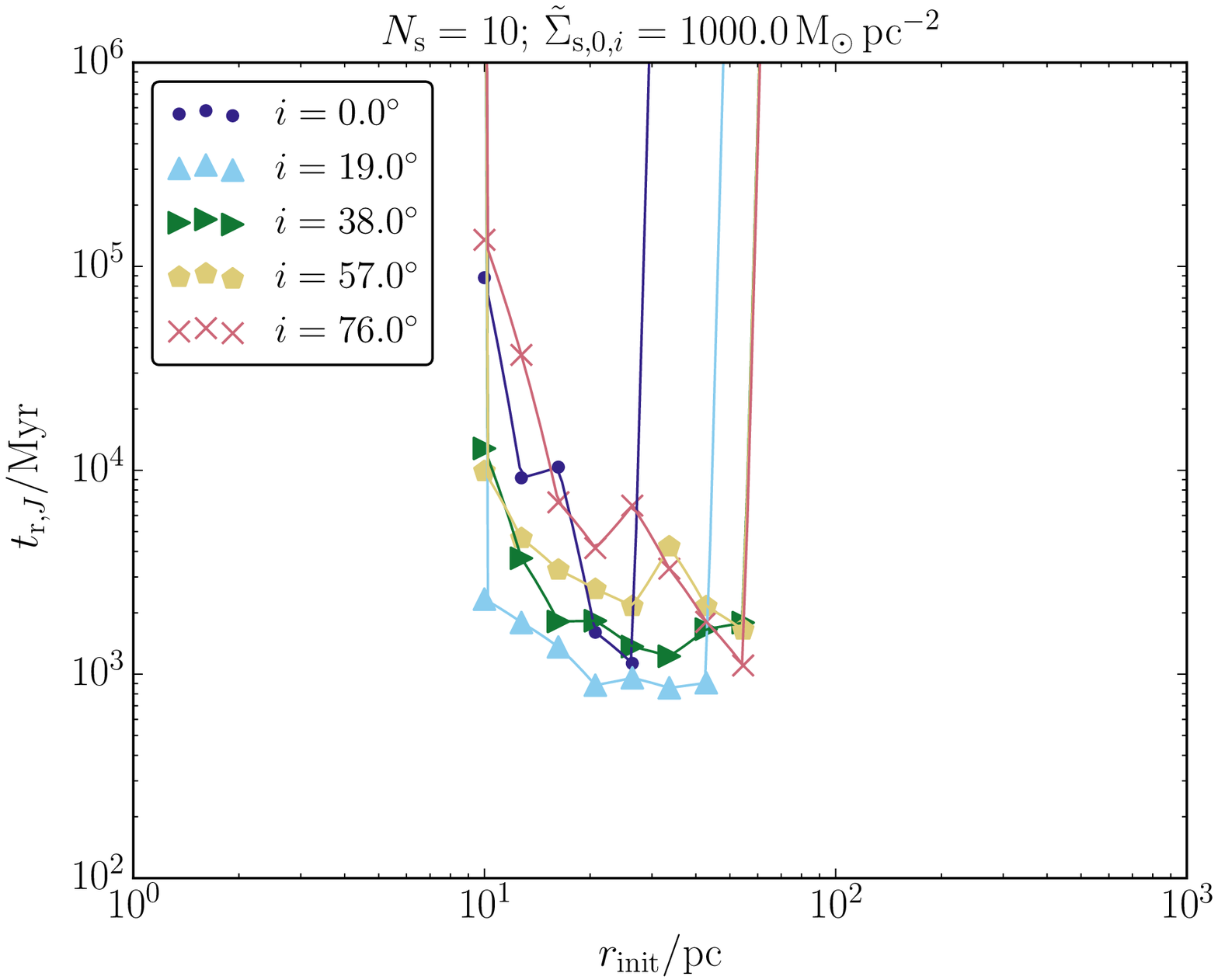}
}
\caption{\small The angular-momentum relaxation time-scales as a function of $r_\mathrm{init}$. In each panel, the different symbols correspond to different inclinations. Each panel corresponds to a different combination of $N_\mathrm{s}$ and $\tilde{\Sigma}_{\mathrm{s},0,i}$, indicated in the panel title. Results are shown for prograde orbits only. The solid lines show linear interpolations between the data points.}
\label{fig:t_r_J_radii_various_inclinations}
\end{figure*}

\begin{figure}
\center
\iftoggle{ApJFigs}{
\includegraphics[scale = 0.46, trim = 15mm 0mm 0mm 0mm]{t_r_J_radii_various_inclinations_index_par4_0_index_par5_0_test06R.eps}
}{
\includegraphics[scale = 0.46, trim = 15mm 0mm 0mm 0mm]{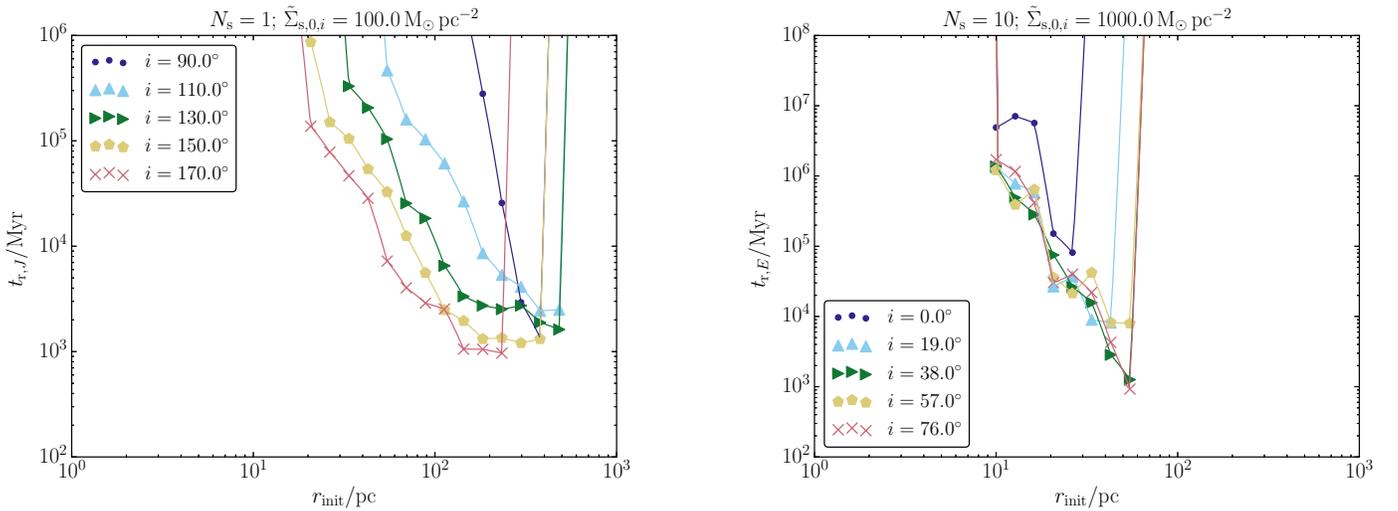}
}
\caption{\small Similar to \F\,\ref{fig:t_r_J_radii_various_inclinations}, now showing, for one combination of $N_\mathrm{s}$ and $\tilde{\Sigma}_{\mathrm{s},0,i}$, the dependence for retrograde orbits. }
\label{fig:t_r_J_radii_various_inclinationsR}
\end{figure}

\begin{figure}
\center
\iftoggle{ApJFigs}{
\includegraphics[scale = 0.46, trim = 5mm 0mm 0mm 0mm]{t_r_E_radii_various_inclinations_index_par4_1_index_par5_2_test06.eps}
}{
\includegraphics[scale = 0.46, trim = 5mm 0mm 0mm 0mm]{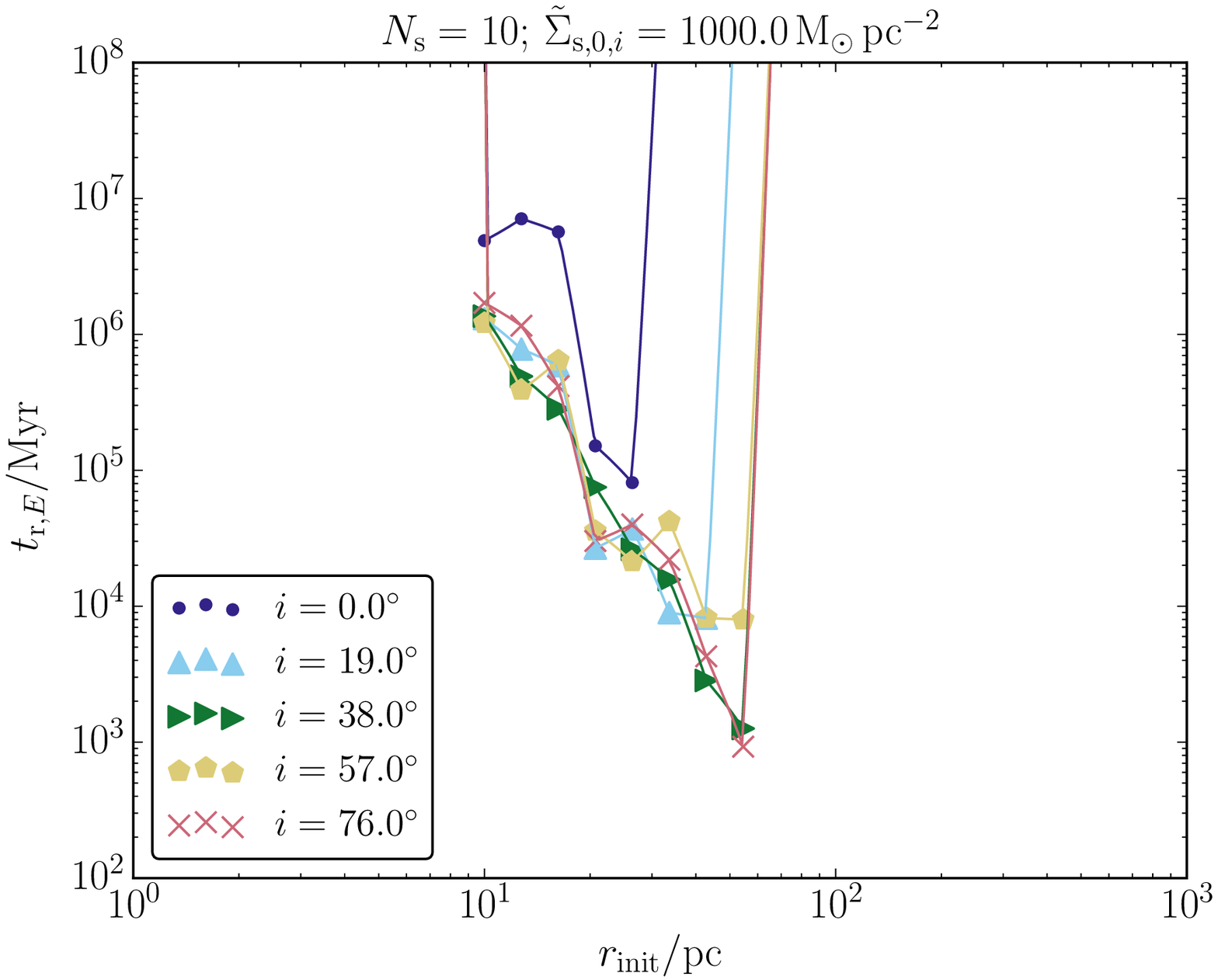}
}
\caption{\small Similar to \F\,\ref{fig:t_r_J_radii_various_inclinations}, now showing the energy relaxation time-scales as a function of $r_\mathrm{init}$ for a single value of $N_\mathrm{s}$ and $\tilde{\Sigma}_{\mathrm{s},0,i}$. }
\label{fig:t_r_E_radii_various_inclinations}
\end{figure}

\subsection{Results}
\label{sect:dif:results}
Although the diffusion coefficients are used in the rate calculations in \S\,\ref{sect:rates}, a (perhaps) physically more intuitive quantity is the relaxation time-scale, $t_{\mathrm{r},i} \equiv 1/\mathcal{D}_i$, where $i$ is either $\Ec$ or $J$. In this section, results are shown in terms of the relaxation time-scales. In \F\,\ref{fig:t_r_J_radii_various_inclinations}, the angular-momentum relaxation time-scales $t_{\mathrm{r},J}$ are shown as a function of $r_\mathrm{init}$. In each panel, the different symbols correspond to different inclinations, restricting to prograde inclinations (retrograde inclinations are included in \F\,\ref{fig:t_r_J_radii_various_inclinationsR}). Each panel corresponds to a different combination of $N_\mathrm{s}$ and $\tilde{\Sigma}_{\mathrm{s},0,i}$. The time-scales are averaged over the initial $\dot{\phi}$, or, equivalently, the initial $J/J_\mathrm{c}$. Similarly, in \F\,\ref{fig:t_r_E_radii_various_inclinations}, the energy relaxation time-scales are plotted, in this case for a single value of $N_\mathrm{s}$ and $\tilde{\Sigma}_{\mathrm{s},0,i}$. Typically, the energy relaxation time-scales are roughly an order-of-magnitude longer than the angular-momentum time-scales. Therefore, in the following, we will consider angular-momentum relaxation only. 

\begin{figure*}
\center
\iftoggle{ApJFigs}{
\includegraphics[scale = 0.46, trim = 15mm 0mm 0mm 0mm]{unbound_fractions_index_par4_0_index_par5_0_test06.eps}
\includegraphics[scale = 0.46, trim = 15mm 0mm 0mm 0mm]{unbound_fractions_index_par4_0_index_par5_2_test06.eps}
\includegraphics[scale = 0.46, trim = 15mm 0mm 0mm 0mm]{unbound_fractions_index_par4_1_index_par5_0_test06.eps}
\includegraphics[scale = 0.46, trim = 15mm 0mm 0mm 0mm]{unbound_fractions_index_par4_1_index_par5_2_test06.eps}
}{
\includegraphics[scale = 0.46, trim = 15mm 0mm 0mm 0mm]{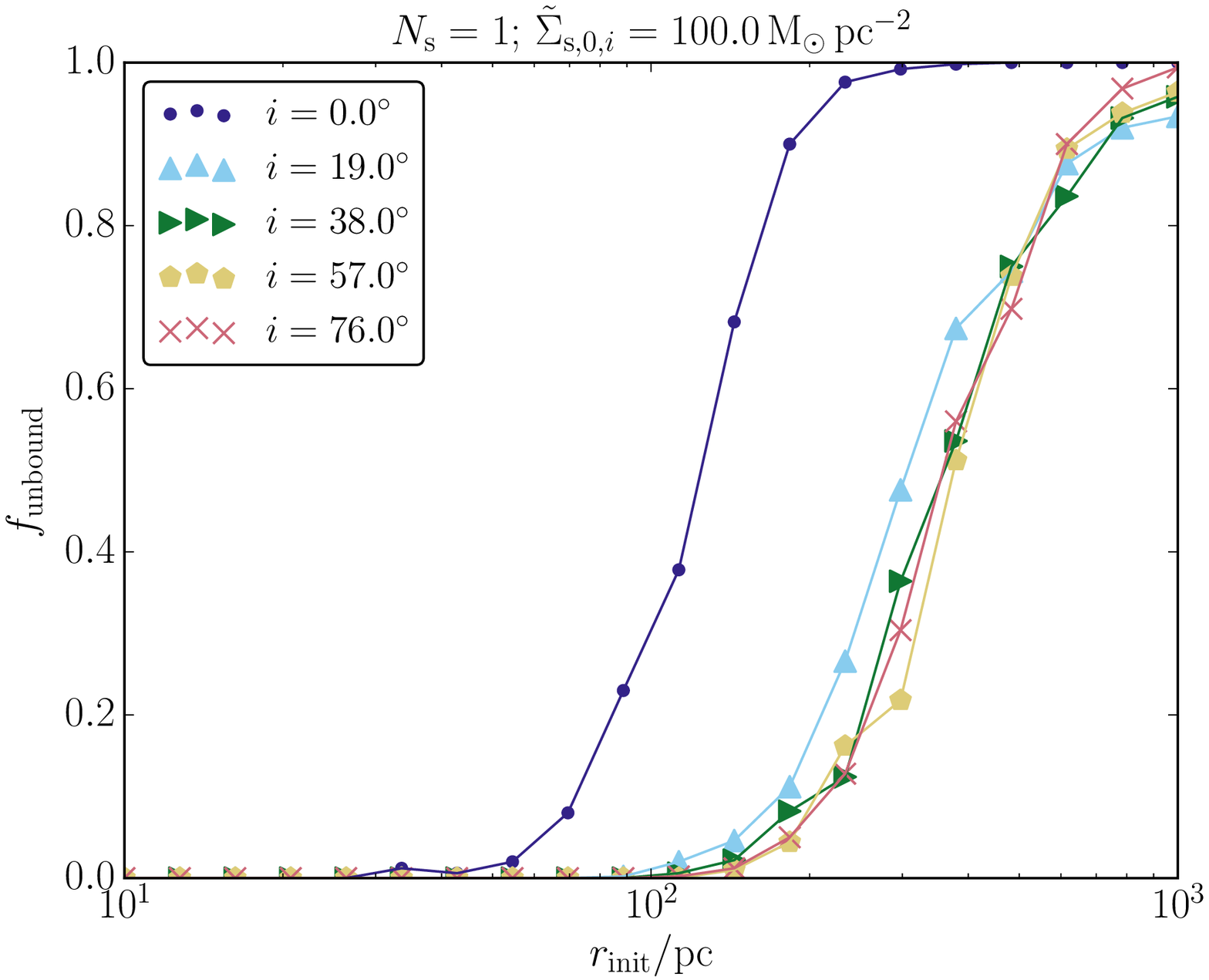}
\includegraphics[scale = 0.46, trim = 15mm 0mm 0mm 0mm]{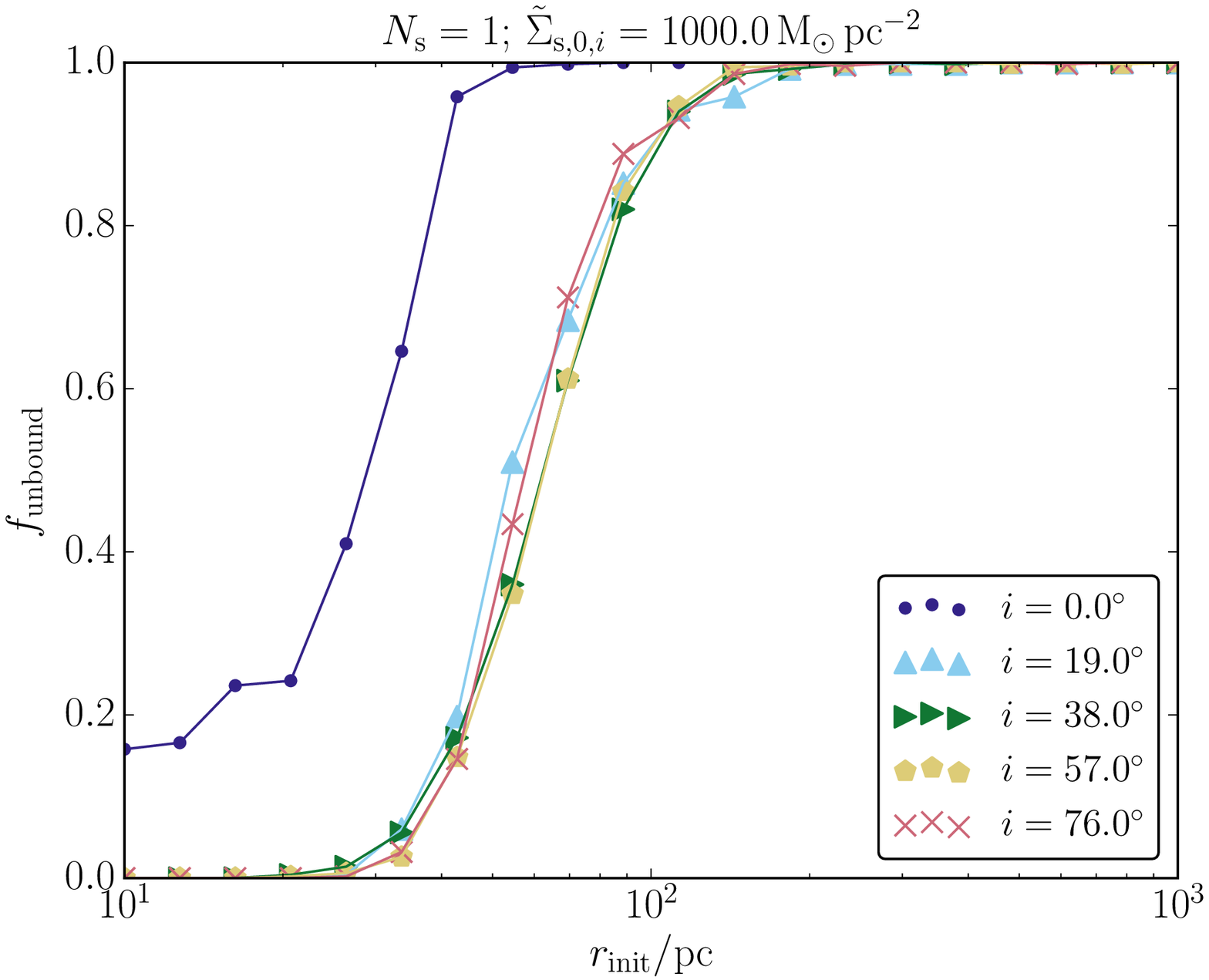}
\includegraphics[scale = 0.46, trim = 15mm 0mm 0mm 0mm]{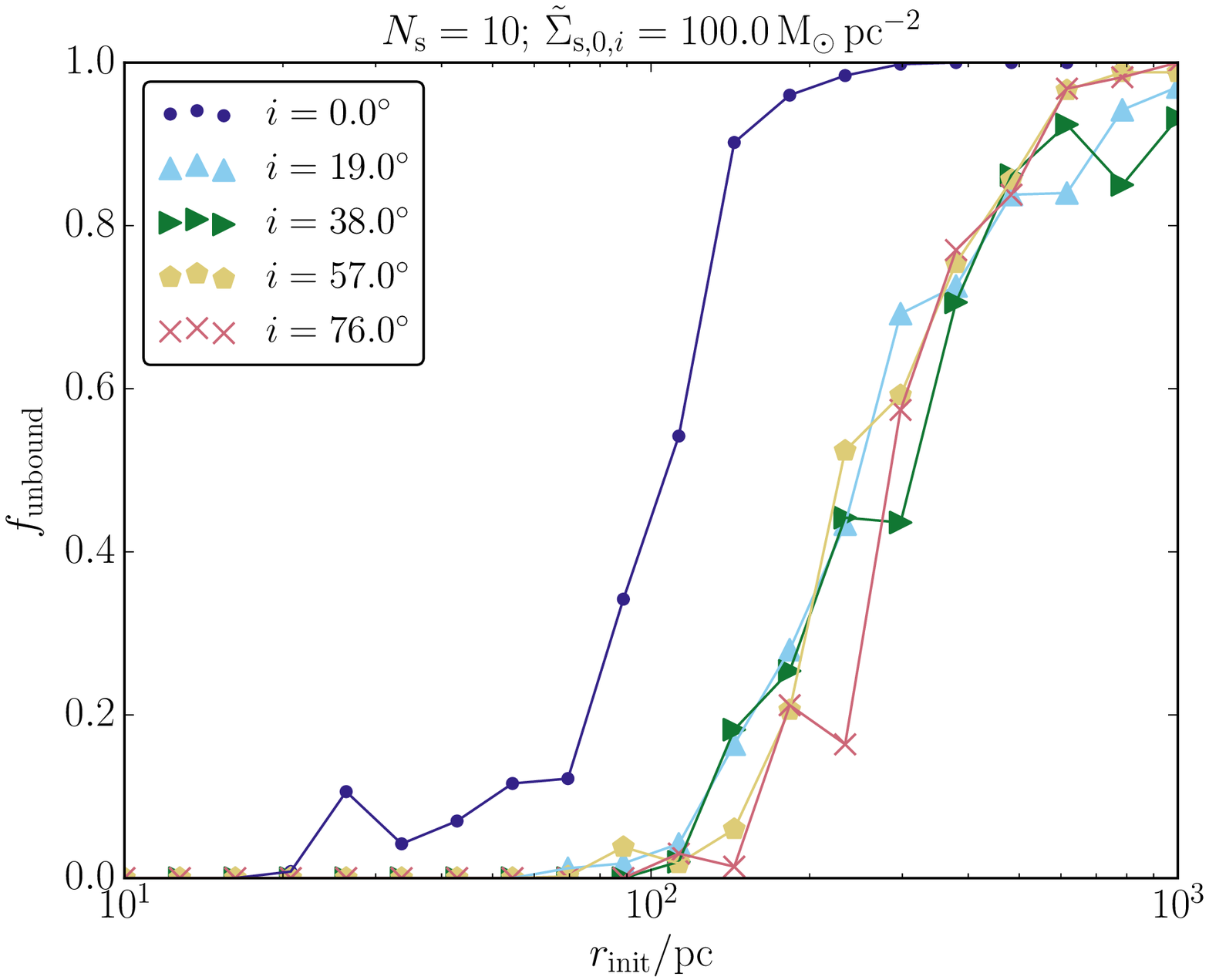}
\includegraphics[scale = 0.46, trim = 15mm 0mm 0mm 0mm]{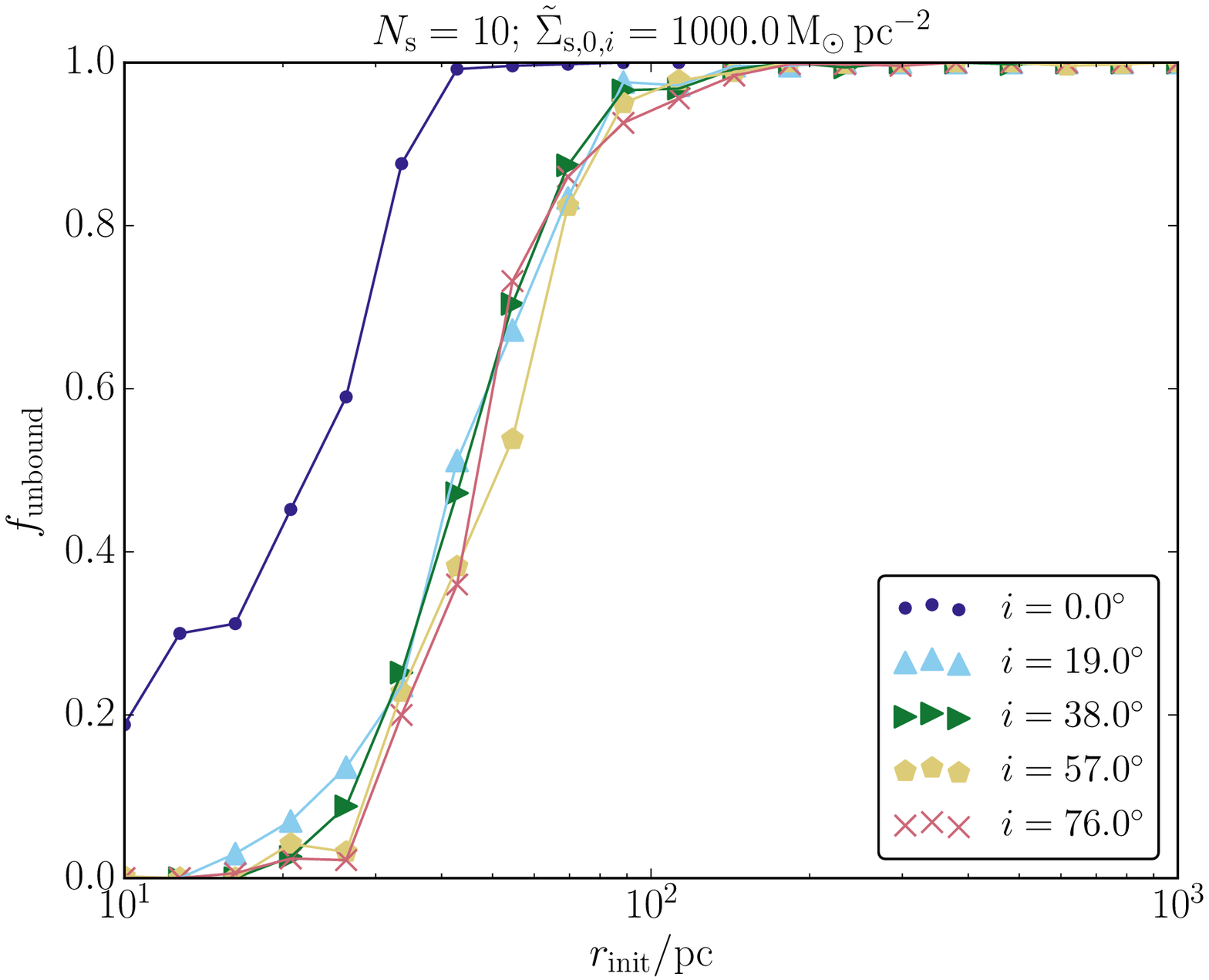}
}
\caption{\small Similar in format to \F\,\ref{fig:t_r_J_radii_various_inclinations}, now showing on the vertical axes the fractions of orbits that become unbound in the integrations. Unbound is defined as $r>10^3\,\mathrm{pc}$ and an orbital speed larger than the local escape speed at the end of the simulation. The fractions are computed over the $N_\mathrm{MC}=100$ realizations, and averaged over the initial $J/J_\mathrm{c}$. }
\label{fig:unbound}
\end{figure*}

In addition, we show in \F\,\ref{fig:unbound} the fractions of unbound orbits in the simulations as a function of the initial distance to the GC. Here, we define an orbit to be unbound if $r>10^3\,\mathrm{pc}$ and the orbital speed is larger than the local escape speed at the end of the integration. The fractions are computed over the $N_\mathrm{MC}=100$ realizations, and averaged over the initial $J/J_\mathrm{c}$. 

Generally, the following trends are revealed in Figures\,\ref{fig:t_r_J_radii_various_inclinations} and \ref{fig:unbound}.
\begin{enumerate}
\item Relaxation by spirals is inefficient close to the center ($r\lesssim 10\,\mathrm{pc}$), where the MBH and bulge potentials dominate. At larger distances($r\sim 300 \mathrm{pc}$), $t_{\mathrm{r},J}$ can be as short as $\sim 500 \, \mathrm{Myr}$.
\item Relaxation by spirals increases in efficiency for larger $\tilde{\Sigma}_{\mathrm{s},0,i}$ (larger overall spiral gas surface density). In addition, relaxation is more effective for a larger number of spiral arm events ($N_\mathrm{s}=10$ vs. 1) with a shorter typical duration (10 vs. 50 Myr). However, as the strength of the spiral perturbations increases, the probability for becoming unbound increases as well. In fact, for all values of $\tilde{\Sigma}_{\mathrm{s},0,i}$ considered here, particles become unbound for $r_\mathrm{init}>r_\mathrm{init,crit}$, where the critical initial radius depends on $i$, $N_\mathrm{s}$ and $\tilde{\Sigma}_{\mathrm{s},0,i}$. For this reason, larger spiral perturbations do not necessarily lead to higher disruption rates (for the latter, we require the orbit to remain bound). This effect will be quantified in \S\,\ref{sect:rates}.
\item Spirals are very efficient at strongly perturbing orbits at zero inclination $(i=0^\circ$), in which case many test particles become unbound at typically several tens of pc. At larger inclinations, perturbations are weaker and particles remain bound at larger radii. 
\end{enumerate}

\section{Disruption rates}
\label{sect:rates}
Having computed diffusion coefficients associated with nuclear spiral arms in \S\,\ref{sect:dif}, we here proceed to compute the disruption rates by the MBH. First, for reference, we briefly review some aspects of loss-cone theory (see also, e.g., \citealt{2013degn.book.....M}).

\begin{figure}
\center
\iftoggle{ApJFigs}{
\includegraphics[scale = 0.46, trim = 5mm 0mm 0mm 0mm]{r_lc.eps}
}{
\includegraphics[scale = 0.46, trim = 5mm 0mm 0mm 0mm]{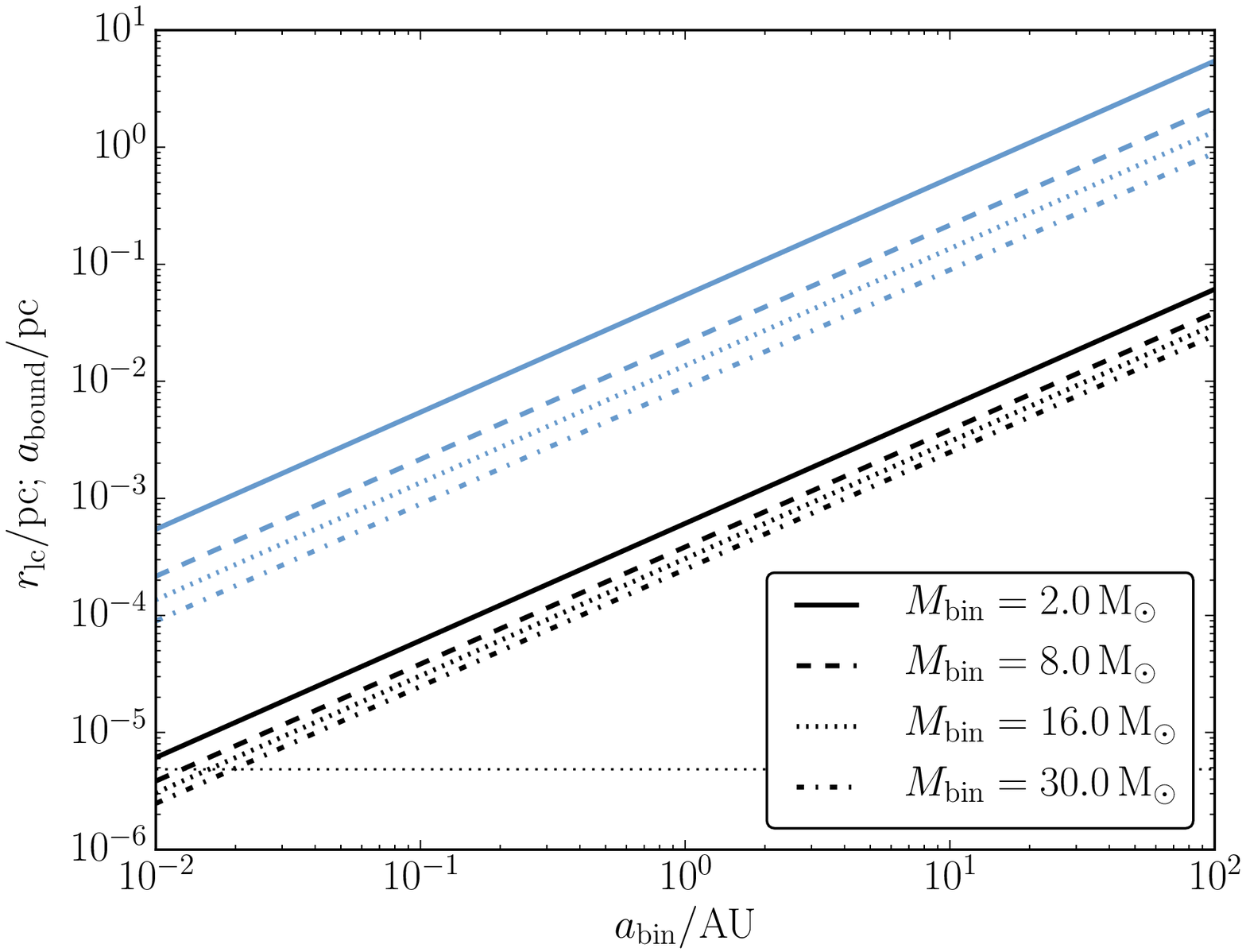}
}
\caption{\small The loss cone radii $r_\mathrm{lc}$ (black lines, cf. equation~\ref{eq:r_lc}) and the semimajor axis $a_\mathrm{bound}$ of the bound star after binary tidal disruption by the MBH (blue lines, cf. equation~\ref{eq:a_bound}) as a function of $a_\mathrm{bin}$. Several values of $M_\mathrm{bin}$ are assumed (corresponding to the binary models adopted in \S\,\ref{sect:rates}), indicated in the legend. The black horizontal dotted line shows $r_\mathrm{lc}=1\,\mathrm{AU}$, the assumed loss-cone radius for single stars. }
\label{fig:r_lc}
\end{figure}

Below, we assume that objects (i.e., single stars or binaries) are disrupted when they pass within the tidal disruption or loss cone radius,
\begin{align}
\label{eq:r_lc}
r_\mathrm{lc} \approx R \, (M_\bullet/M)^{1/3}.
\end{align}
For single stars, $M$ is the stellar mass and $R$ the stellar radius, whereas for binaries, $M=M_\mathrm{bin}=M_1+M_2$ is the binary mass and $R$ is the binary semimajor axis $a_\mathrm{bin}$. Any object initially on an orbit with a pericenter distance less than $r_\mathrm{lc}$ is rapidly depleted (within the orbital time $P$); therefore, the loss cone needs to be refilled in order to drive steady disruption. For reference, $r_\mathrm{lc}$ is plotted as a function of $a_\mathrm{bin}$ with the black lines in \F\,\ref{fig:r_lc} for several values of $M_\mathrm{bin}$.

Close to the MBH, the orbital time-scale is short but relaxation is typically inefficient (this applies both to relaxation by stars, and by nuclear spiral arms). This implies that the loss-cone orbits are depleted, and refilling is driven by the slow process of relaxation (empty loss-cone regime). Far away from the MBH, relaxation (by stars) occurs more efficiently, but the orbital time-scale is longer as well. Rather than being determined by relaxation, loss cone refilling is limited by the orbital time-scale (full loss-cone regime). Therefore, the disruption rate in the full loss-cone regime cannot be enhanced by relaxation processes such as nuclear spiral arms. Any enhancement must originate from the empty loss-cone regime, where refilling is limited by relaxation. Typically, the flux of objects into the loss cone peaks near the transition radius between the empty and full loss cones. 

In the case of the GC and assuming that relaxation is driven by stars, the loss-cone flux for single stars peaks around the sphere of influence of the MBH, i.e., at a few pc. As shown in \S\,\ref{sect:dif}, this is inside the regime where nuclear spiral arms are effective. Therefore, for {\it single} stars, nuclear spiral arms can only increase the relaxation rate in the full loss-cone regime, and therefore do not lead to significantly higher disruption rates. For {\it binaries}, $r_\mathrm{lc}$ can be much larger, depending on the binary semimajor axis. This shifts the transition from the empty to full-loss-cone regimes to much larger radii. Consequently, if there is enhanced relaxation by nuclear spiral arms (compared to relaxation by single stars) within the empty loss cone regime, then the disruption rates can be substantially increased. The aim of \S\,\ref{sect:rates} is to calculate by how much the rates are increased, using the numerical integrations of \S\,\ref{sect:dif}.

\subsection{Methodology}
\label{sect:rates:meth}
The disruption rates are computed using standard loss-cone theory, and assuming spherical symmetry. The latter is not self-consistent with the integrations in \S\,\ref{sect:dif}, in which the assumed geometry was nonspherical. Rather than modifying the standard loss-cone theory to non-spherical geometries, we here choose an approximate approach in which the rates are computed assuming spherical symmetry, taking different values of the relaxation rate by nuclear spiral arms depending on the (initial) inclination. Despite this inconsistency in the geometry, we expect that our approach still captures the most important aspects of the dependence of the disruption rates on the inclination. A self-consistent, anisotropoic loss-cone calculation is left for future work.

Let $\Ec \equiv - \frac{1}{2}v^2 + \psi(r)$ be the (negative) energy, where $\psi(r)$ is the negative of the potential. In our disruption rate calculations, we include in the potential the contributions from the MBH and the spherical bulge potential, i.e.,
\begin{align}
\psi(r) = -(\Phi_\bullet + \Phi_\mathrm{bulge}),
\end{align}
with $\epsilon_\mathrm{soft} = 0$ and $\gamma_x=\gamma_y=\gamma_z=1$ (cf. \S\,\ref{sect:orbit:pot}). The density $\rho(r)$ of single stars is computed from $\Phi_\mathrm{bulge}$ using the Poisson equation, and we assume the same mass for all background stars, $M_\star=1\,\msun$, i.e., the number density $n(r) = \rho(r)/M_\star$.

We adopt the Cohn-Kulsrud boundary layer formalism \citep{1978ApJ...226.1087C} which describes the angular-momentum flux into the loss cone, i.e., $F_\mathrm{lc}(\Ec)$, the number of stars lost per unit time and energy $\Ec$. The formalism is based on matching the non-averaged and orbit-averaged solutions to the Fokker-Planck equation in angular-momentum space. The resulting steady-state angular-momentum flux into the loss cone is
\begin{align}
F_\mathrm{lc}(\Ec) \approx 4\pi^2 P_\mathrm{r}(\Ec) J_\mathrm{c}^2(\Ec) \bar{\mu}(\Ec) \left \{ \ln \left [\Rc_0(\Ec)^{-1} \right ] \right \}^{-1} f(\Ec).
\label{eq:F_lc}
\end{align}
Here, $f(\Ec)$ is the distribution function in energy space of the scattered population (single stars or binaries). It is computed from the number density $n(r)$ using Eddington's formula \citep{1916MNRAS..76..572E}, i.e.,
\begin{align}
f(\Ec) = \frac{\sqrt{2}}{4\pi^2} \frac{\partial}{\partial \Ec} \int_{-\infty}^{\psi^{-1}(\Ec)} \frac{\mathrm{d}r}{\sqrt{\Ec-\psi(r)}} \frac{\mathrm{d}n(r)}{\mathrm{d}r}.
\label{eq:distr_edd}
\end{align}
The (radial) orbital period is given by
\begin{align}
P_\mathrm{r}(\Ec)=\int_0^{\psi^{-1}(\Ec)} \frac{\mathrm{d} r}{v_\mathrm{r}(r,\Ec,0)},
\end{align}
where $v_\mathrm{r}(r,\Ec,\Rc) = \{ 2[\psi(r)-\mathcal{E}] - \Rc \,J_\mathrm{c}^2(\mathcal{E})/r^2 \}^{1/2}$ is the radial orbital speed as a function of the angular-momentum variable $\Rc \equiv (J/J_\mathrm{c})^2$. In equation~(\ref{eq:F_lc}), the dependence of the distribution function of the scattered population in angular-momentum space (i.e., $\Rc$), is assumed to be logarithmic with $\Rc$ as  $\Rc\rightarrow0$. 

The quantity $\bar{\mu}(\Ec)$ is the orbit-averaged angular-momentum diffusion coefficient that describes the diffusion rate in angular-momentum space. For diffusion by stars, it can be computed explicitly from the distribution function $f(\Ec)$ and the potential $\psi(r)$; for completeness, the expression for $\bar{\mu}_\mathrm{stars}$ is given in Appendix\,\ref{app:dif_stars}. 

In the case of relaxation by nuclear spiral arms, the associated $\bar{\mu}(\Ec)$ is computed from the numerical integrations of \S\,\ref{sect:dif}, which yielded $\mathcal{D}_J(r)$, the angular-momentum diffusion coefficient due to spiral arms as a function of distance to the GC. As mentioned in \S\,\ref{sect:dif:results}, diffusion in energy space is much less efficient than diffusion in angular momentum, therefore, we neglect the former. We orbit average $\mathcal{D}_J(r)$ over radius according to (e.g., \citealt[5.5.2]{2013degn.book.....M})
\begin{align}
\label{eq:D_J_av}
\overline{\mathcal{D}}_J(\Ec) = \frac{4}{p(\Ec)} \int_0^{\psi^{-1}(\Ec)} \mathrm{d} r \, r^2 v(r,\Ec) \, \mathcal{D}_J(r),
\end{align}
where $v(r,\Ec) = \sqrt{2[\psi(r)-\Ec]}$ is the orbital speed at radius $r$ for an orbit with energy $\Ec$, and $p(\Ec)$ is the phase-space volume element per unit energy\footnote{The factor 4 in equations~(\ref{eq:D_J_av}) and (\ref{eq:p_e}) is adopted from \citet{2013degn.book.....M}; it is omitted by some authors, e.g., \citet{1987degc.book.....S}.},
\begin{align}
\label{eq:p_e}
p(\Ec) = 4 \int_0^{\psi^{-1}(\Ec)} \, \mathrm{d} r \, r^2 v(r,\Ec).
\end{align}
Subsequently, $\bar{\mu}(\Ec)$ is computed from
\begin{align}
\bar{\mu}(\Ec) = \bar{\mu}_\mathrm{stars}(\Ec) +  \overline{\mathcal{D}}_J(\Ec),
\end{align}
i.e., we include both relaxation by stars and spirals; any enhancement of the disruption rate by stars is due to nuclear spiral arms.

\begin{figure}
\center
\iftoggle{ApJFigs}{
\includegraphics[scale = 0.42, trim = 5mm 0mm 0mm 0mm]{DFC_J_averaged_paper_index_par2_9_index_par4_1_index_par5_2_test06.eps}
}{
\includegraphics[scale = 0.42, trim = 5mm 0mm 0mm 0mm]{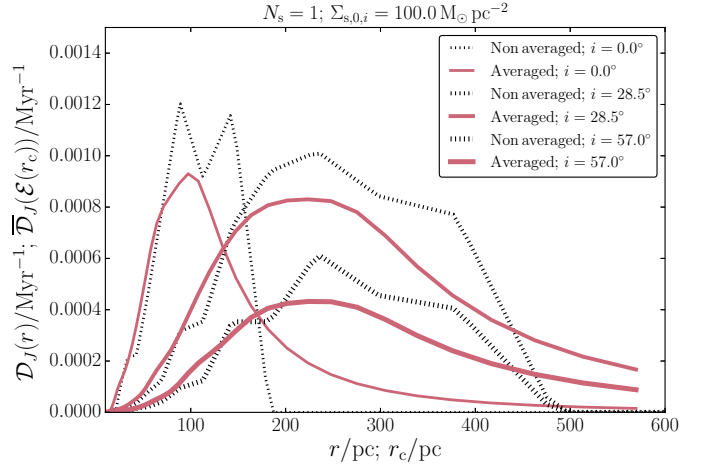}
}
\caption{\small The non-averaged angular-momentum diffusion coefficient $\mathcal{D}_J(r)$ as a function of $r$ (black dotted lines), and the averaged $\overline{\mathcal{D}}_J(\Ec)$ as a function of $r_\mathrm{c} = r_\mathrm{c}(\Ec)$ (solid red lines), for three different inclinations (different line thicknesses), and fixed values of $N_\mathrm{s}$ and $\tilde{\Sigma}_{\mathrm{s},0,i}$ (indicated in the top). }
\label{fig:DFC_J_averaged}
\end{figure}

For illustration, we show in \F\,\ref{fig:DFC_J_averaged} the non-averaged $\mathcal{D}_J(r)$ as a function of $r$ (black dotted lines), and the averaged $\overline{\mathcal{D}}_J(\Ec)$  as a function of $r_\mathrm{c} = r_\mathrm{c}(\Ec)$ (solid red lines), for three different inclinations, and fixed values of $N_\mathrm{s}$ and $\tilde{\Sigma}_{\mathrm{s},0,i}$. We recall that $r_\mathrm{c}(\Ec)$ is the radius of a circular orbit (cf. equation~\ref{eq:r_c_def}). The orbit-averaging procedure tends to smooth out sharp kinks in the non-averaged diffusion coefficients. 

The factor in equation~(\ref{eq:F_lc}) depending on $\Rc_0(\Ec)$ takes into account the empty and full loss-cone regimes. It is given by
\begin{align}
\Rc_0(\Ec)= \Rc_\mathrm{lc}(\Ec) \exp \left [-\frac{q(\Ec)}{\xi(q(\Ec))}  \right ]\leq\Rc_\mathrm{lc}(\Ec),
\end{align}
where $\Rc_\mathrm{lc}(\Ec)$ is the value of $\Rc$ corresponding to the loss cone; it can be found by setting $v_\mathrm{r}(r,\Ec,\Rc_\mathrm{lc}) = 0$ (the orbit just grazing the loss cone), i.e.,  
\begin{align}
\Rc_\mathrm{lc}(\Ec)= \left [2r_\mathrm{lc}^2/J_\mathrm{c}^2(\Ec) \right ][\psi(r_\mathrm{lc})-\Ec].
\end{align}
The function $q(\Ec)$, defined as
\begin{align}
q(\Ec)\equiv \frac{P_\mathrm{r}(\Ec) \bar{\mu}(\Ec)}{\Rc_\mathrm{lc}(\Ec)},
\end{align}
characterizes the full and empty loss-cone regimes, and the function $\xi$ of $q$ is given by
\begin{align}
\xi(q) \equiv 1 - 4 \sum_{m=1}^\infty \frac{\exp \left (-\alpha_m^2 q/4 \right )}{\alpha_m^2}.
\end{align}
Here, $\alpha_m$ is the $m^\mathrm{th}$ zero of the Bessel function of the first kind, $J_0(\alpha)$. For $q\ll1$, $\xi(q) \approx (2/\sqrt{\pi}) \sqrt{q}$, whereas for $q\gg1$, $\xi(q) \approx 1$.

For illustration, we briefly discuss the two limits of $q$ in equation\,(\ref{eq:F_lc}). In the empty loss-cone regime ($q\ll1$, or, more precisely, $q\ll - \ln \Rc_\mathrm{lc}$), $F_\mathrm{lc} \approx 4 \pi^2 P_\mathrm{r}(\Ec) J_\mathrm{c}^2(\Ec) \bar{\mu}(\Ec) f(\Ec) (-\ln\Rc_\mathrm{lc})^{-1}$. Using $p(\Ec) \approx J_\mathrm{c}^2 (\Ec) P_\mathrm{r}(\Ec)$ and $N(\Ec) = 4 \pi^2 p(\Ec) f(\Ec)$ \citep[][3.2.1]{2013degn.book.....M}, this can be written as
\begin{align}
\label{eq:F_lc_approx_empty}
F_\mathrm{lc}(\Ec) \approx \frac{N(\Ec) }{2 \ln [J_\mathrm{c}(\Ec)/J_\mathrm{lc}(\Ec)] \, \overline{t}_\mathrm{r}(\Ec)} \qquad (\mathrm{empty \, loss \, cone}),
\end{align}
where $N(\Ec)$ is the number of stars per unit energy, and $\overline{t}_\mathrm{r}(\Ec) \equiv 1/\bar{\mu}(\Ec)$ is the relaxation time-scale. 

In the full loss-cone regime ($q\gg1$, or, more precisely, $q\gg - \ln \Rc_\mathrm{lc}$), $F_\mathrm{lc} \approx 4 \pi^2 J_\mathrm{lc}^2(\Ec) f(\Ec)$, which can be written as
\begin{align}
\label{eq:F_lc_approx_full}
F_\mathrm{lc}(\Ec) \approx \frac{J_\mathrm{lc}^2(\Ec)}{J_\mathrm{c}^2(\Ec)} \frac{N(\Ec)}{P_\mathrm{r}(\Ec)} \qquad (\mathrm{full \, loss \, cone}).
\end{align}
Equations~(\ref{eq:F_lc_approx_empty}) and (\ref{eq:F_lc_approx_full}) are the well-known relations for the empty and full loss-cone fluxes (e.g., equations 6.8 and 6.9 of \citealt{2005PhR...419...65A}).

From the loss-cone flux computed using the Cohn-Kulsrud formalism (equation~\ref{eq:F_lc}), we compute the total disruption rate by integrating over all energies and binary semimajor axes $a_\mathrm{bin}$ (in the case of binaries), i.e.,
\begin{align}
\label{eq:gamma_int}
\Gamma = \int_{a_\mathrm{bin,min}}^{a_\mathrm{bin,max}} \mathrm{d} a_\mathrm{bin} \int \mathrm{d} \Ec \, F_\mathrm{lc}(\Ec) g_\mathrm{bin}(\Ec,a_\mathrm{bin}).
\end{align}
Here, the function $g_\mathrm{bin}(\Ec,a_\mathrm{bin})$ takes into account the binary occurrence rate, the binary semimajor axis distribution, and the process of binary evaporation (with a similar approach as \citealt{2007ApJ...656..709P}). For single stars, formally $g_\mathrm{bin}(\Ec,a_\mathrm{bin}) = \delta(a_\mathrm{bin} - r_\mathrm{lc})$, where $\delta$ is the delta function, and we set $r_\mathrm{lc}=1\,\mathrm{AU}$. For binaries, 
\begin{align}
\label{eq:g_bin}
g_\mathrm{bin}(\Ec,a_\mathrm{bin}) = f_\mathrm{PDMF} f_\mathrm{bin} \, \frac{\mathrm{d} N}{\mathrm{d} a_\mathrm{bin}} \, \mathrm{min} \left [ 1, \frac{\overline{t}_\mathrm{evap}(\Ec,a_\mathrm{bin})}{\mathrm{min} (t_\mathrm{H},t_\star)} \right ],
\end{align}
where $f_\mathrm{PDMF}$ is the fraction of stars with respect to the assumed background stellar distribution $n(r)$ for the mass range of a binary model given the present-day mass function (PDMF; see \S\,\ref{sect:rates:bin} below), $f_\mathrm{bin}$ is the fraction of binaries with respect to $n(r)$, $\mathrm{d}N/\mathrm{d} a_\mathrm{bin}$ is the initial normalized binary semimajor axis distribution, $t_\mathrm{H}=10 \,\mathrm{Gyr}$ is an (approximate) Hubble time, $t_\star$ is the stellar lifetime (depending on the binary component masses), and $\overline{t}_\mathrm{evap}(\Ec,a_\mathrm{bin})$ is an orbit-averaged evaporation time-scale. The latter time-scale captures the effect of perturbations of binaries by the stellar background distribution -- soft binaries tend to become softer due to this process, eventually unbinding them \citep{1975MNRAS.173..729H,1983ApJ...268..319H,1983ApJ...268..342H}. The local evaporation time-scale is computed from \citep[S7.5.7]{2008gady.book.....B}
\begin{align}
\label{eq:t_evap_local}
t_\mathrm{evap}(r,a_\mathrm{bin}) = \frac{M_\mathrm{bin}}{M_\star} \frac{\sigma(r)}{16 \sqrt{\pi} \rho(r) G a_\mathrm{bin} \ln \Lambda_\mathrm{bin}(r,a_\mathrm{bin})}.
\end{align}
Here, the velocity dispersion $\sigma(r)$ is computed from the isotropic Jeans equation,
\begin{align}
\label{eq:sigma}
n(r) \sigma^2(r) = \int_r^\infty \mathrm{d} r' \, \frac{G M(r') n(r')}{{r'}^2},
\end{align}
where the enclosed mass is
\begin{align}
M(r) = M_\bullet + 4 \pi M_\star \int_0^r n(r') r'^2 \, \mathrm{d} r'.
\label{eq:M_r}
\end{align}
The binary Coulomb factor $\Lambda_\mathrm{bin}$ is taken to be
\begin{align}
\Lambda_\mathrm{bin}(r,a_\mathrm{bin}) = a_\mathrm{bin} \sigma^2(r)/(4G M_\star).
\end{align}
For hard binaries ($\Lambda_\mathrm{bin} < 1$), we assume $t_\mathrm{evap} = 10^8\,\mathrm{Gyr}$, i.e., essentially no evaporation; we neglect the process of binary hardening. Subsequently, the local evaporation time-scale equation~(\ref{eq:t_evap_local}) is orbit averaged (cf. equation~\ref{eq:D_J_av}, with the angular-momentum diffusion coefficient $\mathcal{D}_J$ replaced by the evaporation time-scale), yielding $\overline{t}_\mathrm{evap}(\Ec,a_\mathrm{bin})$. The limits of the integration over $a_\mathrm{bin}$ are discussed below, in \S\,\ref{sect:rates:bin}.

In practice, the calculations are carried out on a grid in energy; the energies in this grid are computed according to $\Ec=\psi(r)$ from a logarithmic grid in radius with $5\times10^{-2}\,\mathrm{pc}<r<10^3\,\mathrm{pc}$. Integrals over radius are carried out over the range $10^{-8}\,\mathrm{pc} < r < 10^{10}\,\mathrm{pc}$.

\begin{table*}
\centering
\begin{tabular}{ccccccccccc}
\toprule
Model & Description & $M_\mathrm{bin}$ & $M_\mathrm{1,low}$ & $M_\mathrm{1,up}$ & $f_\mathrm{PDMF}$ &$f_\mathrm{bin}$ & $t_\star$ &  Period & $a_\mathrm{bin,min}$ & $a_\mathrm{bin,max}$ \\
& & ($\msun$) & ($\msun$) & ($\msun$) & & & ($\mathrm{Myr}$) & distribution & ($\mathrm{AU}$) & ($\mathrm{AU}$) \\
\midrule
M1 & Solar type 	& 2 & 1 & 2 & 		$9.41 \times 10^{-1}$ 	& 0.4 & $1.1\times10^4$ 	& lognormal 	& $3.35\times 10^{-2}$ & 50 \\
M2 & HVS 		& 8 & 2.5 & 4.5 & 	$2.15 \times 10^{-2}$	& 0.6 & 179.2 			& MDS1 		& $5.32\times 10^{-2}$ & 0.984\\
M3 & S-star 		& 16 & 5 & 15 & 	$1.79 \times 10^{-3}$	& 0.8 & 37.2 			& MDS1 		& $6.70\times 10^{-2}$ & 6 \\
M4 & O star 		& 30 & 15 & 120 &	$3.66 \times 10^{-6}$ 	& 1.0 & 12.8 			& MDS1 		& $8.27\times 10^{-2}$ & 50 \\
\bottomrule
\end{tabular}
\caption{ An overview of the binary models adopted for the rate calculations in \S\,\ref{sect:rates}. }
\label{table:binary_models}
\end{table*}

\subsection{Binary star models}
\label{sect:rates:bin}

We adopt four models representing different types of binary stars. An overview is given in Table~\ref{table:binary_models}. The parameters included in each model are the PDMF fraction $f_\mathrm{PDMF}$, the total binary mass $M_\mathrm{bin}$, the semimajor axis/orbital period distribution, the fraction $f_\mathrm{bin}$ of binaries with respect to the assumed background stellar distribution $n(r)$, and the stellar main-sequence lifetime $t_\star$. For simplicity, we assume equal-mass-ratio binaries, and we use a `typical' mass, $M_\mathrm{bin}$, to compute the evaporation time-scales (cf. equation~\ref{eq:t_evap_local}) and the limits on the integration over binary separations (see below), whereas we assume a range of primary star masses when computing $f_\mathrm{PDMF}$. 

The background distribution $n(r)$ is assumed to consist of $1\,\msun$ stars, but the binaries considered here are more massive, and, therefore, less common. We take the relative frequency of the stars of the binaries into account with the fraction $f_\mathrm{PDMF}$. This fraction is computed assuming a multicomponent power law Miller-Scalo initial mass function (IMF; \citealt{1979ApJS...41..513M}) which is converted to the PDMF assuming a constant star-formation history (\citealt{2004ApJ...601..319F}, see also \citealt{2010RvMP...82.3121G}) over $t_\mathrm{H}=10 \,\mathrm{Gyr}$, and using the main-sequence lifetime $t_\star$. For a given $M_\mathrm{bin}$, we compute $t_\star$, the main-sequence life-time of a star with mass $M_\mathrm{bin}/2$ and metallicity $z=0.02$, using \textsc{SSE} \citep{2000MNRAS.315..543H} within the \textsc{AMUSE} framework \citep{2013CoPhC.183..456P,2013A&A...557A..84P}.

The relative frequency of stars with respect to the $1\,\msun$ population is then given by
\begin{align}
\nonumber f_\mathrm{PDMF} &= \left [ \int_{M_\mathrm{1,low}}^{M_\mathrm{1,up}} \mathrm{d} M \, \left ( \frac{\mathrm{d} N}{\mathrm{d} M} \right )_{\mathrm{IMF}} h(M)\right ]  \\
&\quad \times \left [ \int_{M_\star}^{M_{\star,\mathrm{up}}} \mathrm{d} M \, \left ( \frac{\mathrm{d} N}{\mathrm{d} M} \right )_{\mathrm{IMF}} h(M) \right ]^{-1},
\end{align}
where $M_\mathrm{1,low}$ and $M_\mathrm{1,up}$ are the adopted lower and upper ranges on the mass of the primary star of the binary (depending on the binary model), $M_\star = 1 \, \msun$, $M_{\star,\mathrm{up}} = 125 \, \msun$, and $h(M) = t_\star(M)/t_\mathrm{H}$ if $t_\star(M) < t_\mathrm{H}$, and $h(M) = 1$ otherwise. The masses $M_\mathrm{1,low}$ and $M_\mathrm{1,up}$ and the fractions $f_\mathrm{PDMF}$ are included in Table~\ref{table:binary_models}. 

The binary models include Solar-type binaries, binaries capable of producing HVSs and S-stars, and O-star binaries. For Solar-type stars, we adopt a lognormal period distribution with mean $\mu_{\log_{10}(P/\mathrm{d})}=5$ and standard deviation $\sigma_{\log_{10}(P/\mathrm{d})} = 2.3$ (\citealt{1991A&A...248..485D,2010ApJS..190....1R}; \citealt{2016arXiv160605347M}, hereafter MDS16). For the more massive binaries, we adopt the analytic fits from MDS16 (cf. equations 20-23 from that paper) for the distribution of $\log_{10}(P/\mathrm{d})$ as a function of primary mass $M_1=M_\mathrm{bin}/2$. The binary fractions $f_\mathrm{bin}$ are adopted from visual inspection of Fig. 37 of MS16. 

The following limits for $a_\mathrm{bin}$ are adopted in equation~(\ref{eq:gamma_int}). The lower limit $a_\mathrm{bin,min}$ is computed assuming an orbital period of $\log_{10}(P/\mathrm{d}) = 0.2$, the lowest value considered in \citet{2016arXiv160605347M}. With the exception of our HVS and S-star models, the upper limit $a_\mathrm{bin,max}$ is taken to be $a_\mathrm{bin,max}=50 \,\mathrm{AU}$, approximately the largest semimajor axis for which, after tidal disruption by the MBH, the bound star orbits the MBH within several pc. The semimajor axis of the bound orbit is approximately given by \citep{1988Natur.331..687H} 
\begin{align}
\label{eq:a_bound}
a_\mathrm{bound} \approx 2^{-3/2} \, a_\mathrm{bin} \, (M_\mathrm{bin}/M_2) (M_\bullet/M_\mathrm{bin})^{2/3}.
\end{align}
Here, we assume $M_2=M_\mathrm{bin}/2$. For reference, we plot equation~(\ref{eq:a_bound}) for our assumed binary models with the blue lines in \F\,\ref{fig:r_lc}. For  $a_\mathrm{bin}=50 \,\mathrm{AU}$, $a_\mathrm{bound}$ is on the order of 1 pc (with some dependence on $M_\mathrm{bin}$).  

For our HVS model (cf. model `M2' in Table~\ref{table:binary_models}), we impose a smaller $a_\mathrm{bin,max}$ because we are only interested in stars that can escape the Galaxy potential. The escape speed is approximately given by \citep{1988Natur.331..687H}
\begin{align}
\label{eq:v_esc}
v_\infty^2 \approx 2^{1/2} \, (GM_\mathrm{bin}/a_\mathrm{bin}) ( M_\bullet/M_\mathrm{bin} )^{1/3}.
\end{align}
Assuming an escape speed from the Galactic Bulge of $900\,\mathrm{km\,s^{-1}}$ \citep{2006MmSAI..77..653H}, equation~(\ref{eq:v_esc}) for our model `M2' implies a maximum binary separation of $a_\mathrm{bin,max} \approx 0.984 \, \mathrm{AU}$. We impose this maximum separation for the HVS model in equation~(\ref{eq:gamma_int}).

For our S-star model (cf. model `M3' in Table~\ref{table:binary_models}), we impose $a_\mathrm{bin,max} = 6\, \mathrm{AU}$, corresponding to a maximum semimajor axis of the bound orbit around the MBH of $\approx 0.044\, \mathrm{pc}$ (cf. equation~\ref{eq:a_bound} and \F\,\ref{fig:r_lc}), approximately consistent with the S-star orbits \citep{2003ApJ...594..812G,2005ApJ...628..246E,2008ApJ...689.1044G,2009ApJ...692.1075G}.

\subsection{Results}
\label{sect:rates:res}

\begin{figure}
\center
\iftoggle{ApJFigs}{
\includegraphics[scale = 0.45, trim = 5mm 0mm 0mm 0mm]{LC_flux_index_binary_model_3_index_r_lc_13_index_par2_0_index_par4_0_index_par5_0_test06_analysis_name_an01.eps}
}{
\includegraphics[scale = 0.45, trim = 5mm 0mm 0mm 0mm]{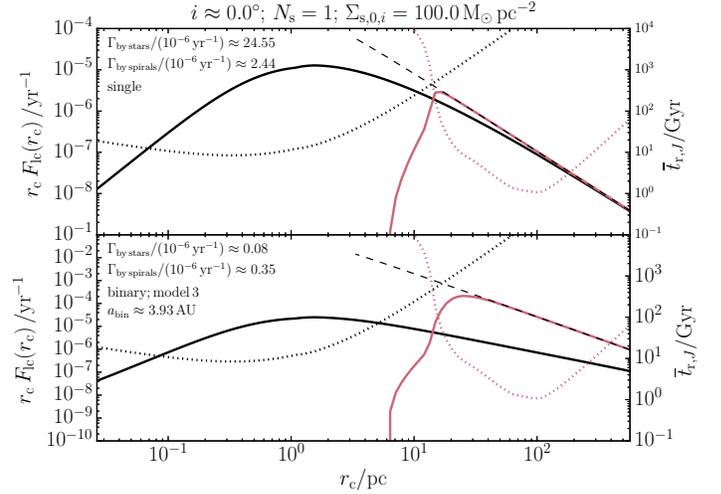}
}
\caption{\small Solid lines: the differential loss-cone rate $r_\mathrm{c} \, F_\mathrm{lc}(r_\mathrm{c})$ as a function of $r_\mathrm{c} = r_\mathrm{c}(\Ec)$ for one combination of the grid parameters $(i,N_\mathrm{s},\tilde{\Sigma}_{\mathrm{s},0,i})$ indicated in the title. Top panel: disruption of single stars; bottom panel: disruption of binary stars. Binary model `M3' (i.e., S-stars) is adopted, and $a_\mathrm{bin} \approx 3.93 \, \mathrm{AU}$. The associated relaxation time-scales $\overline{t}_\mathrm{r}(\Ec) = 1/\overline{\mathcal{D}}_J(\Ec)$ are shown with the black and red dotted lines; in this case, the right-hand axes apply. The black dashed lines correspond to the full lone-cone regime (cf. equation~\ref{eq:F_lc_approx_full}).}
\label{fig:LC_flux1}
\end{figure}

\subsubsection{Differential loss rates}
\label{sect:rates:res:dif_lc}
\begin{figure}
\center
\iftoggle{ApJFigs}{
\includegraphics[scale = 0.45, trim = 5mm 0mm 0mm 0mm]{LC_flux_index_binary_model_3_index_r_lc_13_index_par2_7_index_par4_0_index_par5_0_test06_analysis_name_an01.eps}
}{
\includegraphics[scale = 0.45, trim = 5mm 0mm 0mm 0mm]{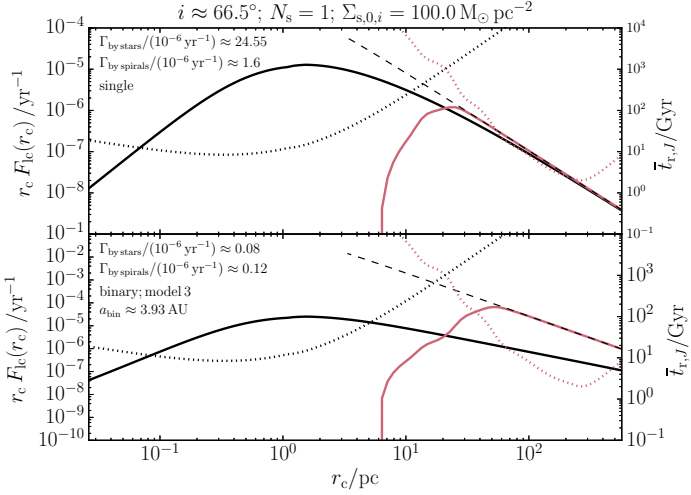}
}
\caption{\small Similar to \F\,\ref{fig:LC_flux1}, now for a higher inclination of $66.5^\circ$. }
\label{fig:LC_flux2}
\end{figure}

To illustrate the enhancement of the loss rates due to relaxation by nuclear spiral structure in the empty loss-cone regime, we show in Figs.~\ref{fig:LC_flux1} and \ref{fig:LC_flux2}, for two combinations of the grid parameters $(i,N_\mathrm{s},\tilde{\Sigma}_{\mathrm{s},0,i})$, the loss-cone flux $F_\mathrm{lc}$. For the case of the disruption of binary stars, we here do not yet integrate over the binary semimajor axis distribution although we do take into account the PDMF and the binary fraction (cf. $f_\mathrm{PDMF}$ and $f_\mathrm{bin}$ in equation~\ref{eq:g_bin}). Rather than plotting $F_\mathrm{lc}(\Ec)$ as a function of energy $\Ec$, we plot the loss-cone flux as a function of distance to the GC. We define $F_\mathrm{lc}(r)$ as the differential flux per unit distance, i.e., $F_\mathrm{lc}(\Ec) \, \mathrm{d} \Ec = F_\mathrm{lc}(r) \, \mathrm{d} r$. Also, we set $r=r_\mathrm{c}(\Ec)$, the radius of a circular orbit, and multiply $F_\mathrm{lc}(r_\mathrm{c})$ by $r_\mathrm{c}$ to get the loss rate per unit $\ln r_\mathrm{c}$ (i.e., $r_\mathrm{c}\,F_\mathrm{lc}(r_\mathrm{c})$ has dimensions of one over time). 

The top panel in \F\,\ref{fig:LC_flux1} applies to the disruption of single stars ($r_\mathrm{lc}=1\,\mathrm{AU}$). The solid black (red) lines correspond to relaxation by stars (nuclear spiral arms). For reference, the associated relaxation time-scales $\overline{t}_\mathrm{r}(\Ec) = 1/\overline{\mathcal{D}}_J(\Ec)$ are shown with the black and red dotted lines; in this case, the right-hand axes apply. For relaxation by single stars, the differential rate peaks at several pc, near the radius of influence. In the case of relaxation by nuclear spiral arms, the peak occurs at larger radii, where the relaxation rate is enhanced compared to relaxation by single stars. The relaxation time-scale by spirals is shorter at radii beyond the radius of the peak; however, the latter regime lies within the full loss-cone regime for single stars. Therefore, the rates are not enhanced, and the integrated rate due to spirals only, $\Gamma_\mathrm{by\,spirals} \approx 2 \times 10^{-6}\,\mathrm{yr^{-1}}$, is actually lower in this example (by an order of magnitude) compared to the integrated rate due to stellar relaxation, $\Gamma_\mathrm{by\,stars} \approx 2.5\times10^{-5}\,\mathrm{yr^{-1}}$ (these rates are indicated in the top panel in \F\,\ref{fig:LC_flux1}; note that the rates given below in \S\,\ref{sect:rates:res:tot_lc} always include relaxation by stars). 

The situation is different for binaries. In the bottom panel of \F\,\ref{fig:LC_flux1}, we consider the disruption of binaries (binary model `M3, i.e., representing S-stars) with a semimajor axis of $\approx 3.93\,\mathrm{AU}$, for which the loss-cone radius is $r_\mathrm{lc} \approx 1.2 \times 10^{-3}\, \mathrm{pc}$ (cf. equation\,\ref{eq:r_lc} or \F\,\ref{fig:r_lc}). For relaxation by single stars, the integrated rate over energies (indicated in the panels; note that these rates are not integrated over $a_\mathrm{bin}$) would be larger compared to the case of disruptions of single stars if the PDMF were ignored. In \F\,\ref{fig:LC_flux1}, the PDMF is taken into account, however; consequently, the integrated binary disruption rate is lower compared to the single star disruption rate. Nuclear spirals do significantly enhance the binary disruption rate compared to single stars. Comparing relaxation by nuclear spirals to relaxation by stars, the binary disruption rate is higher by a factor of $\approx 4.4$. 

For the disruption of binaries and radii $r_\mathrm{c} \gtrsim 30 \, \mathrm{pc}$, spirals sufficiently enhance the relaxation rate to drive relaxation into the full loss-cone regime. This is clear from the solid red line in the bottom panel of \F\,\ref{fig:LC_flux1}, which approaches the black dashed line (representing the full lone-cone regime, cf. equation~\ref{eq:F_lc_approx_full}) for $r_\mathrm{c} \gtrsim 30 \, \mathrm{pc}$. In contrast, the loss-cone flux for relaxation by stars (black solid line in the bottom panel) does not reach the full loss-cone value for the radial range shown. 

A similar figure is shown in \F\,\ref{fig:LC_flux2}, now for a larger, nonzero inclination. The spiral relaxation time-scales at intermediate radii ($\sim 30\, \mathrm{pc}$) are longer compared to the coplanar case, but at large radii ($\gtrsim 50\, \mathrm{pc}$), the loss cone is still full. The integrated rates are lower compared to the case of $i=0^\circ$ in \F\,\ref{fig:LC_flux1}.

\begin{table}
\centering
\scriptsize
\begin{tabular}{cccccccc}
\toprule
& & & \multicolumn{5}{c}{$\Gamma/(10^{-6} \, \mathrm{yr^{-1}})$} \\
\midrule
& & & Single & \multicolumn{4}{c}{Binary} \\
& & & & M1 & M2 & M3 & M4 \\
\midrule
\multicolumn{3}{l}{By stars} & 24.5 & 6.8 & 0.130 & 0.031 & 0.002 \\
\midrule
\multicolumn{3}{l}{By stars and spirals} \\
$i$ & $N_\mathrm{s}$ & $\tilde{\Sigma}_{\mathrm{s},0,i}$ & & & & & \\
(deg) & & $(\msun\,\mathrm{pc^{-2}}$) \\
\midrule
0.0 & 1 & 500.0 & 29.5 & 504.6 & 0.518 & 0.227 & 0.027\ \\
9.5 & 10 & 1000.0 & 30.4 & 673.8 & 0.636 & 0.298 & 0.038\ \\
90.0 & 1 & 500.0 & 26.3 & 251.7 & 0.269 & 0.100 & 0.011\ \\
180.0 & 1 & 500.0 & 28.2 & 470.6 & 0.384 & 0.160 & 0.021\ \\
\midrule
\multicolumn{3}{l}{Maximum enhancement} & 1.2 & 99.1 & 4.9 & 9.6 & 24.6\ \\
\midrule
\midrule
& & & & &  \multicolumn{2}{c}{$N_\star = \Gamma \, t_\star$} \\
\midrule
\multicolumn{3}{l}{By stars} & & &  23 & 1 \\
\midrule
\multicolumn{3}{l}{By stars and spirals} \\
\midrule
0.0 & 1 & 500.0 & & & 93 & 8 & \\
9.5 & 10 & 1000.0 & & & 114 & 11 & \\
90.0 & 1 & 500.0 & & & 48 & 4 & \\
180.0 & 1 & 500.0 & & & 69 & 6 & \\
\midrule
\multicolumn{3}{l}{Observed/estimated number} & & & $\sim 300$ & $\sim 20$ & \\
\bottomrule
\end{tabular}
\caption{ Top part: disruption rates $\Gamma$ of single stars (`Single') and binary stars (`Binary') due to relaxation driven by stars (first data row), and driven by nuclear spiral arms (in addition to stars; all other data rows). In the latter case, each row corresponds to a different inclination $i$ (expressed in degrees), number of spiral arm events $N_\mathrm{s}$, and (constant) spiral arm surface density $\tilde{\Sigma}_{\mathrm{s},0,i}$ (in units of $\msun \, \mathrm{pc^{-2}}$).  For binary stars, four models, `M1' through `M4', are assumed (cf. Table~\ref{table:binary_models}). Details of the rate calculations are given in \S\,\ref{sect:rates:meth}. The largest enhancement factor of the rates with respect to relaxation by single stars is given in the row `Maximum enhancement'. Only a small selection of the parameters is shown here; the complete list of the disruption rates is given in Table~\ref{table:ratesall}. Bottom part: the expected number $N_\star$ of HVSs and S-stars (models `M2' and `M3'). The bottom row shows the observed number (for the S-stars; e.g., \citealt{2010RvMP...82.3121G}) and the estimated number (for HVSs; e.g., \citealt{2015ARA&A..53...15B}). The complete list of numbers is given in Table~\ref{table:numbersall}. }
\label{table:rates}
\end{table}

\subsubsection{Integrated loss rates}
\label{sect:rates:res:tot_lc}
For each grid parameter $(i,N_\mathrm{s},\tilde{\Sigma}_{\mathrm{s},0,i})$, the loss-cone flux, equation~(\ref{eq:F_lc}), is integrated over energy, and, for binaries, over the binary semimajor axis taking into account the PDMF, the semimajor axis distribution, the finite stellar lifetime, and binary evaporation (cf. equation~\ref{eq:gamma_int}). An abbreviated list of the rates is given in the top part of Table~\ref{table:rates}; a complete list is given in Table\,\ref{table:ratesall} in  Appendix~\ref{app:rates}. We include results taking into account relaxation by stars only (first data row), and taking into account relaxation driven by nuclear spiral arms in addition to stars (all other data rows). For the disruption of binary stars, the four assumed models are included in the right four columns. 

The highest binary disruption rate obtained is $\Gamma \approx 6\times 10^{-4} \, \mathrm{yr^{-1}}$ for binary model `M1' with $i=9.5^\circ$, $N_\mathrm{s}=10$, and $\tilde{\Sigma}_{\mathrm{s},0,i}=1000\,\msun\,\mathrm{pc^{-2}}$. This rate is a factor $\approx 100$ higher compared to relaxation by stars only, the largest enhancement compared to relaxation by single stars. For the other binary disruption models, nuclear spirals can enhance the rates by factors of $\sim$ 5, 10 and 25 for `M2', `M3' and `M4', respectively (cf. the bottom rows in Table~\ref{table:rates}).

Although less significantly, nuclear spiral arms also enhance the disruption rate of single stars. Our highest TDE rate is $\approx3.0\times 10^{-5} \, \mathrm{yr^{-1}}$, which is a factor of 1.2 higher compared to relaxation by stars only. 

\begin{figure}
\center
\iftoggle{ApJFigs}{
\includegraphics[scale = 0.48, trim = 10mm 0mm 0mm 0mm]{Gamma_inclination_combined_index_binary_model_2_test06R_analysis_name_an01.eps}
}{
\includegraphics[scale = 0.48, trim = 10mm 0mm 0mm 0mm]{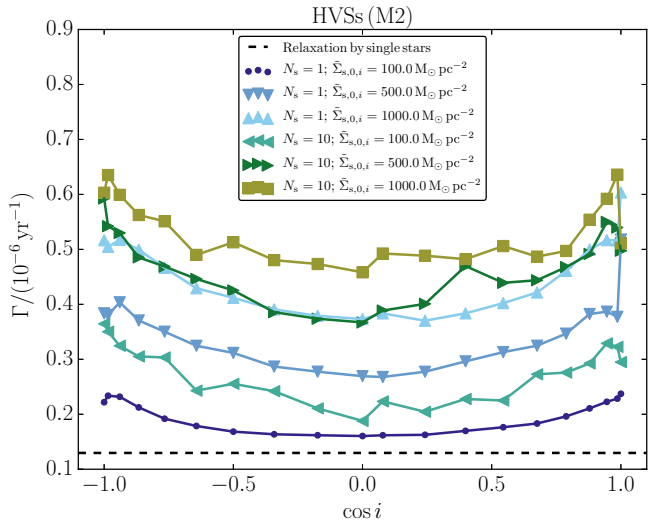}
}
\caption{\small The disruption rates by spirals and stars  as a function of the initial orbital inclination, assuming binary model `M2' (representing HVSs). Different colors and symbols correspond to different values of $N_\mathrm{s}$ and $\tilde{\Sigma}_{\mathrm{s},0,i}$, indicated in the legend. The horizontal black dashed line shows the corresponding rate for relaxation by stars only (cf. Table~\ref{table:rates}). }
\label{fig:Gamma_inclination}
\end{figure}

In \F\,\ref{fig:Gamma_inclination}, the disruption rates by spirals and stars are plotted as a function of $\cos\,i$, assuming binary model `M2' (representing HVSs). Different colors and symbols correspond to different values of $N_\mathrm{s}$ and $\tilde{\Sigma}_{\mathrm{s},0,i}$. There is a clear dependence of $\Gamma$ on $i$, with the overall normalization depending on $N_\mathrm{s}$ and $\tilde{\Sigma}_{\mathrm{s},0,i}$. For large $N_\mathrm{s}$ and/or $\tilde{\Sigma}_{\mathrm{s},0,i}$, the rates for coplanar binaries ($i=0^\circ$) are {\it lower} compared to the rates for slightly inclined binaries ($i\approx 10^\circ$). For more inclined binaries, the rates decline again. This dependence on $i$ be understood by noting that for $i=0^\circ$, many binaries become unbound due to spiral perturbations (cf. \S\,\ref{sect:dif:results}). For larger $i$, the unbinding effect becomes less dominant, whereas the diffusion rate also decreases, resulting in a maximum of the disruption rate. If $N_\mathrm{s}$ and/or $\tilde{\Sigma}_{\mathrm{s},0,i}$ are small, the overall strength of the spiral perturbations is smaller. Therefore, orbits can remain bound if $i=0^\circ$. 

Apart the above intricacy of the $i$-dependence for different $N_\mathrm{s}$ and $\tilde{\Sigma}_{\mathrm{s},0,i}$, there is generally a clear trend of decreasing $\Gamma$ with less inclined orbits. Nevertheless, even for high inclinations, the rates are still significantly higher (by factors of a few) compared to relaxation by stars only (in which $\Gamma \approx 1.3 \times 10^{-7} \, \mathrm{yr^{-1}}$, cf. Table\,\ref{table:rates} and the horizontal black dashed line in \F\,\ref{fig:Gamma_inclination}). The disruption rates are approximately symmetric around $\cos \, i =0$.

\subsubsection{Expected number of HVSs and S-stars}
\label{sect:rates:res:number}
In the bottom part of Table~\ref{table:rates}, we give the expected number of HVSs and S-stars computed from $N_\star = \Gamma \, t_\star$, where the lifetime $t_\star$ is taken from Table~\ref{table:binary_models}. A complete list is given in Table~\ref{table:numbersall} in Appendix~\ref{app:rates}. The number of HVSs in our models ranges between $\sim 50$ to $\sim 100$; the estimated number of HVSs within 100 kpc of the Galaxy is $\sim 300$ \citep{2015ARA&A..53...15B}, consistent within a factor of a few. Our predicted S-star numbers are a few, up to $\sim 10$, consistent within a factor of a few with the observed number of $\sim 20$ (e.g., \citealt{2010RvMP...82.3121G}).

\section{Discussion}
\label{sect:discussion}

\subsection{Overall rates}
\label{sect:discussion:rates}
In our calculations, the disruption rate of single stars driven by relaxation by stars is $\approx 2.5 \times 10^{-5}\,\mathrm{yr^{-1}}$, which is consistent with previous studies in which rates of $\sim 10^{-5} \, \mathrm{yr^{-1}}$ were found \citep{1999MNRAS.306...35S,1999MNRAS.309..447M}. In our models with nuclear spiral arms included as well, the disruption rate of single stars is at most $\approx 3.0\times 10^{-5} \, \mathrm{yr^{-1}}$, which is $20\%$ higher compared to relaxation by single stars only. Although the enhancement by nuclear spiral arms is not as significant as for binaries, it complements other models for enhanced TDE rates such as triaxial nuclei \citep{2013ApJ...774...87V}.

For our model `M1' (Solar-type stars), the binary disruption rate driven by stars is $\approx 0.7 \times 10^{-5} \, \mathrm{yr^{-1}}$, which is lower compared to the rate of \citealt{2003ApJ...599.1129Y}, who found, for $1\, \msun$ stars, a rate of $10^{-5} (\eta/0.1) \, \mathrm{yr}^{-1} = 4 \times 10^{-5} \, \mathrm{yr^{-1}}$, where $\eta$ is the binary fraction, which we took to be 0.4 (cf. Table\,\ref{table:binary_models}). For more massive binaries ($M_\mathrm{bin}=4\,\msun$ and higher), we found rates of at most $10^{-7} \, \mathrm{yr^{-1}}$. \citet{2013ApJ...768..153Z} found rates of $\sim 10^{-4}$ to $10^{-5} \, \mathrm{yr^{-1}}$, for various binary injection models and stellar masses between 3 and 15 $\msun$. The much higher rates found by the latter authors could be explained by noting that \citet{2013ApJ...768..153Z} assumed an IMF and not the PDMF. In addition, \citet{2013ApJ...768..153Z} also considered top-heavy IMFs. 

For our HVS and S-star model binaries, our disruption rates for relaxation by nuclear spiral arms are on the order of $10^{-6}\,\mathrm{yr}$, implying numbers of HVSs and S-stars which are roughly consistent with the observed numbers. Assuming relaxation by stars only, the expected numbers are inconsistent with the observed numbers. This result is similar to relaxation by massive perturbers \citep{2007ApJ...656..709P}. Nuclear spiral arms can be considered as a complementary channel to supply binaries to the loss cone of an MBH, with a roughly equal contribution compared to relaxation driven by massive perturbers.

\subsection{Implications for HVSs, and caveats}
\label{sect:discussion:HVS}

We assumed that the nuclear spiral arm structures are confined to the plane of the Galaxy. Consequently, the disruption rates, in particular for binaries able to produce HVSs (binary model `M2'), are dependent on the inclination $i$ of the (original) binary with respect to the Galactic plane; this dependence was shown in \F\,\ref{fig:Gamma_inclination}. The difference in disruption rates between a nearly coplanar and nearly perpendicular orbit is modest, i.e., a factor of $\sim 2$. Even for nearly perpendicular orbits, the disruption rates are still higher compared to relaxation by stars. 

An anisotropic distribution of HVSs was also considered by \citet{2010ApJ...709.1356L}, who suggested that HVSs originate from the inner clockwise-rotating stellar disk within half a parsec of the MBH in the GC. This implies an angular distribution which is distinct from ours: our inclination distribution is not linked to the clockwise disk but to the Galactic plane, suggesting that HVSs should be less common far above the this plane, compared to within it.

\begin{figure}
\center
\iftoggle{ApJFigs}{
\includegraphics[scale = 0.48, trim = 10mm 0mm 0mm 0mm]{delta_i_index_par4_0_index_par5_0_test06.eps}
\includegraphics[scale = 0.48, trim = 10mm 0mm 0mm 0mm]{delta_i_index_par4_1_index_par5_0_test06.eps}
}{
\includegraphics[scale = 0.48, trim = 10mm 0mm 0mm 0mm]{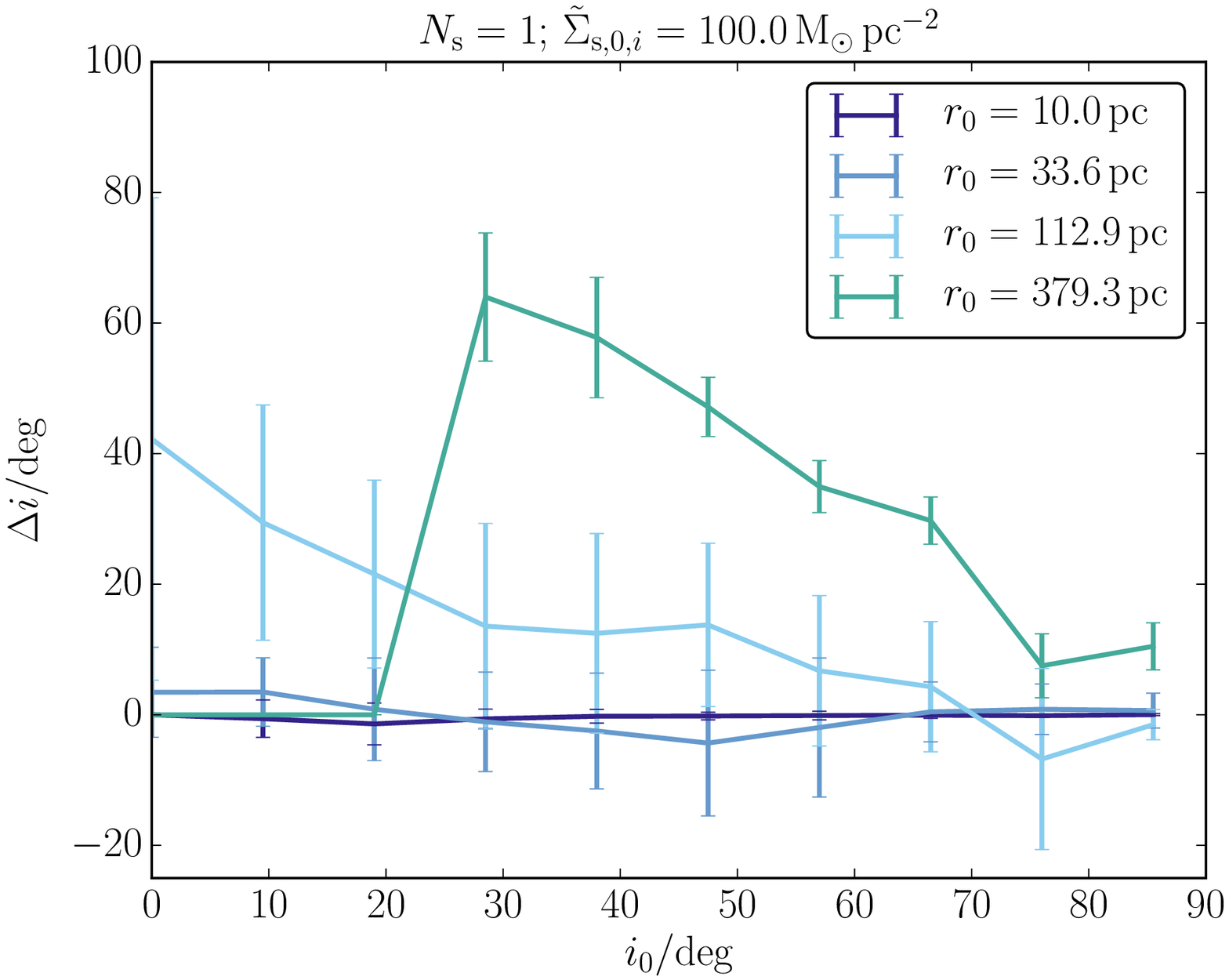}
\includegraphics[scale = 0.48, trim = 10mm 0mm 0mm 0mm]{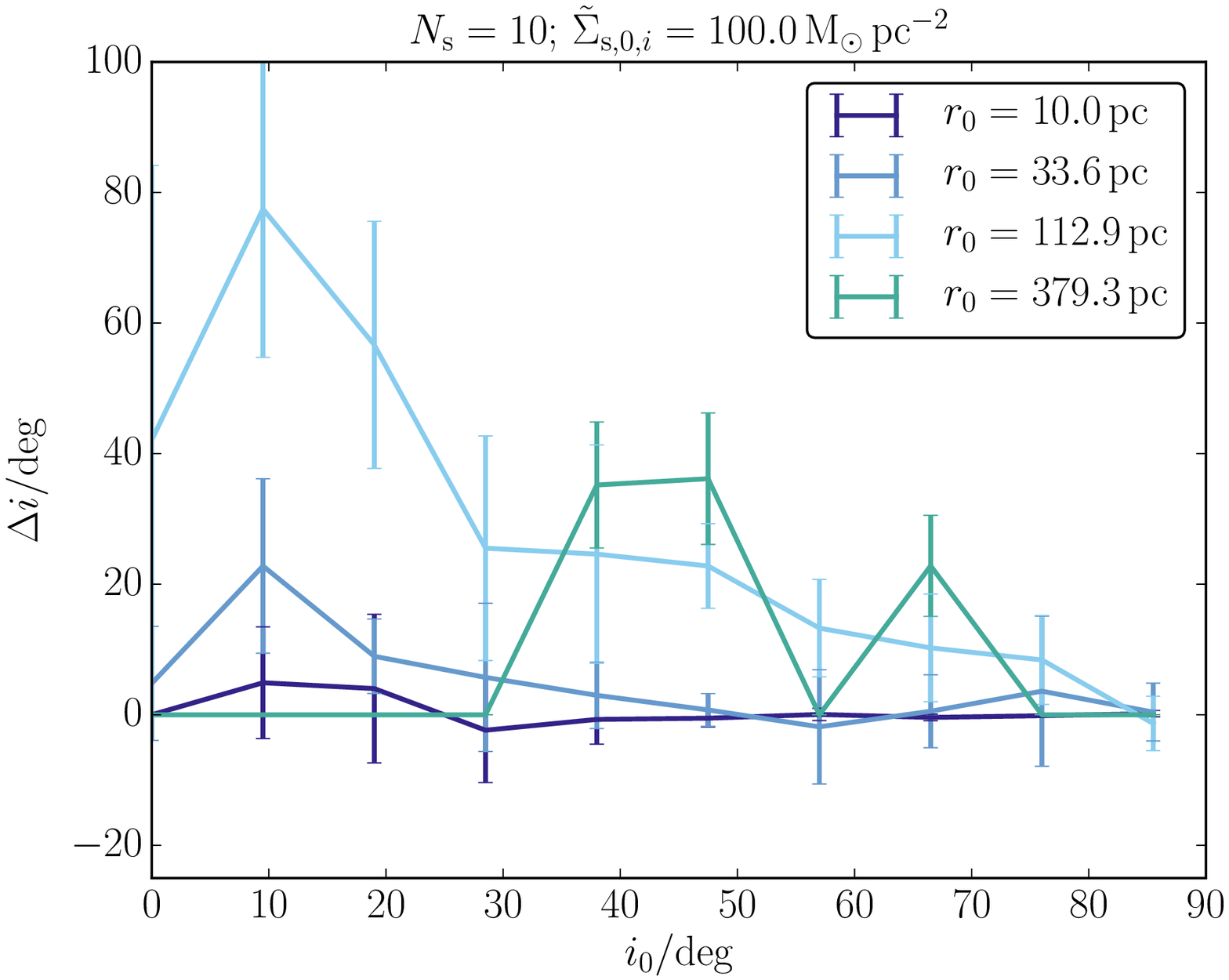}
}
\caption{\small The change of the orbital inclination over the integration averaged over the $N_\mathrm{MC}=100$ realizations, as a function of the initial inclination $i_0$, for initially prograde orbits. The top and bottom panels correspond to different parameters, indicated in the titles. The standard deviations of the change of the inclination are shown with the error bars. Results for different initial radii are shown, indicated in the legends. }
\label{fig:delta_i}
\end{figure}

A caveat in our results is that our quoted inclination is the {\it initial} orbital inclination of the binary with respect to the Galactic plane. In addition to stochastically changing the orbital angular momentum, transient nuclear spiral arms can also affect the orbital inclination (see, e.g., \F\,\ref{fig:examples_i_test01}). To simplify our analysis, a changing inclination was not taken into account. 

In \F\,\ref{fig:delta_i}, we show, for two values of the parameter $N_\mathrm{s}$, the  change of the orbital inclination over the integration averaged over the $N_\mathrm{MC}=100$ realizations, as a function of the initial inclination $i_0$, for initially prograde orbits. The standard deviations of the change of the inclination are shown with the error bars. The data are averaged over the initial orbital angular momentum, $J/J_\mathrm{c}$, and results for different initial radii are shown. 

Depending on the parameters, $\Delta i$ can be large, even approaching $90^\circ$, although the standard deviation can be substantial as well. There is a strong dependence on the initial radius and inclination, with larger changes for larger radii, and smaller initial inclinations. These effects could modify the true inclination dependence of the disruption rates, since objects on initially nearly coplanar can be transferred to highly inclined orbits, for which the disruption rates are lower. Effectively, this could smear out the inclination dependence shown in \F\,\ref{fig:Gamma_inclination}. 

Another caveat, mentioned in \S\,\ref{sect:rates:meth}, is that spherical symmetry was assumed in the rate computations, which is clearly not self-consistent with the underlying nuclear spiral potential (the latter was derived assuming a razor-thin surface density distribution). Also, the binary was treated as a point mass, i.e., the quadrupole moment of the binary was neglected. These caveats should be addressed in future, more detailed work. Another aspect that merits future investigation is the effect of triaxial potentials. 

\subsection{Implications for S-stars}
Binary model `M3' represents the origin of S-stars (B-type stars with masses of $\sim 8 \, \msun$). Their disruption rates by nuclear spiral arms are similar to the HVS model. The dependence of the rates on the inclination is the same compared to HVS, apart from an overall different normalization factor. This may appear to be inconsistent with the S-stars, which are observed to be highly isotropic in their orbital distribution (e.g., \citealt{2005ApJ...628..246E,2008ApJ...689.1044G}). However, the time-scale for the S-stars orbits to randomize their orientation, i.e., the `vector' or  `2d' resonant relaxation time-scale (e.g., \citealt[eq. 5.237]{2013degn.book.....M}) is
\begin{align}
t_\mathrm{VRR} = t_\mathrm{2dRR} \approx \frac{1}{2} \frac{M_\bullet}{M_\star \sqrt{N}} P \sim 3\, \mathrm{Myr},
\end{align}
which is short compared to the typical S-star lifetime of $\sim 40\,\mathrm{Myr}$ (e.g., Table~\ref{table:binary_models}). Here, we assumed a semimajor axis of $a=10\,\mathrm{mpc}$, a background stellar mass of $M_\star=1\,\msun$, and $N=10^3$. Therefore, our result of the inclination dependence of the disruption rate is not inconsistent with the observed isotropy of the S-star orbits.

\subsection{Binary MBHs}
\label{sect:discussion:binary_MBH}
Relaxation by nuclear spiral arms can also be important for binary MBH systems. Stars approaching a binary MBH at a close distance statistically extract energy from the binary MBH orbit, gradually shrinking the latter, whereas the stars are ejected through the slingshot effect (\citealt{1974ApJ...190..253S}). In the standard model, this process halts when the binary MBH reaches a separation of order 1 parsec, giving rise to the `final parsec problem' (see,.e.g, \citealt{2005LRR.....8....8M} for a review). Among the proposed mechanisms to solve this problem (see, e.g., \citealt{2014arXiv1411.1762V} for an overview), is enhanced relaxation by massive perturbers \citep{2008ApJ...677..146P}. Similarly, nuclear spiral arms can drive stars into the `loss cone' of the binary MBH, thereby accelerating its coalescence. 

\begin{figure}
\center
\iftoggle{ApJFigs}{
\includegraphics[scale = 0.45, trim = 5mm 0mm 0mm 0mm]{LC_flux_index_binary_SBH_1_index_r_lc_0_index_par2_0_index_par4_0_index_par5_1_test06_analysis_name_an03.eps}
}{
\includegraphics[scale = 0.45, trim = 5mm 0mm 0mm 0mm]{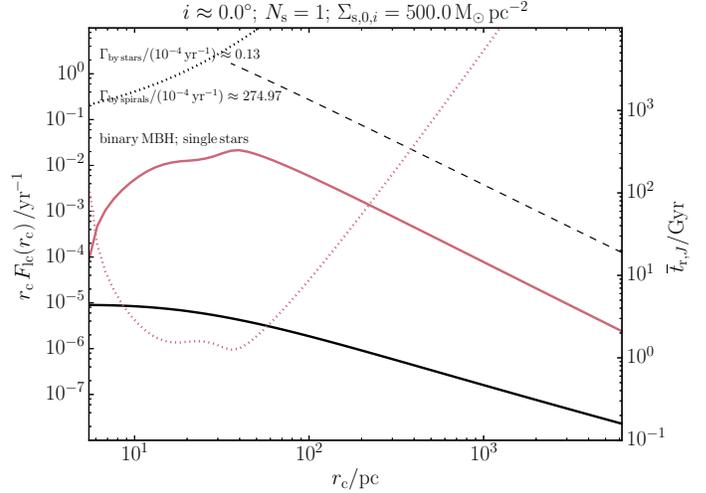}
}
\caption{\small The differential loss-cone rate for encounters of single stars with a binary MBH, assuming a loss-cone radius of $r_\mathrm{lc} = 0.5 \, \mathrm{pc}$. As in Figs\,\ref{fig:LC_flux1} and \ref{fig:LC_flux2}, black (red) lines apply to relaxation by single stars (nuclear spiral arms). The associated relaxation time-scales $\overline{t}_\mathrm{r}(\Ec) = 1/\overline{\mathcal{D}}_J(\Ec)$ are shown with the black and red dotted lines; in this case, the right-hand axes apply. The black dashed lines correspond to the full lone-cone regime (cf. equation~\ref{eq:F_lc_approx_full}); note that the full loss-cone regime is not attained even in the case of relaxation by nuclear spiral arms.}
\label{fig:LC_flux3}
\end{figure}

Here, we only briefly consider the effects of nuclear spiral arms, and do not consider the evolution of the binary MBH in response to stellar encounters. Also, for simplicity, we assume the same background population of stars as before (cf. \S\,\ref{sect:rates:meth}), whereas in a more realistic situation, the stellar distribution is dynamic (in particular, it is depleted near the MBH). Note, however, that in our case the enhancement by nuclear spiral arms is at large radii of several hundred parsec, at which stars could be supplied from the Galactic Bulge. Here, we adopt a loss-cone radius of $r_\mathrm{lc} = 0.5 \, \mathrm{pc}$ (approximately the separation of a `stalled' binary MBH).

In \F\,\ref{fig:LC_flux3}, we show the differential rates at which stars are supplied to the binary MBH, assuming relaxation driven by stars (solid black line), and relaxation driven by nuclear spiral arms (solid red line). The integrated rates are $\approx 1.3\times 10^{-5}$ and $\approx 2.7 \times 10^{-2}\, \mathrm{yr^{-1}}$ for relaxation by stars and spirals, respectively. This suggests that nuclear spiral arms are extremely efficient at supplying stars to the binary MBH, although it remains to be seen how this rate is affected when binary MBH evolution and the depletion of stars are taken into account.

\subsection{Other implications}
\label{sect:discussion:other}
In principle, the perturbations from nuclear spiral arms act not only on stars and stellar binaries, but also on any object or system in similar regions as the nuclear spiral arms. In particular, the orbits of stellar clusters like the Quintuplet cluster could be excited to high eccentricities, leading to the breakup of clusters and building up nuclear star clusters \citep{1975ApJ...196..407T}. Similarly, gas clouds could be excited to near-radial orbits, supplying nuclear star clusters with gas for star formation (e.g., \citealt{2004ApJ...605L..13M,2006ApJ...650L..37M}). In the GC in particular, such processes may have contributed to the formation of the innermost O stars, which are concentrated in a disk-like structure (e.g., \citealt{2000MNRAS.317..348G,2003ApJ...590L..33L,2006ApJ...641..891T,2006ApJ...643.1011P,2009ApJ...690.1463L,2009ApJ...697.1741B}).

Also, planets and planetesimals could be driven to loss-cone orbits, giving rise to their tidal breakup, and relatively low-intensity flares (\citealt{2012MNRAS.421.1315Z}; in the case of rocky planetesimals, the injected energy arises from friction of small tidally disrupted fragments with the surrounding ambient gas). In the case of the GC, relaxation of planetesimals by stars in this model is, however, already sufficient to account for the observed flaring rate of about once per day, irrespective whether the planetesimals formed in a large-scale disk around the MBH, or in disks around stars \citep{2015MNRAS.446..710H}.

\section{Conclusions}
\label{sect:conclusions}
We have considered diffusion of single and binary stars into the loss cone of an MBH, driven by perturbations from transient nuclear spiral arm structures which are observed in the centers of galaxies. Disruptions of single stars are believed to be observable as TDEs. Disruptions of binary stars are thought to result in a highly eccentric star on a tight orbit around the MBH, and a star with high velocity escaping from the galaxy \citep{1988Natur.331..687H}. The latter mechanism is currently favored to explain the origin of the S-stars in the GC, and HVSs. 

Focussing on the GC, we adopted a simple model for the potential associated with nuclear spiral arms, and we carried out test-particle integrations within this potential to model the dynamics of single and binary stars in regions of the central few hundred pc around the GC. From these integrations, we extracted energy and angular-momentum diffusion coefficients and relaxation time-scales due to transient nuclear spiral arms, for different combinations of plausible parameters. Using the angular-momentum diffusion coefficients, we computed disruption rates of single stars and binaries, giving estimates of the formation rates of S-stars and HVSs due nuclear spiral arms in the GC. Our conclusions are as follows.

\medskip \noindent 1. Nuclear spiral arms are ineffective at driving orbital diffusion in the regions of the central $\lesssim 10\,\mathrm{pc}$ around the MBH, where the potential is dominated by the MBH and the stellar bulge. At larger radii, angular-momentum relaxation time-scales can be as short as $\sim 500 \, \mathrm{Myr}$ depending on the inclination, the spiral arm surface density $\tilde{\Sigma}_{\mathrm{s},0,i}$, and the number of spiral arm events $N_\mathrm{s}$. Relaxation in energy is typically an order-of-magnitude less efficient compared to relaxation in angular momentum.

\medskip \noindent 2. The relaxation rate due to nuclear spiral arms depends on the inclination of the orbit of the binary with respect to the plane of the Galaxy (we assumed that the nuclear spiral arm structures are confined within the plane of the Galaxy). Generally, a higher inclination implies less efficient relaxation. However, if the inclination is low and spiral perturbations are strong, then the binary can become unbound from the GC, in which case the binary can no longer be disrupted by the MBH. Similarly, if the spiral arm surface density $\tilde{\Sigma}_{\mathrm{s},0,i}$ is high and/or the number of spiral arm events $N_\mathrm{s}$ is large, then spiral perturbations can drive the unbinding of the binary from the GC. Therefore, the disruption rates do not always increase monotonically with these parameters. 

\medskip \noindent 3. The calculated disruption rates of massive binaries (binary masses of $\sim 8\,\msun$ and higher) are typically a few times $10^{-7} \, \mathrm{yr^{-1}}$, which is a factor of $\sim 10$ higher compared to relaxation by stars only (cf. Table~\ref{table:rates}). The largest enhancement is a factor of $\sim 25$, assuming an inclination of $i=9.5^\circ$, $N_\mathrm{s}=1$ spiral arm events, and a surface density of $\tilde{\Sigma}_{\mathrm{s},0,i}=1000\,\msun\,\mathrm{pc^{-2}}$. Our rates are similar to the rates found for relaxation by massive perturbers \citep{2007ApJ...656..709P}, indicating that nuclear spiral arms can act in conjunction with massive perturbers, in particular GMCs, to increase the injection rate of binaries to MBHs by orders of magnitude. The numbers of predicted HVSs and S-stars in our models are consistent with the observed numbers within factors of a few (assuming a binary fraction of 0.6 for HVS progenitors, and 0.8 for S-star progenitors).

\medskip \noindent 4. The disruption rates are moderately dependent on the inclination of the progenitor binary, peaking around $10^\circ$ (cf. \F\,\ref{fig:Gamma_inclination}). The difference in the disruption rate between nearly coplanar and perpendicular orbits is a factor of $\sim 2$. Our model therefore provides a novel potential source for anisotropy in the distribution of HVSs, although further, more detailed study, is still warranted. There are no implications for S-star-like stars, since the reorientation time-scale for the latter is short compared to the typical S-star lifetime. 

\medskip \noindent 5. Our TDE rates are up to $20\%$ higher compared to relaxation by single stars only. Although the enhancement by nuclear spiral arms is not as significant as for binaries, it complements other models for enhanced TDE rates.

\medskip \noindent 6. In addition to enhancing TDE, S-star, and HVS rates, nuclear spiral arms could also accelerate the coalescence of binary MBHs. Other implications include the disruption of clusters, building up nuclear star clusters, and supplying the inner regions of nuclear stars cluster with gas, triggering high-mass star formation. Also, the disruption rate of planets and planetesimals could be enhanced.

\section*{Acknowledgements}
We thank Scott Tremaine, Ben Bar-Or and Nicholas Stone for stimulating discussions and comments on the manuscript, and the referee for a thorough review. ASH and HBP acknowledge support from the Israeli I-CORE center for astrophysics under ISF grant 1829/12; HBP acknowledges support from the Asher space research institute in the Technion. ASH gratefully acknowledges support from the Institute for Advanced Study, and NASA grant NNX14AM24G.

\bibliographystyle{apj}
\bibliography{literature}

\appendix

\section{Loss-cone diffusion coefficients for relaxation driven by stars}
\label{app:dif_stars}
As discussed in \S\,\ref{sect:rates:meth}, the orbit-averaged angular-momentum diffusion coefficient that describes the diffusion rate in angular-momentum space by stars, $\bar{\mu}_\mathrm{stars}(\Ec)$, can be computed from the distribution function $f(\Ec)$ and the potential $\psi(r)$. The formal definition of the diffusion coefficient is
\begin{align}
\bar{\mu}_\mathrm{stars}(\Ec) &= \frac{2}{P_\mathrm{r}(\Ec)} \int_0^{\psi^{-1}(\Ec)} \frac{\mathrm{d}r}{v_\mathrm{r}(r,\Ec,0)} \lim_{\Rc\rightarrow0} \frac{ \langle (\Delta \Rc)^2 \rangle}{(2\Rc)},
\end{align}
where $\langle (\Delta \Rc)^2 \rangle$ is the second-order diffusion coefficient in $\Rc\equiv (J/J_\mathrm{c})^2$ describing angular-momentum relaxation by stars. Using standard expressions for $\langle (\Delta \Rc)^2 \rangle$ (e.g., \citealt{1978ApJ...226.1087C}), $\bar{\mu}_\mathrm{stars}(\Ec)$ can be expressed in terms of $f(\Ec)$ as
\begin{align}
\bar{\mu}_\mathrm{stars}(\Ec) &= \frac{32 \pi^2 G^2}{3P_\mathrm{r}(\Ec)J_\mathrm{c}^2(\Ec)} m_\star^2 \ln(\Lambda) \left [ 3 \bar{I}_{1/2}(\Ec) + 2 \bar{I}_{0}(\Ec) - \bar{I}_{3/2}(\Ec) \right ],
\label{eq:mu_e}
\end{align}
where $\ln (\Lambda) = \ln[M_\bullet/(2M_\star)]$ is the Coulomb logarithm, and the integrals are given by 
\begin{align}
\nonumber \bar{I}_{0}(\Ec) &= \int_0^{\psi^{-1}(\Ec)} \frac{\mathrm{d} r \, r^2}{\sqrt{2[\psi(r)-\Ec)]}} \int_{-\infty}^\Ec \mathrm{d} \Ec' \, f (\Ec'); \\
\bar{I}_{n/2}(\Ec) &= \int_0^{\psi^{-1}(\Ec)} \frac{\mathrm{d} r \, r^2}{\sqrt{2[\psi(r)-\Ec]}} \int_{\Ec}^{\psi(r)} \mathrm{d} \Ec' \, \left ( \frac{\psi(r)-\Ec'}{\psi(r)-\Ec} \right )^{n/2} f (\Ec' ).
\label{eq:mu_e_I}
\end{align}

\section{Disruption rates and expected number of HVSs and S-stars}
\label{app:rates}
Table~\ref{table:ratesall} gives a complete list of the disruption rates (cf. the top part of Table~\ref{table:rates}). A complete list of the expected number of HVSs and S-stars (cf. the bottom part of Table~\ref{table:rates}) in given in Table~\ref{table:numbersall}.

\begin{table*}
\begin{threeparttable}
\scriptsize
\begin{minipage}{.45\linewidth}
\begin{tabular}{cccccccc}
\multicolumn{8}{c}{\bf{Prograde}} \\
\toprule
& & & \multicolumn{5}{c}{$\Gamma/(10^{-6} \, \mathrm{yr^{-1}})$} \\
\midrule
& & & Single & \multicolumn{4}{c}{Binary} \\
& & & & M1 & M2 & M3 & M4 \\
\midrule
\multicolumn{3}{l}{By stars} & 24.5 & 6.8 & 0.130 & 0.031 & 0.002\ \\
\midrule
\multicolumn{3}{l}{By stars and spirals} \\
$i$ & $N_\mathrm{s}$ & $\tilde{\Sigma}_{\mathrm{s},0,i}$ & & & & & \\
(deg) & & $(\msun\,\mathrm{pc^{-2}}$) \\
\midrule
0.0 & 1 & 100.0 & 25.5 & 259.3 & 0.237 & 0.091 & 0.011\ \\
0.0 & 1 & 500.0 & 29.5 & 504.6 & 0.518 & 0.227 & 0.027\ \\
0.0 & 1 & 1000.0 & 30.3 & 459.8 & 0.604 & 0.264 & 0.029\ \\
0.0 & 10 & 100.0 & 26.1 & 358.6 & 0.295 & 0.123 & 0.016\ \\
0.0 & 10 & 500.0 & 29.4 & 384.8 & 0.498 & 0.211 & 0.023\ \\
0.0 & 10 & 1000.0 & 29.8 & 356.5 & 0.511 & 0.214 & 0.023\ \\
9.5 & 1 & 100.0 & 26.0 & 181.4 & 0.228 & 0.077 & 0.008\ \\
9.5 & 1 & 500.0 & 28.0 & 457.4 & 0.377 & 0.159 & 0.020\ \\
9.5 & 1 & 1000.0 & 29.5 & 601.1 & 0.516 & 0.231 & 0.030\ \\
9.5 & 10 & 100.0 & 27.6 & 359.4 & 0.323 & 0.126 & 0.015\ \\
9.5 & 10 & 500.0 & 29.7 & 636.2 & 0.540 & 0.246 & 0.033\ \\
9.5 & 10 & 1000.0 & 30.4 & 673.8 & 0.636 & 0.298 & 0.038\ \\
19.0 & 1 & 100.0 & 26.0 & 164.8 & 0.223 & 0.074 & 0.007\ \\
19.0 & 1 & 500.0 & 28.2 & 446.9 & 0.387 & 0.161 & 0.020\ \\
19.0 & 1 & 1000.0 & 29.5 & 583.0 & 0.517 & 0.231 & 0.030\ \\
19.0 & 10 & 100.0 & 27.7 & 311.5 & 0.329 & 0.125 & 0.014\ \\
19.0 & 10 & 500.0 & 29.7 & 568.2 & 0.550 & 0.244 & 0.030\ \\
19.0 & 10 & 1000.0 & 30.1 & 610.2 & 0.592 & 0.271 & 0.034\ \\
28.5 & 1 & 100.0 & 25.8 & 146.9 & 0.211 & 0.069 & 0.007\ \\
28.5 & 1 & 500.0 & 28.1 & 405.4 & 0.382 & 0.157 & 0.019\ \\
28.5 & 1 & 1000.0 & 29.5 & 519.6 & 0.500 & 0.215 & 0.026\ \\
28.5 & 10 & 100.0 & 27.1 & 260.9 & 0.293 & 0.108 & 0.012\ \\
28.5 & 10 & 500.0 & 29.3 & 494.0 & 0.492 & 0.210 & 0.025\ \\
28.5 & 10 & 1000.0 & 30.0 & 535.2 & 0.554 & 0.244 & 0.029\ \\
38.0 & 1 & 100.0 & 25.5 & 127.3 & 0.196 & 0.062 & 0.006\ \\
38.0 & 1 & 500.0 & 27.7 & 361.1 & 0.347 & 0.139 & 0.016\ \\
38.0 & 1 & 1000.0 & 29.1 & 474.0 & 0.461 & 0.197 & 0.024\ \\
38.0 & 10 & 100.0 & 26.8 & 218.4 & 0.276 & 0.099 & 0.010\ \\
38.0 & 10 & 500.0 & 29.1 & 455.6 & 0.468 & 0.198 & 0.023\ \\
38.0 & 10 & 1000.0 & 29.4 & 486.6 & 0.497 & 0.216 & 0.026\ \\
47.5 & 1 & 100.0 & 25.3 & 110.8 & 0.183 & 0.057 & 0.005\ \\
47.5 & 1 & 500.0 & 27.5 & 315.2 & 0.325 & 0.126 & 0.014\ \\
47.5 & 1 & 1000.0 & 28.7 & 401.8 & 0.421 & 0.174 & 0.020\ \\
47.5 & 10 & 100.0 & 26.9 & 212.0 & 0.273 & 0.096 & 0.010\ \\
47.5 & 10 & 500.0 & 28.9 & 422.1 & 0.444 & 0.184 & 0.021\ \\
47.5 & 10 & 1000.0 & 29.3 & 450.0 & 0.487 & 0.208 & 0.024\ \\
57.0 & 1 & 100.0 & 25.2 & 100.8 & 0.176 & 0.054 & 0.005\ \\
57.0 & 1 & 500.0 & 27.1 & 291.1 & 0.313 & 0.121 & 0.013\ \\
57.0 & 1 & 1000.0 & 28.4 & 404.1 & 0.402 & 0.167 & 0.019\ \\
57.0 & 10 & 100.0 & 26.0 & 184.3 & 0.225 & 0.078 & 0.008\ \\
57.0 & 10 & 500.0 & 28.8 & 385.0 & 0.439 & 0.178 & 0.019\ \\
57.0 & 10 & 1000.0 & 29.7 & 449.3 & 0.506 & 0.212 & 0.024\ \\
66.5 & 1 & 100.0 & 25.0 & 95.6 & 0.170 & 0.051 & 0.005\ \\
66.5 & 1 & 500.0 & 26.9 & 259.9 & 0.296 & 0.112 & 0.012\ \\
66.5 & 1 & 1000.0 & 28.2 & 383.4 & 0.384 & 0.157 & 0.018\ \\
66.5 & 10 & 100.0 & 25.8 & 183.0 & 0.228 & 0.079 & 0.008\ \\
66.5 & 10 & 500.0 & 29.2 & 415.3 & 0.470 & 0.194 & 0.021\ \\
66.5 & 10 & 1000.0 & 29.5 & 401.1 & 0.482 & 0.199 & 0.022\ \\
76.0 & 1 & 100.0 & 24.9 & 85.1 & 0.163 & 0.048 & 0.004\ \\
76.0 & 1 & 500.0 & 26.6 & 245.6 & 0.277 & 0.102 & 0.011\ \\
76.0 & 1 & 1000.0 & 28.1 & 362.8 & 0.370 & 0.151 & 0.017\ \\
76.0 & 10 & 100.0 & 25.3 & 151.6 & 0.205 & 0.071 & 0.007\ \\
76.0 & 10 & 500.0 & 28.6 & 372.0 & 0.401 & 0.160 & 0.018\ \\
76.0 & 10 & 1000.0 & 29.6 & 453.8 & 0.488 & 0.205 & 0.023\ \\
85.5 & 1 & 100.0 & 24.9 & 79.4 & 0.162 & 0.047 & 0.004\ \\
85.5 & 1 & 500.0 & 26.3 & 245.1 & 0.268 & 0.098 & 0.011\ \\
85.5 & 1 & 1000.0 & 28.2 & 381.8 & 0.383 & 0.156 & 0.018\ \\
85.5 & 10 & 100.0 & 25.5 & 155.3 & 0.223 & 0.077 & 0.007\ \\
85.5 & 10 & 500.0 & 27.8 & 380.1 & 0.389 & 0.162 & 0.018\ \\
85.5 & 10 & 1000.0 & 29.5 & 438.6 & 0.492 & 0.205 & 0.023\ \\
\midrule
\multicolumn{3}{l}{Max. enhancement} & 1.2 & 99.1 & 4.9 & 9.6 & 24.6\ \\
\bottomrule
\end{tabular}
\end{minipage}
\begin{minipage}{.45\linewidth}
\begin{tabular}{cccccccc}
\multicolumn{8}{c}{\bf{Retrograde}} \\
\toprule
& & & \multicolumn{5}{c}{$\Gamma/(10^{-6} \, \mathrm{yr^{-1}})$} \\
\midrule
& & & Single & \multicolumn{4}{c}{Binary} \\
& & & & M1 & M2 & M3 & M4 \\
\midrule
\multicolumn{3}{l}{By stars} & 24.5 & 6.8 & 0.130 & 0.031 & 0.002\ \\
\midrule
\multicolumn{3}{l}{By stars and spirals} \\
$i$ & $N_\mathrm{s}$ & $\tilde{\Sigma}_{\mathrm{s},0,i}$ & & & & & \\
(deg) & & $(\msun\,\mathrm{pc^{-2}}$) \\
\midrule
90.0 & 1 & 100.0 & 24.8 & 72.2 & 0.160 & 0.047 & 0.004\ \\
90.0 & 1 & 500.0 & 26.3 & 251.7 & 0.269 & 0.100 & 0.011\ \\
90.0 & 1 & 1000.0 & 28.1 & 356.0 & 0.373 & 0.151 & 0.017\ \\
90.0 & 10 & 100.0 & 25.3 & 174.7 & 0.188 & 0.063 & 0.007\ \\
90.0 & 10 & 500.0 & 28.0 & 370.0 & 0.367 & 0.149 & 0.017\ \\
90.0 & 10 & 1000.0 & 29.1 & 388.4 & 0.458 & 0.191 & 0.021\ \\
100.0 & 1 & 100.0 & 24.9 & 77.8 & 0.162 & 0.047 & 0.004\ \\
100.0 & 1 & 500.0 & 26.4 & 261.2 & 0.277 & 0.104 & 0.011\ \\
100.0 & 1 & 1000.0 & 28.2 & 378.1 & 0.379 & 0.154 & 0.018\ \\
100.0 & 10 & 100.0 & 25.6 & 145.7 & 0.211 & 0.069 & 0.007\ \\
100.0 & 10 & 500.0 & 28.0 & 369.3 & 0.374 & 0.152 & 0.017\ \\
100.0 & 10 & 1000.0 & 29.2 & 429.9 & 0.473 & 0.197 & 0.022\ \\
110.0 & 1 & 100.0 & 24.9 & 91.3 & 0.164 & 0.049 & 0.004\ \\
110.0 & 1 & 500.0 & 26.7 & 267.9 & 0.287 & 0.108 & 0.012\ \\
110.0 & 1 & 1000.0 & 28.3 & 370.7 & 0.391 & 0.158 & 0.018\ \\
110.0 & 10 & 100.0 & 26.1 & 166.7 & 0.242 & 0.081 & 0.008\ \\
110.0 & 10 & 500.0 & 28.1 & 382.5 & 0.386 & 0.157 & 0.018\ \\
110.0 & 10 & 1000.0 & 29.2 & 427.9 & 0.480 & 0.201 & 0.023\ \\
120.0 & 1 & 100.0 & 25.0 & 97.9 & 0.168 & 0.051 & 0.005\ \\
120.0 & 1 & 500.0 & 27.1 & 283.0 & 0.312 & 0.119 & 0.013\ \\
120.0 & 1 & 1000.0 & 28.6 & 394.4 & 0.412 & 0.169 & 0.019\ \\
120.0 & 10 & 100.0 & 26.3 & 203.2 & 0.255 & 0.090 & 0.009\ \\
120.0 & 10 & 500.0 & 28.6 & 410.6 & 0.425 & 0.179 & 0.020\ \\
120.0 & 10 & 1000.0 & 29.8 & 432.6 & 0.513 & 0.212 & 0.023\ \\
130.0 & 1 & 100.0 & 25.2 & 108.6 & 0.179 & 0.055 & 0.005\ \\
130.0 & 1 & 500.0 & 27.4 & 316.5 & 0.325 & 0.128 & 0.014\ \\
130.0 & 1 & 1000.0 & 28.8 & 408.0 & 0.429 & 0.178 & 0.020\ \\
130.0 & 10 & 100.0 & 26.5 & 197.7 & 0.243 & 0.084 & 0.009\ \\
130.0 & 10 & 500.0 & 29.2 & 405.1 & 0.446 & 0.181 & 0.020\ \\
130.0 & 10 & 1000.0 & 29.4 & 420.1 & 0.490 & 0.206 & 0.023\ \\
140.0 & 1 & 100.0 & 25.5 & 122.2 & 0.192 & 0.060 & 0.006\ \\
140.0 & 1 & 500.0 & 27.8 & 341.3 & 0.350 & 0.138 & 0.016\ \\
140.0 & 1 & 1000.0 & 29.2 & 453.5 & 0.466 & 0.197 & 0.023\ \\
140.0 & 10 & 100.0 & 27.4 & 239.9 & 0.303 & 0.110 & 0.011\ \\
140.0 & 10 & 500.0 & 29.1 & 434.7 & 0.469 & 0.194 & 0.022\ \\
140.0 & 10 & 1000.0 & 30.0 & 458.2 & 0.551 & 0.236 & 0.026\ \\
150.0 & 1 & 100.0 & 25.8 & 148.5 & 0.212 & 0.069 & 0.007\ \\
150.0 & 1 & 500.0 & 28.0 & 389.0 & 0.371 & 0.151 & 0.018\ \\
150.0 & 1 & 1000.0 & 29.4 & 498.1 & 0.500 & 0.216 & 0.026\ \\
150.0 & 10 & 100.0 & 27.4 & 256.7 & 0.305 & 0.110 & 0.011\ \\
150.0 & 10 & 500.0 & 29.1 & 483.5 & 0.485 & 0.211 & 0.025\ \\
150.0 & 10 & 1000.0 & 29.9 & 528.9 & 0.563 & 0.250 & 0.030\ \\
160.0 & 1 & 100.0 & 26.0 & 185.2 & 0.232 & 0.079 & 0.008\ \\
160.0 & 1 & 500.0 & 28.3 & 443.1 & 0.404 & 0.169 & 0.021\ \\
160.0 & 1 & 1000.0 & 29.5 & 550.8 & 0.517 & 0.227 & 0.028\ \\
160.0 & 10 & 100.0 & 27.5 & 309.5 & 0.325 & 0.123 & 0.014\ \\
160.0 & 10 & 500.0 & 29.8 & 564.2 & 0.530 & 0.232 & 0.029\ \\
160.0 & 10 & 1000.0 & 30.3 & 585.8 & 0.599 & 0.271 & 0.033\ \\
170.0 & 1 & 100.0 & 26.1 & 201.4 & 0.234 & 0.081 & 0.008\ \\
170.0 & 1 & 500.0 & 28.1 & 457.7 & 0.381 & 0.160 & 0.021\ \\
170.0 & 1 & 1000.0 & 29.3 & 569.8 & 0.505 & 0.226 & 0.029\ \\
170.0 & 10 & 100.0 & 28.0 & 339.5 & 0.350 & 0.133 & 0.015\ \\
170.0 & 10 & 500.0 & 29.7 & 578.2 & 0.542 & 0.241 & 0.030\ \\
170.0 & 10 & 1000.0 & 30.4 & 644.3 & 0.635 & 0.296 & 0.037\ \\
180.0 & 1 & 100.0 & 26.0 & 193.8 & 0.222 & 0.075 & 0.008\ \\
180.0 & 1 & 500.0 & 28.2 & 470.6 & 0.384 & 0.160 & 0.021\ \\
180.0 & 1 & 1000.0 & 29.5 & 596.9 & 0.517 & 0.231 & 0.030\ \\
180.0 & 10 & 100.0 & 28.4 & 348.0 & 0.364 & 0.138 & 0.015\ \\
180.0 & 10 & 500.0 & 30.1 & 639.8 & 0.593 & 0.270 & 0.034\ \\
180.0 & 10 & 1000.0 & 30.2 & 640.0 & 0.603 & 0.279 & 0.036\ \\
\midrule
\multicolumn{3}{l}{Maximum enhancement} & 1.2 & 94.8 & 4.9 & 9.6 & 24.0\ \\
\bottomrule
\end{tabular}
\end{minipage}
\centering
\caption{ Disruption rates $\Gamma$ of single stars (`Single') and binary stars (`Binary') due to relaxation driven by stars (first data row), and driven by nuclear spiral arms (in addition to stars; all other data rows). In the latter case, each row corresponds to a different inclination $i$ (expressed in degrees), number of spiral arm events $N_\mathrm{s}$, and (constant) spiral arm surface density $\tilde{\Sigma}_{\mathrm{s},0,i}$ (in units of $\msun \, \mathrm{pc^{-2}}$). The left (right) table applies to prograde (retrograde) orbits. For binary stars, four models, `M1' through `M4', are assumed (cf. Table~\ref{table:binary_models}). Details of the rate calculations are given in \S\,\ref{sect:rates:meth}. Bottom rows: the largest enhancement factor of the rates with respect to relaxation by single stars. }
\label{table:ratesall}
\end{threeparttable}
\end{table*}

\begin{table}
\centering
\begin{threeparttable}
\scriptsize
\begin{minipage}{.45\linewidth}
\begin{tabular}{ccccc}
\multicolumn{5}{c}{\bf{Prograde}} \\
\toprule
& & & \multicolumn{2}{c}{$N_\star = \Gamma \, t_\star$} \\
\midrule
& & & \multicolumn{2}{c}{Binary} \\
& & & M2 & M3 \\
\midrule
\multicolumn{3}{l}{By stars} & 23 & 1\ \\
\midrule
\multicolumn{3}{l}{By stars and spirals} \\
$i$ & $N_\mathrm{s}$ & $\tilde{\Sigma}_{\mathrm{s},0,i}$ & & \\
(deg) & & $(\msun\,\mathrm{pc^{-2}}$) \\
\midrule
0.0 & 1 & 100.0  & 43 & 3\ \\
0.0 & 1 & 500.0  & 93 & 8\ \\
0.0 & 1 & 1000.0  & 108 & 10\ \\
0.0 & 10 & 100.0  & 53 & 5\ \\
0.0 & 10 & 500.0  & 89 & 8\ \\
0.0 & 10 & 1000.0  & 92 & 8\ \\
9.5 & 1 & 100.0  & 41 & 3\ \\
9.5 & 1 & 500.0  & 68 & 6\ \\
9.5 & 1 & 1000.0  & 93 & 9\ \\
9.5 & 10 & 100.0  & 58 & 5\ \\
9.5 & 10 & 500.0  & 97 & 9\ \\
9.5 & 10 & 1000.0  & 114 & 11\ \\
19.0 & 1 & 100.0  & 40 & 3\ \\
19.0 & 1 & 500.0  & 69 & 6\ \\
19.0 & 1 & 1000.0  & 93 & 9\ \\
19.0 & 10 & 100.0  & 59 & 5\ \\
19.0 & 10 & 500.0  & 99 & 9\ \\
19.0 & 10 & 1000.0  & 106 & 10\ \\
28.5 & 1 & 100.0  & 38 & 3\ \\
28.5 & 1 & 500.0  & 69 & 6\ \\
28.5 & 1 & 1000.0  & 90 & 8\ \\
28.5 & 10 & 100.0  & 52 & 4\ \\
28.5 & 10 & 500.0  & 88 & 8\ \\
28.5 & 10 & 1000.0  & 99 & 9\ \\
38.0 & 1 & 100.0  & 35 & 2\ \\
38.0 & 1 & 500.0  & 62 & 5\ \\
38.0 & 1 & 1000.0  & 83 & 7\ \\
38.0 & 10 & 100.0  & 49 & 4\ \\
38.0 & 10 & 500.0  & 84 & 7\ \\
38.0 & 10 & 1000.0  & 89 & 8\ \\
47.5 & 1 & 100.0  & 33 & 2\ \\
47.5 & 1 & 500.0  & 58 & 5\ \\
47.5 & 1 & 1000.0  & 76 & 6\ \\
47.5 & 10 & 100.0  & 49 & 4\ \\
47.5 & 10 & 500.0  & 80 & 7\ \\
47.5 & 10 & 1000.0  & 87 & 8\ \\
57.0 & 1 & 100.0  & 32 & 2\ \\
57.0 & 1 & 500.0  & 56 & 4\ \\
57.0 & 1 & 1000.0  & 72 & 6\ \\
57.0 & 10 & 100.0  & 40 & 3\ \\
57.0 & 10 & 500.0  & 79 & 7\ \\
57.0 & 10 & 1000.0  & 91 & 8\ \\
66.5 & 1 & 100.0  & 30 & 2\ \\
66.5 & 1 & 500.0  & 53 & 4\ \\
66.5 & 1 & 1000.0  & 69 & 6\ \\
66.5 & 10 & 100.0  & 41 & 3\ \\
66.5 & 10 & 500.0  & 84 & 7\ \\
66.5 & 10 & 1000.0  & 86 & 7\ \\
76.0 & 1 & 100.0  & 29 & 2\ \\
76.0 & 1 & 500.0  & 50 & 4\ \\
76.0 & 1 & 1000.0  & 66 & 6\ \\
76.0 & 10 & 100.0  & 37 & 3\ \\
76.0 & 10 & 500.0  & 72 & 6\ \\
76.0 & 10 & 1000.0  & 88 & 8\ \\
85.5 & 1 & 100.0  & 29 & 2\ \\
85.5 & 1 & 500.0  & 48 & 4\ \\
85.5 & 1 & 1000.0  & 69 & 6\ \\
85.5 & 10 & 100.0  & 40 & 3\ \\
85.5 & 10 & 500.0  & 70 & 6\ \\
85.5 & 10 & 1000.0  & 88 & 8\ \\
\bottomrule
\end{tabular}
\end{minipage}
\begin{minipage}{.45\linewidth}
\begin{tabular}{ccccc}
\multicolumn{5}{c}{\bf{Retrograde}} \\
\toprule
& & & \multicolumn{2}{c}{$N_\star = \Gamma \, t_\star$} \\
\midrule
& & & \multicolumn{2}{c}{Binary} \\
& & & M2 & M3 \\
\midrule
\multicolumn{3}{l}{By stars} & 23 & 1\ \\
\midrule
\multicolumn{3}{l}{By stars and spirals} \\
$i$ & $N_\mathrm{s}$ & $\tilde{\Sigma}_{\mathrm{s},0,i}$ & & \\
(deg) & & $(\msun\,\mathrm{pc^{-2}}$) \\
\midrule
90.0 & 1 & 100.0  & 29 & 2\ \\
90.0 & 1 & 500.0  & 48 & 4\ \\
90.0 & 1 & 1000.0  & 67 & 6\ \\
90.0 & 10 & 100.0  & 34 & 2\ \\
90.0 & 10 & 500.0  & 66 & 6\ \\
90.0 & 10 & 1000.0  & 82 & 7\ \\
100.0 & 1 & 100.0  & 29 & 2\ \\
100.0 & 1 & 500.0  & 50 & 4\ \\
100.0 & 1 & 1000.0  & 68 & 6\ \\
100.0 & 10 & 100.0  & 38 & 3\ \\
100.0 & 10 & 500.0  & 67 & 6\ \\
100.0 & 10 & 1000.0  & 85 & 7\ \\
110.0 & 1 & 100.0  & 29 & 2\ \\
110.0 & 1 & 500.0  & 51 & 4\ \\
110.0 & 1 & 1000.0  & 70 & 6\ \\
110.0 & 10 & 100.0  & 43 & 3\ \\
110.0 & 10 & 500.0  & 69 & 6\ \\
110.0 & 10 & 1000.0  & 86 & 7\ \\
120.0 & 1 & 100.0  & 30 & 2\ \\
120.0 & 1 & 500.0  & 56 & 4\ \\
120.0 & 1 & 1000.0  & 74 & 6\ \\
120.0 & 10 & 100.0  & 46 & 3\ \\
120.0 & 10 & 500.0  & 76 & 7\ \\
120.0 & 10 & 1000.0  & 92 & 8\ \\
130.0 & 1 & 100.0  & 32 & 2\ \\
130.0 & 1 & 500.0  & 58 & 5\ \\
130.0 & 1 & 1000.0  & 77 & 7\ \\
130.0 & 10 & 100.0  & 44 & 3\ \\
130.0 & 10 & 500.0  & 80 & 7\ \\
130.0 & 10 & 1000.0  & 88 & 8\ \\
140.0 & 1 & 100.0  & 34 & 2\ \\
140.0 & 1 & 500.0  & 63 & 5\ \\
140.0 & 1 & 1000.0  & 84 & 7\ \\
140.0 & 10 & 100.0  & 54 & 4\ \\
140.0 & 10 & 500.0  & 84 & 7\ \\
140.0 & 10 & 1000.0  & 99 & 9\ \\
150.0 & 1 & 100.0  & 38 & 3\ \\
150.0 & 1 & 500.0  & 66 & 6\ \\
150.0 & 1 & 1000.0  & 90 & 8\ \\
150.0 & 10 & 100.0  & 55 & 4\ \\
150.0 & 10 & 500.0  & 87 & 8\ \\
150.0 & 10 & 1000.0  & 101 & 9\ \\
160.0 & 1 & 100.0  & 42 & 3\ \\
160.0 & 1 & 500.0  & 72 & 6\ \\
160.0 & 1 & 1000.0  & 93 & 8\ \\
160.0 & 10 & 100.0  & 58 & 5\ \\
160.0 & 10 & 500.0  & 95 & 9\ \\
160.0 & 10 & 1000.0  & 107 & 10\ \\
170.0 & 1 & 100.0  & 42 & 3\ \\
170.0 & 1 & 500.0  & 68 & 6\ \\
170.0 & 1 & 1000.0  & 90 & 8\ \\
170.0 & 10 & 100.0  & 63 & 5\ \\
170.0 & 10 & 500.0  & 97 & 9\ \\
170.0 & 10 & 1000.0  & 114 & 11\ \\
180.0 & 1 & 100.0  & 40 & 3\ \\
180.0 & 1 & 500.0  & 69 & 6\ \\
180.0 & 1 & 1000.0  & 93 & 9\ \\
180.0 & 10 & 100.0  & 65 & 5\ \\
180.0 & 10 & 500.0  & 106 & 10\ \\
180.0 & 10 & 1000.0  & 108 & 10\ \\
\bottomrule
\end{tabular}
\end{minipage}
\caption{ Similar to Table\,\ref{table:ratesall}, here showing the expected number of HVSs (model `M2') and S-stars (model `M3').}
\label{table:numbersall}
\end{threeparttable}
\end{table}

\end{document}